\documentclass[11pt]{article}
\usepackage{dsfont}

\usepackage{amsmath,amssymb,graphicx}
\usepackage{diagbox}
\usepackage{hyperref}
\usepackage[nosort]{cite}

\usepackage{tikz}
\usetikzlibrary{calc}
\usetikzlibrary{patterns}
\usetikzlibrary{decorations.pathreplacing}
\usetikzlibrary{decorations.markings}
\usetikzlibrary{decorations.pathmorphing}
\usetikzlibrary{positioning}
\usetikzlibrary{arrows.meta}

\usepackage{color,xcolor}
\definecolor{darkred}{rgb}{0.65,0.15,0}
\definecolor{newgreen}{rgb}{0.2,0.62,0.14}
\hypersetup{pdfborder={0 0 0},colorlinks=true,urlcolor=blue,citecolor=blue,linkcolor=darkred,linktocpage=true}

\numberwithin{equation}{section}

\def\nn{\nonumber}

\def\spa#1.#2{\left\langle#1\,#2\right\rangle}
\def\spb#1.#2{\left[#1\,#2\right]}

\def\ep{\epsilon}

\def\ap{\alpha'}

\newcommand{\Bhat}{\widehat B}
\newcommand{\Jhat}{\widehat J}

\newcommand{\Yhat}{\widehat Y}

\newcommand{\EBR}[3]{{\cal E}\! \left[\begin{smallmatrix}#1\\#2\end{smallmatrix};#3\right]}
\newcommand{\EBRno}[2]{{\cal E}\! \left[\begin{smallmatrix}#1\\#2\end{smallmatrix}\right]}
\newcommand{\EsvBR}[3]{{\cal E}^{\rm sv}\!  \left[\begin{smallmatrix}#1\\#2\end{smallmatrix};#3\right]}

\newcommand{\betaBR}[3]{\beta \! \left[\begin{smallmatrix}#1\\#2\end{smallmatrix};#3\right]}
\newcommand{\betaBRno}[2]{\beta \! \left[\begin{smallmatrix}#1\\#2\end{smallmatrix}\right]}
\newcommand{\betasv}[1]{\beta^{\rm sv}\! \left[\begin{smallmatrix}#1\end{smallmatrix}\right]}
\newcommand{\bsvBR}[3]{\beta^{\rm sv} \! \left[\begin{smallmatrix}#1\\#2\end{smallmatrix};#3\right]}
\newcommand{\bsvBRno}[2]{\beta^{\rm sv}\! \left[\begin{smallmatrix}#1\\#2\end{smallmatrix}\right]}
\newcommand{\alphaBR}[3]{\alpha\! \left[\begin{smallmatrix}#1\\#2\end{smallmatrix};#3\right]}
\newcommand{\alphaBRno}[2]{\alpha\! \left[\begin{smallmatrix}#1\\#2\end{smallmatrix}\right]}

\font\tenshuffle=shuffle10 \font\sevenshuffle=shuffle7 \font\fiveshuffle=shuffle7 at 5pt
\def\shuffle{{%
\def\Dshuffle{\mathbin{\hbox{\tenshuffle\char'001}}}%
\def\Sshuffle{\mathbin{\hbox{\sevenshuffle\char'001}}}%
\def\SSshuffle{\mathbin{\hbox{\fiveshuffle\char'001}}}%
\mathchoice{\Dshuffle}{\Dshuffle}{\Sshuffle}{\SSshuffle}}}

\def\beq{\begin{equation}}
\def\eeq{\end{equation}}
\let\Re\relax
\let\Im\relax
\DeclareMathOperator{\Re}{Re}
\DeclareMathOperator{\Im}{Im}

\newcommand{\eq}{\begin{equation}}
\newcommand{\eqe}{\end{equation}}
\newcommand{\eqa}{\begin{eqnarray}}
\newcommand{\eqae}{\end{eqnarray}}

\newcommand{\bea}{\begin{eqnarray}}
\newcommand{\eea}{\end{eqnarray}}
\newcommand{\dd}{\mathrm{d}}
\newcommand{\vph}{\varphi}

\newcommand{\RR}{\mathbb R}

\newcommand{\CC}{\mathbb C}

\newcommand{\ZZ}{\mathbb Z}
\newcommand{\QQ}{\mathbb Q}

\newcommand{\cform}[1]{\,{\cal C}\!\left[\protect\begin{smallmatrix}#1\protect\end{smallmatrix}\right]}

\newbox\charbox
\newbox\slabox
\def\s#1{{      
        \setbox\charbox=\hbox{$#1$}
        \setbox\slabox=\hbox{$/$}
        \dimen\charbox=\ht\slabox
        \advance\dimen\charbox by -\dp\slabox
        \advance\dimen\charbox by -\ht\charbox
        \advance\dimen\charbox by \dp\charbox
        \divide\dimen\charbox by 2
        \raise-\dimen\charbox\hbox to \wd\charbox{\hss/\hss}
        \llap{$#1$}
}}

\usepackage[shadow,textwidth=2.7cm]{todonotes}
\usepackage{ifthen}
\reversemarginpar
\newcounter{todocounter}

\colorlet{jgcolor}{green!40!white}

\newcommand{\jginline}[2][]{
  \ifthenelse { \equal {#1} {} }
    { \def\temp {#2} }  
    { \def\temp {#1} }   
  \refstepcounter{todocounter}\todo[color=jgcolor,inline,caption={\textbf{\thetodocounter. JG} \temp}]{\textbf{\thetodocounter. JG:} #2}{}}

\colorlet{oscolor}{blue!20!white}

\newcommand{\osinline}[2][]{
  \ifthenelse { \equal {#1} {} }
    { \def\temp {#2} }  
    { \def\temp {#1} }   
  \refstepcounter{todocounter}\todo[color=oscolor,inline,caption={\textbf{\thetodocounter. OS} \temp}]{\textbf{\thetodocounter. OS:} #2}{}}

\colorlet{akcolor}{yellow!40!white}

\newcommand{\akinline}[2][]{
  \ifthenelse { \equal {#1} {} }
    { \def\temp {#2} }  
    { \def\temp {#1} }   
  \refstepcounter{todocounter}\todo[color=akcolor,inline,caption={\textbf{\thetodocounter. AK} \temp}]{\textbf{\thetodocounter. AK:} #2}{}}

\colorlet{bvcolor}{violet!50!white}

\newcommand{\bvinline}[2][]{
  \ifthenelse { \equal {#1} {} }
    { \def\temp {#2} }  
    { \def\temp {#1} }   
  \refstepcounter{todocounter}\todo[color=bvcolor,inline,caption={\textbf{\thetodocounter. BV} \temp}]{\textbf{\thetodocounter. BV:} #2}{}}

\begin{document}

 {\flushright UUITP--43/20\\[15mm]}

\begin{center}

{\LARGE \bf \sc  Towards closed strings as single-valued\\[2mm] open strings at genus one}\\[5mm]

\vspace{6mm}
\normalsize
{\large  Jan E. Gerken${}^{1,2}$, Axel Kleinschmidt${}^{2,3}$, Carlos R. Mafra${}^4$,\\[2mm] Oliver Schlotterer${}^5$ and Bram Verbeek${}^5$}

\vspace{10mm}
${}^{1}${\it Department of Mathematical Sciences\\
Chalmers University of Technology, 41296 Gothenburg, Sweden}
\vskip 1 em
${}^2${\it Max-Planck-Institut f\"{u}r Gravitationsphysik (Albert-Einstein-Institut)\\
Am M\"{u}hlenberg 1, DE-14476 Potsdam, Germany}
\vskip 1 em
${}^3${\it International Solvay Institutes\\
ULB-Campus Plaine CP231, BE-1050 Brussels, Belgium}
\vskip 1 em
${}^4${\it STAG Research Centre and Mathematical Sciences,} \\ {\sl  University of Southampton, Highfield, Southampton SO17 1BJ, UK}

\vskip1em
${}^5${\it  Department of Physics and Astronomy\\
Uppsala University, 75108 Uppsala, Sweden}

\vspace{10mm}

\hrule

\vspace{5mm}

 \begin{tabular}{p{14cm}}

We relate the low-energy expansions of world-sheet integrals in genus-one amplitudes of
open- and closed-string states. The respective expansion coefficients are elliptic multiple zeta values
in the open-string case and non-holomorphic modular forms dubbed ``modular graph forms''
for closed strings. By inspecting the differential equations and degeneration limits of
suitable generating series of genus-one integrals, we identify formal substitution rules mapping the elliptic multiple zeta values of open strings
to the modular graph forms of closed strings.
Based on the properties of these rules, we refer to them as an elliptic single-valued map which generalizes the genus-zero notion of a single-valued map acting on multiple zeta values
seen in tree-level relations between the open and closed string.
\end{tabular}

\vspace{6mm}
\hrule
\end{center}

\thispagestyle{empty}

\newpage
\setcounter{page}{1}

\setcounter{tocdepth}{2}
\tableofcontents


\bigskip

\section{Introduction}

One-loop amplitudes in string theories are computed from integrals over moduli spaces
of punctured genus-one world-sheets. For open and closed strings, the punctures are integrated
over a cylinder boundary and the entire torus, respectively, which is often done
in a low-energy expansion, i.e.\ order by order in the inverse string tension $\alpha'$.
The coefficients of such $\alpha'$-expansions involve special numbers and functions
which have triggered fruitful interactions between number theorists, particle phenomenologists
and string theorists. For instance, elliptic polylogarithms \cite{Lev, BrownLev} and
{\it elliptic multiple zeta values} (eMZVs) \cite{Enriquez:Emzv} were identified to form the number-theoretic
backbone of genus-one open-string integrals \cite{Broedel:2014vla, Broedel:2017jdo, Broedel:2019vjc}.

For the closed string, the analogous genus-one integrals involve non-holomorphic modular
forms \cite{Green:1999pv, Green:2008uj, DHoker:2015gmr} dubbed {\it modular graph forms} (MGFs) \cite{DHoker:2015wxz,DHoker:2016mwo} which inspired
mathematical research lines \cite{Zerbini:2015rss, Brown:2017qwo,Brown:2017qwo2, Panzertalk, Zagier:2019eus}. As a unifying building
block shared by open and closed strings, both eMZVs \cite{Enriquez:Emzv, Broedel:2015hia}
and MGFs \cite{DHoker:2015wxz,DHoker:2016mwo, Gerken:2020yii} can be reduced
to iterated integrals over holomorphic Eisenstein series, or {\it iterated Eisenstein
integrals}. Similar iterated integrals over holomorphic modular forms play a key
role in recent progress on the evaluation of Feynman integrals \cite{Bloch:2013tra,Bloch:2014qca,Adams:2017ejb,Ablinger:2017bjx,Remiddi:2017har,Bourjaily:2017bsb,Broedel:2017kkb,Broedel:2017siw,Adams:2018yfj,Broedel:2018iwv,Adams:2018bsn,Adams:2018kez,Broedel:2018tgw,Blumlein:2018aeq,Broedel:2019hyg,Bogner:2019lfa,Broedel:2019kmn,Duhr:2019rrs,Abreu:2019fgk,Bogner:2019vhn,Walden:2020odh}.
As a main result of this work, we identify infinite families of closed-string
integrals, where the appearance of iterated Eisenstein integrals is in precise correspondence
with those in open-string $\alpha'$-expansions.

More specifically, we give an explicit proposal for a single-valued map at genus one,
mapping individual eMZVs to combinations of iterated Eisenstein integrals and their complex conjugates
which should be contained in Brown's single-valued iterated Eisenstein integrals \cite{Brown:2017qwo,Brown:2017qwo2}.
This generalizes the genus-zero result that the sphere integrals in closed-string tree
amplitudes are single-valued versions of the disk integrals in open-string tree amplitudes
\cite{Schlotterer:2012ny,Stieberger:2013wea,Stieberger:2014hba,Schlotterer:2018abc,Vanhove:2018elu,Brown:2019wna}. The notion of single-valued periods \cite{Brown:2013gia, FrancisLecture} and single-valued integration \cite{Schnetz:2013hqa, Brown:2018omk}
is very general, and in the case of multiple zeta values (MZVs) amounts to evaluating single-valued
polylogarithms \cite{svpolylog} at unit argument. While the single-valued map for the MZVs in tree-level
$\alpha'$-expansions has been pinpointed in \cite{Schnetz:2013hqa, Brown:2013gia}, the genus-one studies
of single-valued maps from mathematical \cite{Zerbini:2015rss, Brown:2017qwo,Brown:2017qwo2} and physical \cite{Broedel:2018izr, Gerken:2018jrq} viewpoints\footnote{See \cite{DHoker:2017zhq, DHoker:2019txf, DHoker:2019xef, Zagier:2019eus, Gerken:2020yii, Vanhove:2020qtt} for recent progress in identifying single-valued MZVs in the degeneration of dihedral MGFs from closed-string genus-one integrals at the cusp.}
have not yet led to a consensus for the single-valued version of individual eMZVs.

Our proposal for single-valued eMZVs can be seen as a correspondence between integration
cycles and antimeromorphic forms that is akin to Betti--deRham duality \cite{betti1,betti2, Brown:2018omk}.
In a tree-level context, Betti--deRham duality relates the ordering of open-string punctures
on a disk boundary to Parke--Taylor factors \cite{Schlotterer:2012ny,Stieberger:2013wea,Stieberger:2014hba,Schlotterer:2018abc,Vanhove:2018elu,Brown:2019wna} -- cyclic products of propagators $(\bar z_i- \bar z_j)^{-1}$ on the sphere. As a genus-one generalization, we spell
out certain antielliptic (i.e.\ antimeromorphic and
doubly-periodic) functions on the torus which
will be referred to as the Betti--deRham
duals\footnote{We shall use this terminology at genus one even though we are not aware of any
explicitly worked out notion of Betti--deRham duality beyond genus zero.}  of integration cycles on a cylinder boundary.

It will be important to collect the various eMZVs and MGFs in generating series
similar to those in~\cite{Mafra:2019ddf,Mafra:2019xms,Gerken:2019cxz,Gerken:2020yii} as the genus-one single-valued map ${\rm SV}$ is most conveniently described at the level of these generating series.
The $\alpha'$-expansion of genus-one closed-string integrals -- using the techniques of \cite{Gerken:2020yii} -- yields an explicit form of the proposed single-valued map of the eMZVs in open-string integrals.
The open-string punctures on a cylinder boundary are ordered according
to the cycle which is Betti--deRham dual to the additional antielliptic functions in the closed-string integrand.
For the purpose of this work, it will be sufficient to place all the open-string punctures on the same cylinder boundary which corresponds to planar genus-one amplitudes: As will be discussed in future work, single-valued non-planar open-string integrals yield the same collection of MGFs as the planar ones. Apart from a characterization via iterated Eisenstein integrals,
we will arrive at a closed formula for the single-valued versions of any convergent eMZV that straightforwardly yields the familiar lattice-sum representations of MGFs.

\begin{figure}[t!]
\centering
\begin{tikzpicture}
\draw[rounded corners=5pt] (-2,5.5) rectangle (2,-1);
\draw (0,5) node {open string};
\draw[rounded corners=5pt] (6,5.5) rectangle (10,-1);
\draw (8,5) node {closed string};
\draw [thick,->] (1,3.7)--(7,3.7);
\draw (4,3.9) node[anchor=south] {SV};
\draw (0,4) node[anchor=north] {$B^\tau_{\vec\eta}$};
\draw (8,4) node[anchor=north] {$J^\tau_{\vec\eta}$};
\draw [thick,->] (-0.2,3.25)--(-0.2,2.2);
\draw [thick,<-] (0.2,3.25)--(0.2,2.2);
\draw (-0.9,2.7) node {\footnotesize expand};
\draw (0.9,2.7) node {\footnotesize isolate};
%
\draw [thick,->] (7.8,3.25)--(7.8,2.2);
\draw [thick,<-] (8.2,3.25)--(8.2,2.2);
\draw (7.1,2.7) node {\footnotesize expand};
\draw (8.9,2.7) node {\footnotesize isolate};
\draw (0,2) node[anchor=north] {eMZV};
\draw (8,2) node[anchor=north] {MGF};
\draw [thick,dashed,->] (1,1.7)--(7,1.7);
\draw (4,1.8) node[anchor=south] {SV};
\draw (0,0) node[anchor=north] {MZV};
\draw (8,0) node[anchor=north] {svMZV};
\draw [thick,->] (0,1.25)--(0,0.2);
\draw [thick,->] (8,1.25)--(8,0.2);
\draw (-0.2,.7) node[anchor=east] {\footnotesize $\tau{\to}i \infty$};
\draw (8.2,.7) node[anchor=west] {\footnotesize $\tau{\to}i \infty$};
\draw [thick,->] (1,-0.3)--(7,-0.3);
\draw (4,-0.1) node[anchor=south] {sv};
\end{tikzpicture}
\caption{\textit{Diagram illustrating the various pieces involved in constructing the proposal ${\rm SV}$ for an elliptic single-valued map with open-string quantities on its left-hand side and closed-string quantities on its right-hand side. The generating series $B^\tau_{\vec\eta}$  of the open string contains  elliptic multiple zeta values (eMZVs) in its $\alpha'$- and $\eta_j$-expansion. Conversely, a given eMZV can be isolated as a specific component of the generating series. 
The $\tau$-dependent eMZVs contain multiple zeta values (MZVs) in their degeneration limit $\tau\to i\infty$.
Similarly, the closed-string generating series $J^\tau_{\vec\eta}$ yields modular graph forms (MGFs) upon expansion and MGFs can be isolated as specific components in this expansion. 
The degeneration limit $\tau\to i\infty$ of MGFs is expected to only contain single-valued multiple zeta values (svMZVs) that are related to the MZV by the known single-valued map ${\rm sv}$.
Instead of attempting a direct construction of an elliptic ${\rm SV}$-map from eMZVs to MGFs,
we exploit the differential equations of the generating series $B_{\vec{\eta}}^\tau$  and $J_{\vec{\eta}}^\tau$ together with their boundary values from $\tau\to i \infty$ to describe the map ${\rm SV}$ at the level of generating series, see~\eqref{introduc.1}. From this one can extract the map ${\rm SV}:{\rm eMZV} \to {\rm MGF}$ by inspecting individual orders in the $\alpha'$- and $\eta_j$-expansions.}}
\label{fig:struc}
\end{figure}
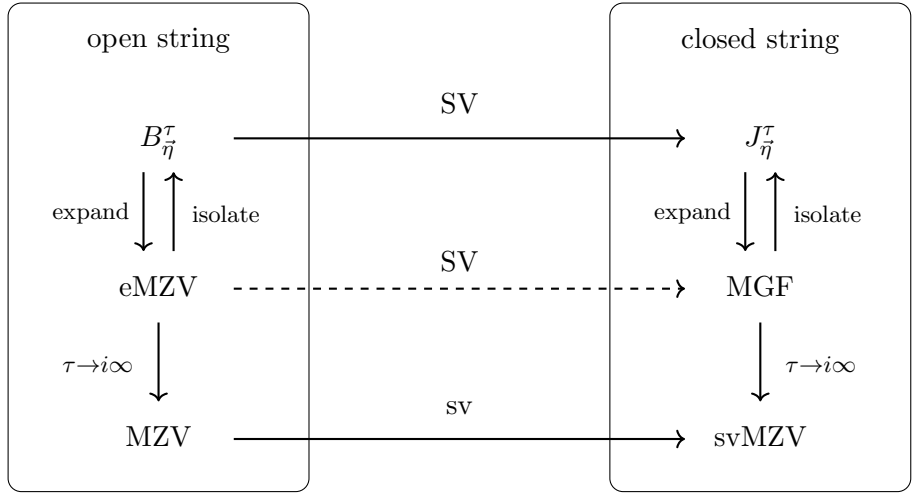

The main evidence for our proposal for an elliptic single-valued map stems from its consistency with
holomorphic derivatives in the modular parameters $\tau$ of the surfaces and the degeneration
$\tau \rightarrow i\infty$ of the torus to a nodal sphere. Compatibility with the holomorphic derivative
is a simple consequence of recent results on the differential equations of genus-one open-string
integrals \cite{Mafra:2019ddf, Mafra:2019xms} and closed-string integrals \cite{Gerken:2019cxz} in $\tau$. Our antielliptic integrands on the
torus ensure that the closed-string differential equations match those of the open string apart from
the disappearance of $\zeta_2$ as expected from the single-valued map of MZVs. Moreover, the
anti\-elliptic integrands are engineered such as to reproduce Parke--Taylor factors in the degeneration
$\tau \rightarrow i\infty$. Hence, compatibility of the single-valued maps at genus zero and one is
supported by the identification of sphere integrals as single-valued disk integrals \cite{Schlotterer:2012ny,Stieberger:2013wea,Stieberger:2014hba,Schlotterer:2018abc,Vanhove:2018elu,Brown:2019wna}. 
The logic of our construction is illustrated in figure~\ref{fig:struc}.

\subsection{Summary of main results}

The main result of this work is the proposal
\beq
J^\tau_{\vec{\eta}}   = {\rm SV} \, B^\tau_{\vec{\eta}}
\label{introduc.1}
\eeq
for a single-valued map SV at genus one which
relates generating series $B^\tau_{\vec{\eta}}$ and $J^\tau_{\vec{\eta}}$
of open- and closed-string integrals, respectively, see (\ref{newBB.20}). 
As summarized in figure~\ref{fig:struc},
this induces an SV action on the eMZVs in the $\ap$-expansion of the
cylinder integrals $B^\tau_{\vec{\eta}}$ to be defined in (\ref{newBB.1}). By comparing
coefficients of dimensionless Mandelstam invariants $\alpha' k_i\cdot k_j$
and formal expansion variables $\eta_j$, SV maps each eMZV generated by $B^\tau_{\vec{\eta}}$ 
to combinations of MGFs at the same order in the analogous expansion 
of the torus integrals $J^\tau_{\vec{\eta}}$ in (\ref{newBB.11}). The integrands of
$B^\tau_{\vec{\eta}}$ and $J^\tau_{\vec{\eta}}$ are assembled from combinations
of doubly-periodic Kronecker--Eisenstein series $\varphi^\tau_{\vec{\eta}}$ in (\ref{looprev.5}) 
known from \cite{Mafra:2019ddf,Mafra:2019xms,Gerken:2019cxz,Gerken:2020yii} and antielliptic 
functions $\overline{V(\ldots|\tau)}$ that we introduce in (\ref{newBB.6}) as tentative Betti--deRham 
duals of integration cycles on a cylinder boundary.

A key motivation and evidence for this construction stems from the degeneration 
limit $\tau\to i\infty$ of the series $B^\tau_{\vec{\eta}}$ and $J^\tau_{\vec{\eta}}$. Within this limit, genus-zero
integrals similar to those in open- and closed-string tree-level amplitudes are recovered, the latter being related by the single-valued map of MZVs \cite{Schlotterer:2012ny,Stieberger:2013wea,Stieberger:2014hba,Schlotterer:2018abc,Vanhove:2018elu,Brown:2019wna}.
The leading terms of the eMZVs in the $\tau\to i\infty$ limit of $B^\tau_{\vec{\eta}}$ are certain Laurent polynomials in the modular
parameter $\tau$ of the cylinder with MZVs in its coefficients. As visualized in the lower part of
figure \ref{fig:struc}, the known single-valued map sv of MZVs \cite{Schnetz:2013hqa, Brown:2013gia} is 
conjectured to yield the analogous Laurent polynomials in the degeneration limit $\tau\to i\infty$ of the
torus integrals $J^\tau_{\vec{\eta}}$. This is made precise in the conjecture (\ref{newBB.15}) -- a
central prerequisite for (\ref{introduc.1}) -- which generalizes earlier observations in 
\cite{Broedel:2018izr, Gerken:2018jrq} and has been proven at the leading orders in the formal 
expansion variable $\eta_j$ at two points \cite{Zagier:2019eus}.

The earlier proposal for an elliptic single-valued map ``esv'' in \cite{Broedel:2018izr} concerns the full $\tau$-dependence of certain generating series of eMZVs or the composing iterated Eisenstein integrals. This reference associates open-string prototypes
(i.e.\ esv preimages) to the simplest closed-string integrals at genus one whose integrands
are solely built from Green functions involving any number of punctures. On the one
hand, the proposal for the single-valued map in \cite{Broedel:2018izr} is contained in
(\ref{introduc.1}) upon symmetrizing over the integration cycles on its left-hand side and
extracting the lowest order in $\eta_j$. On the other hand, the implementation of the single-valued
map at the level of iterated Eisenstein integrals in the reference is very different from the
proposal in the present work. In comparison to the proposal of \cite{Broedel:2018izr}, our SV 
action on iterated Eisenstein integrals in (\ref{eesv.1}) does not necessarily generate real combinations
and is therefore applicable to imaginary cusp forms and MGFs of different holomorphic and antiholomorphic
modular weights. Moreover, in contrast to esv in \cite{Broedel:2018izr}, the SV map in (\ref{introduc.1}) is observed to be compatible with shuffle multiplication in all known examples. Our SV map additionally introduces combinations of (conjecturally single-valued) MZVs and antiholomorphic terms, which are absent in \cite{Broedel:2018izr}.

At the time of writing, the antiholomorphic admixtures introduced by our SV map on iterated Eisenstein
integrals at depth $\geq 2$ can only be fixed by indirect methods beyond the reach of open-string data.
Instead, the explicit form of the SV action on iterated Eisenstein integrals has so far been extracted
from the reality properties of closed-string generating series in \cite{Gerken:2020yii} that extend the
$J^\tau_{\vec{\eta}}$ series as described below. However, this limitation does not affect the formulation
of our SV map at the level of the lattice-sum representation of MGFs: By virtue of the antielliptic functions
$\overline{V(\ldots|\tau)}$ in (\ref{newBB.6}), the SV image of an arbitrary convergent eMZV given in 
(\ref{nsvemzv.1}) can be straightforwardly expressed in terms of lattice sums using the integration techniques
of \cite{Green:2008uj,DHoker:2015gmr,DHoker:2015wxz,Gerken:2018jrq} and, for certain weights, further simplified using the {\tt Mathematica} package \cite{Gerken:2020aju}.

\subsection{Outline}

This work is organized as follows. We start by reviewing open- and closed-string
integrals at genus zero and genus one as well as the basic definitions of single-valued MZVs,
eMZVs and MGFs in section \ref{sec:2}. Then, section \ref{sec:3} is dedicated to
the modified generating series of open- and closed-string integrals as well as their relation through our
proposed single-valued map at genus one. In particular, the central antielliptic integrands
and the resulting proposal for an elliptic single-valued map can be found in sections \ref{sec:3.2}
and \ref{sec:3.5}, respectively. In section \ref{sec:4}, we set the stage for generating explicit
examples of single-valued eMZVs by introducing a new expansion method for open-string
integrals over B-cycles and relating it to similar closed-string $\alpha'$-expansions.
This leads to the identifications of MGFs as single-valued eMZVs in
section \ref{sec:5}, where examples of the antielliptic integrands are related to earlier
approaches to an elliptic single-valued map in the literature. The resulting
lattice-sum representations of all single-valued convergent eMZVs are discussed in
section \ref{sec:5.8}. In the concluding section \ref{sec:9}, we comment on the relation
of string amplitudes to the generating series of this work and further directions.

\section{Review of genus-zero and genus-one integrals}
\label{sec:2}

In this section, we collect background material on world-sheet integrals at
genus zero  and one, including the genus-zero single-valued map, and review various definitions relevant to the single-valued map at genus one.

\subsection{Genus-zero integrals}
\label{sec:2.1}

We briefly review the basic disk (open-string) and sphere (closed-string) integrals for genus-zero world-sheets and how they are related by the genus-zero single-valued map.

\subsubsection{Definitions of disk and sphere integrals}

Massless tree-level $n$-point amplitudes of the open superstring \cite{Mafra:2011nv} and
the open bosonic string \cite{Azevedo:2018dgo} can be expanded in a basis of iterated integrals
\cite{Zfunctions}
\beq
Z^{\rm tree}(\gamma|\rho ) = \! \!  \int\limits_{\mathfrak D(\gamma)} \! \! \frac{\big( \prod_{j=1}^n \dd z_j \big)}{ {\rm vol} \, {\rm SL}_2(\mathbb R)}   \prod_{1\leq i <j }^n \! \!  |z_{ij}|^{-s_{ij}} {\rm PT}(\rho(1,2,\ldots,n))
\label{treerev.1}
\eeq
over the boundary of a disk which we parametrize through the real line
\beq
\mathfrak D(\gamma) = \{ z_j \in \mathbb R, \ -\infty < z_{\gamma(1)} < z_{\gamma(2)}
<\ldots < z_{\gamma(n)} < \infty \} \, .
\label{treerev.2}
\eeq
The disk integrands involve dimensionless Mandelstam invariants
\beq
s_{ij} = - \frac{\alpha' }{2} k_i \cdot k_j \, , \ \ \ \ \ \ k_j^2=0
\label{treerev.3}
\eeq
and Parke--Taylor factors
\beq
{\rm PT}(\rho(1,2,\ldots,n)) = \frac{1}{ z_{\rho(1) \rho(2)} z_{\rho(2) \rho(3)}  \ldots z_{\rho(n) \rho(1)}  }
\, , \ \ \ \ \ \ z_{ij}=z_i{-}z_j
\, .
\label{treerev.4}
\eeq
The inverse ${\rm vol} \,{\rm SL}_2(\mathbb R)$ in (\ref{treerev.1}) instructs us to
set any triplet of punctures to $0,1,\infty$, where the ${\rm SL}_2(\mathbb R)$
invariance of genus-zero integrands hinges on momentum conservation $\sum_{j=1}^n k_j=0$.
Both the domains and the Parke--Taylor
integrands are indexed via permutations $\gamma,\rho \in S_n$ of
the external legs $1,2,\ldots,n$. One can arrive at smaller bases of $(n{-}3)!$
cycles $\gamma$ and Parke--Taylor orderings $\rho$ via monodromy relations
\cite{BjerrumBohr:2009rd, Stieberger:2009hq}
and integration by parts \cite{Mafra:2011nv, Zfunctions}, respectively.

Closed-string tree amplitudes in turn can be reduced to sphere integrals
\begin{align}
J^{\rm tree}(\gamma|\rho ) &= \frac{1}{\pi^{n-3}} \int\limits_{\CC^{n}}  \frac{\big( \prod_{j=1}^n \dd^2 z_j \big)}{ {\rm vol} \, {\rm SL}_2(\mathbb C)} \prod_{1\leq i <j }^n \! \!  |z_{ij}|^{-2s_{ij}}
 \overline{ {\rm PT}(\gamma(1,2,\ldots,n)) } {\rm PT}(\rho(1,2,\ldots,n))
\label{treerev.5}
\end{align}
involving $\dd^2 z_j = \frac{i}{2} \dd z_j \wedge \dd \bar z_j $ and permutations $\gamma,\rho \in S_n$ of meromorphic and antimeromorphic Parke--Taylor factors subject to the same integration-by-parts
relations as in the open-string case.

\subsubsection{Single-valued map between disk and sphere integrals}
\label{sec:2.2}

The disk and sphere integrals (\ref{treerev.1}) and (\ref{treerev.5}) converge for a suitable range of
the $\Re(s_{ij})$ and they admit a Laurent expansion in $\alpha'$, i.e.\ around the value $s_{ij}=0$ of the dimensionless Mandelstam
invariants (\ref{treerev.3}).
The coefficients in the $\alpha'$-expansions of disk integrals
$Z^{\rm tree}$ are MZVs \cite{Terasoma, Brown:2009qja},
\beq
\zeta_{n_1,n_2,\ldots, n_r} = \sum_{0<k_1<k_2<\ldots<k_r} k_1^{-n_1} k_2^{-n_2} \ldots k_r^{-n_r} \, , \ \ \ \ \ \ n_r\geq 2
\label{defmzvs}
\eeq
whose weight $n_1+n_2+\ldots + n_r$ matches the order in $\alpha'$ beyond the low-energy limit
(i.e.\ beyond the leading order in $\alpha'$). The polynomial structure of the $Z^{\rm tree}$ in $s_{ij}$
can for instance be generated from the Drinfeld associator \cite{Broedel:2013aza} or Berends--Giele
recursions \cite{Mafra:2016mcc}, with explicit results available for download from \cite{MZVWebsite, gitrep}.

When applying the single-valued map \cite{Schnetz:2013hqa, Brown:2013gia} of motivic \cite{BrownTate} MZVs\footnote{Strictly speaking, MZVs need to be replaced by their motivic versions to have
a well-defined single-valued~map.}
\begin{align}
{\rm sv} \, \zeta_{2k} = 0 \, , \ \ \ \
{\rm sv} \, \zeta_{2k+1} = 2 \zeta_{2k+1}\,,\ \ \ \
{\rm sv} \, \zeta_{3,5} = -10 \zeta_3 \zeta_5 \, , \ \ \ \ {\rm etc.}
\label{treerev.6}
\end{align}
order by order in $\alpha'$, the disk and sphere integrals (\ref{treerev.1}) and (\ref{treerev.5})
are related by \cite{Schlotterer:2012ny,Stieberger:2013wea,Stieberger:2014hba,Schlotterer:2018abc,Vanhove:2018elu,Brown:2019wna}
\beq
J^{\rm tree}(\gamma|\rho )=  {\rm sv} \, Z^{\rm tree}(\gamma|\rho ) \, .
\label{treerev.7}
\eeq
The first permutation $\gamma$ in $Z^{\rm tree}$ and $J^{\rm tree}$ refers
to a disk ordering (\ref{treerev.2}) and an antimeromorphic Parke--Taylor factor (\ref{treerev.4}),
respectively, which are connected by a Betti--deRham duality \cite{betti1,betti2, Brown:2018omk}. The key result of
this work is to identify similar pairs of cycles and antimeromorphic functions at genus one.

\subsection{Genus-one integrals}

As a preparation for our proposal of a genus-one single-valued map, we now introduce the basic genus-one world-sheet integrals and the objects appearing in their $\alpha'$-expansion.

\subsubsection{Genus-one open-string A-cycle integrals}
\label{sec:2.3}

\begin{figure}[t]
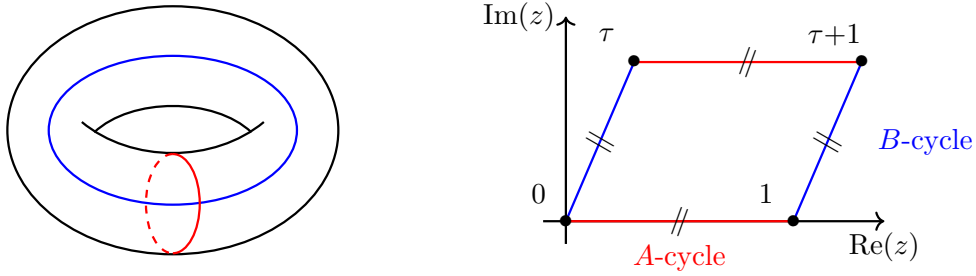

\begin{center}
\tikzpicture[scale=0.55, line width=0.30mm]
\draw(0,0) ellipse  (4cm and 3cm);
\draw(-2.2,0.2) .. controls (-1,-0.8) and (1,-0.8) .. (2.2,0.2);
\draw(-1.9,-0.05) .. controls (-1,0.8) and (1,0.8) .. (1.9,-0.05);
\draw[blue](0,0) ellipse  (3cm and 1.8cm);
\draw[red] (0,-2.975) arc (-90:90:0.65cm and 1.2cm);
\draw[red,dashed] (0,-0.575) arc (90:270:0.65cm and 1.2cm);
\scope[xshift=9.5cm,yshift=-2.2cm,scale=1.1]
\draw[->](-0.5,0) -- (7,0) node[below]{${\rm Re}(z)$};
\draw[->](0,-0.5) -- (0,4.5) node[left]{${\rm Im}(z)$};
\draw(-0.6,0.6)node{$0$};
\draw[blue](0,0) -- (1.5,3.5);
\draw(0.75,1.75)node[rotate=60]{$| \; \! \!|$};
\draw(0.9,4.1)node{$\tau$};
\draw[red](0,0) -- (5,0);
\draw(2.5,0)node[rotate=-20]{$| \; \! \! |$};
\draw (4.4,0.6)node{$1$};
\draw[red](1.5,3.5) -- (6.5,3.5);
\draw(4,3.5)node[rotate=-20]{$| \; \! \! |$};
\draw[blue](5,0) -- (6.5,3.5);
\draw(5.75,1.75)node[rotate=60]{$| \; \! \! |$};
\draw(5.9,4.1)node{$\tau{+}1$};
\draw[red](2.5,-0.8)node{$A$-cycle};
\draw[blue](7.9,1.75)node{$B$-cycle};
\draw (5,0)node{$\bullet$};
\draw(0,0)node{$\bullet$};
\draw (1.5,3.5)node{$\bullet$} ;
\draw(6.5,3.5)node{$\bullet$};
\endscope
\endtikzpicture
\caption{\textit{Parametrization of the torus $\mathfrak T= \frac{\Bbb C}{\Bbb Z {+} \tau \Bbb Z}$
with identifications $z \cong z{+}1 \cong z{+}\tau$ marked by $|\!|$ along the A- and B-cycles. While the torus is drawn for non-vanishing $\Re(\tau)$ to accommodate closed-string amplitudes, the cylinder world-sheets for open-string amplitudes are derived from tori at $\tau \in i\mathbb R^+$ via suitable involutions \cite{Polchinski:1998rq}.}}
\label{figtorus}
\end{center}
\end{figure}

In the same way as disk integrals can be cast into a
Parke--Taylor-type basis (\ref{treerev.1}), the basis integrals for massless genus-one open-string amplitudes
are claimed to be generated by \cite{Mafra:2019ddf, Mafra:2019xms}
\beq
Z^\tau_{\vec{\eta}}(\gamma|\rho ) = \! \!  \int\limits_{ \mathfrak A(\gamma)} \! \! \Big( \prod_{j=2}^n \dd z_j \Big)  \vph^\tau_{\vec{\eta}}(1,\rho(2,\ldots,n)) \! \! \prod_{1\leq i <j }^n \! \!  e^{s_{ij} {\cal G}_{\mathfrak A}(z_{ij},\tau)}\, ,
\label{looprev.1}
\eeq
where we have set $z_1=0$ by translation invariance. In this work we restrict to planar amplitudes with all state insertions on a single cylinder boundary (as opposed to non-planar amplitudes with punctures on both boundaries of the cylinder). We do not impose momentum conservation in a
genus-one context and treat all the $s_{ij}$ with $1\leq i < j \leq n$ as independent. The
ordering of the open-string punctures on a cylinder boundary is encoded in an
integration domain on the A-cycle of a torus (see figure~\ref{figtorus} for the standard parametrization) with $\tau \in i \mathbb R^+$ \cite{Polchinski:1998rq}
\beq
\mathfrak A(\gamma) = \{ z_{j} \in \RR, \
0< z_{\gamma(2)}< z_{\gamma(3)} <\ldots <z_{\gamma(n)}<1\}\,,
\label{looprev.2}
\eeq
with similar integration domains \cite{Green:1987mn} for the non-planar open-string integrals.

The integrand of (\ref{looprev.1}) features the open-string Green function
on an A-cycle (which is chosen to enforce ${\cal G}_{\mathfrak A}(z,\tau)= {\cal G}_{\mathfrak A}(-z,\tau)$ and $\int^1_0 \dd z\, {\cal G}_{\mathfrak A}(z,\tau) =0$ \cite{Broedel:2018izr, Zerbini:2018hgs})
\beq
{\cal G}_{\mathfrak A}(z,\tau) =
 - \log \Big( \frac{ \theta_1(|z|,\tau) }{ \eta(\tau) } \Big) + \frac{ i \pi \tau}{6} + \frac{ i \pi }{2} \, , \ \ \ \ \ \ z\in (-1,1)
\label{looprev.3}
\eeq
and the following combination of the doubly-periodic Kronecker--Eisenstein series \cite{Kronecker}
\begin{align}
\Omega(z,\eta,\tau) &= \exp \Big( 2\pi i \eta\frac{ \Im z }{\Im \tau} \Big)
\frac{ \theta_1'(0,\tau) \theta_1(z{+}\eta,\tau) }{\theta_1(z,\tau)
\theta_1(\eta,\tau)}\label{looprev.4}\\
\vph^\tau_{\vec{\eta}}(1,2,\ldots,n)&=  \Omega(z_{12},\eta_{23\ldots n},\tau)
 \Omega(z_{23},\eta_{3\ldots n},\tau)
   \cdots  \Omega(z_{n-1,n},\eta_{n},\tau)
\label{looprev.5}
\end{align}
with $\eta_{ij\ldots k} = \eta_i+\eta_j+\ldots+\eta_k$.\footnote{Our conventions for the standard odd Jacobi theta function are
\[
\theta_1(z,\tau) = q^{1/8} (e^{i\pi z} - e^{-i\pi z} ) \prod_{n=1}^\infty (1-q^n)(1-e^{2\pi i z} q^n)(1-e^{-2\pi i z} q^n)
\]
and $\eta(\tau)$ is the Dedekind eta function. In order to avoid
confusion with the expansion parameters $\eta_j$, we always spell out the argument $\tau$ of the Dedekind eta function. Representations of the open-string Green function in terms of elliptic polylogarithms
are discussed in \cite{Broedel:2014vla, Broedel:2017jdo, Broedel:2018izr}, and we follow the conventions of \cite{Broedel:2014vla} for regularizing endpoint divergences.}
The permutation $\rho \in S_{n-1}$ in $\vph^\tau_{\vec{\eta}}(1,\rho(2,\ldots,n))$ is
taken to act on both the $z_j$ and the formal expansion variables $\eta_j \in \mathbb C$
in (\ref{looprev.5}). The conjectural basis (\ref{looprev.1}) is a generating function
of the world-sheet integrals over the Kronecker--Eisenstein coefficients~$f^{(w)}$
\beq
\Omega(z,\eta,\tau) =  \sum_{w=0}^{\infty} \eta^{w-1} f^{(w)}(z,\tau)
\label{looprev.6}
\eeq
that occur in the integrands of genus-one open- and closed-string amplitudes
\cite{Dolan:2007eh, Broedel:2014vla, Gerken:2018jrq}, e.g.\
\beq
  f^{(0)}(z{,}\tau)= 1 \, , \ \ \ \
f^{(1)}(z{,}\tau)= \partial_z \log \theta_1(z{,}\tau) + 2\pi i \frac{ \Im z }{\Im \tau}\, .
\label{looprev.7}
\eeq
While the massless four-point genus-one amplitude of the open superstring \cite{Green:1982sw} is proportional
to the most singular $\eta_j^{-3}$-order of $Z^\tau_{\vec{\eta}}(\cdot |1,2,3,4 )$, the analogous amplitude of the
open bosonic string additionally involves contributions of $Z^\tau_{\vec{\eta}}(\cdot |1,2,3,4 )$
(and its permutations in $2,3,4$) at the orders of $\eta_j^{\pm 1}$ \cite{Green:1987mn}\footnote{By using Fay identities and integration by parts, the massless four-point genus-one amplitude of open bosonic
strings in section 8.1.1 of \cite{Green:1987mn} can be rewritten in terms of the coefficients in the $\eta_j$-expansion
of $Z^\tau_{\vec{\eta}}(\cdot |1,2,3,4 )$.}. The short-distance behavior $f^{(1)}(z,\tau) = \frac{1}{z} + {\cal O}(z)$
introduces kinematic poles into the $\alpha'$-expansion of (\ref{looprev.1}),
and the remaining $f^{(w\neq 1)}(z,\tau)$ are regular for any $z \in \mathbb C$.

\subsubsection{Genus-one closed-string integrals}
\label{sec:2.4}

In the same way as (\ref{looprev.1}) is claimed to be a universal basis of genus-one open-string
integrals, the integrals over the torus punctures for massless genus-one amplitudes in type II, heterotic and bosonic
string theories should be generated by \cite{Gerken:2019cxz}
\begin{align}
\label{looprev.8}
Y^\tau_{\vec{\eta}}(\gamma|\rho ) = (2i)^{n{-}1} \!  \! \int_{\mathfrak T^{n{-}1}} \!  \! \Big( \prod_{j=2}^n \dd^2 z_j \Big)  \! \!  \prod_{1\leq i <j }^n \! \!  e^{s_{ij} {\cal G}_{\mathfrak T}(z_{ij},\tau)}
 \overline{ \vph^\tau_{\vec{\eta}}(1,\gamma(2,\ldots,n)) } \vph^\tau_{(\tau-\bar \tau)\vec{\eta}}(1,\rho(2,\ldots,n))
\end{align}
with $z_1=0$. The remaining $z_j$ are integrated over the torus $\mathfrak T = \frac{ \mathbb C }{\tau \mathbb Z + \mathbb Z}$ with modular parameter $\tau \in \mathbb H = \{ \tau \in \mathbb C , \ \Im \tau >0 \}$. The closed-string Green function
\beq
{\cal G}_{\mathfrak T}(z,\tau) = - \log \left| \frac{ \theta_1(z,\tau) }{\eta(\tau)} \right|^2 + \frac{ 2\pi (\Im z)^2 }{\Im \tau}
\label{looprev.9}
\eeq
is chosen to be modular invariant and to obey $\int_{\mathfrak T} \dd ^2 z \,{\cal G}_{\mathfrak T}(z,\tau)=0$,
and its holomorphic derivatives parallel those of the open-string Green function ${\cal G}_{\mathfrak A}(z,\tau)$ in (\ref{looprev.3}),
\begin{align}
\partial_z {\cal G}_{\mathfrak T}(z,\tau) &= - f^{(1)}(z,\tau) \, , &2\pi i \partial_\tau {\cal G}_{\mathfrak T}(u\tau {+} v,\tau) &= - f^{(2)}(u\tau {+} v,\tau)
\hspace{10mm}\text{(fixed $u,v$)}\notag \\
\partial_v {\cal G}_{\mathfrak A}(v,\tau) &= - f^{(1)}(v,\tau) \, , &
2\pi i \partial_\tau {\cal G}_{\mathfrak A}( v,\tau) &= - f^{(2)}( v,\tau) - 2 \zeta_2\, ,
\label{looprev.10}
\end{align}
where $u,v\in \mathbb R$ parametrize the covering space of the torus and the $f^{(w)}(z,\tau)$ with $z= u \tau + v$ are defined by (\ref{looprev.7}).
The second arguments $(\tau{-}\bar \tau)\eta_{j}$ and $\bar \eta_j$ of the
Kronecker--Eisenstein series and their complex conjugates in (\ref{looprev.8})
have been chosen such that each order in the $\eta_j$- and $\alpha'$-expansion
gives rise to modular forms of purely antiholomorphic modular weight\footnote{Functions $F(\tau)$ on the upper half plane with transformations $F(\frac{ \alpha \tau + \beta}{\gamma \tau + \delta}) =(\gamma\tau+\delta)^w (\gamma\bar\tau+\delta)^{\bar{w}}F(\tau)$ under $(\begin{smallmatrix} \alpha &\beta \\ \gamma &\delta \end{smallmatrix} ) \in {\rm SL}_{2}(\ZZ)$ are said to carry holomorphic and antiholomorphic modular weight $w$ and $\bar{w}$, respectively.}.

When assembling genus-one amplitudes of open and closed strings from the
series $Z^\tau_{\vec{\eta}}$ and $Y^\tau_{\vec{\eta}}$, it remains to dress the component integrals in their $\eta_j$-expansions with kinematic factors that carry the dependence on the external polarizations. The latter are
determined from the conformal-field-theory correlators of the vertex operators, see
e.g.\ \cite{Tsuchiya:1988va, Mafra:2018qqe}, and are unaffected by our proposal for the
single-valued map at genus one.

\subsubsection{Differential equations in $\tau$}
\label{sec:2.5}

Based on the differential equations (\ref{looprev.10})
of the Green functions and integration by parts in the $z_j$, the open- and closed-string
integrals (\ref{looprev.1}) and (\ref{looprev.8}) were shown in \cite{Mafra:2019xms}
and \cite{Gerken:2019cxz} to obey the differential equations
\begin{align}
2\pi i \partial_\tau Z^\tau_{\vec{\eta}}(\gamma|\rho ) &= \sum_{k=0}^\infty (1{-}k){\rm G}_k(\tau)
\sum_{\alpha \in S_{n-1}} r_{\vec{\eta}}(\ep_k)_\rho{}^\alpha Z^\tau_{\vec{\eta}}(\gamma|\alpha )
\label{looprev.11}  \\
2\pi i \partial_\tau Y^\tau_{\vec{\eta}}(\gamma|\rho ) &= \sum_{k=0}^\infty (1{-}k)(\tau{-}\bar \tau)^{k-2}{\rm G}_k(\tau)
 \sum_{\alpha \in S_{n-1}} R_{\vec{\eta}}(\ep_k)_\rho{}^\alpha Y^\tau_{\vec{\eta}}(\gamma|\alpha )
\,, \notag
\end{align}
respectively. The right-hand sides involve holomorphic Eisenstein series ${\rm G}_0=-1$ and
\beq
{\rm G}_k(\tau) = \sum_{m,n\in \ZZ \atop{(m,n) \neq (0,0)}} \frac{1}{(m\tau + n)^k} \, , \ \ \ \ k \geq 4
\label{looprev.12}
\eeq
as well as $(n{-}1)! \times (n{-}1)!$ matrices $r_{\vec{\eta}}(\ep_k),R_{\vec{\eta}}(\ep_k)$
independent of $\tau$ that vanish for $k=2$ and $k\in 2\mathbb N{-}1$. This means in particular that ${\rm G}_2(\tau)$ does not appear in~\eqref{looprev.11}.

The two-point instances are
\begin{align}
r_{\eta_2}(\ep_0) &= s_{12}  \bigg( \frac{1}{\eta_2^2} + 2 \zeta_2 - \frac{1}{2} \partial_{\eta_2}^2 \bigg)  \notag \\
r_{\eta_2}(\ep_k) &= R_{\eta_2}(\ep_k) = s_{12} \eta_2^{k-2} \, , \ \ \ \ \ \ k \geq 4
\label{looprev.13} \\
R_{\eta_2}(\ep_0) &= s_{12}  \bigg( \frac{1}{\eta_2^2} - \frac{1}{2} \partial_{\eta_2}^2 \bigg) - 2\pi i \bar \eta_2 \partial_{\eta_2} \, . \notag
\end{align}
The notation $\ep_k$ reflects the expectation that the
$r_{\vec{\eta}}(\ep_k),R_{\vec{\eta}}(\ep_k)$ are matrix representations of
Tsunogai's derivation algebra \cite{Tsunogai} and obey relations such as (see
\cite{LNT, Pollack, Broedel:2015hia} for similar relations at higher weight and depth)
\beq
[ r_{\vec{\eta}}(\ep_{10}) ,  r_{\vec{\eta}}(\ep_{4}) ] - 3  [ r_{\vec{\eta}}(\ep_{8}) ,  r_{\vec{\eta}}(\ep_{6}) ]
=0\, .
\label{derrels}
\eeq
The all-multiplicity formulae
for these $(n{-}1)! \times (n{-}1)!$ representations in \cite{Mafra:2019xms, Gerken:2019cxz} manifest
that the $r_{\vec{\eta}}(\ep_k)$ are linear in the $s_{ij}$, i.e.\ proportional to
$\alpha'$, and their closed-string analogues $R_{\vec{\eta}}(\ep_k)$ additionally
involve terms $\sim  \bar \eta_j \partial_{\eta_j}$ independent of $\alpha'$ (with $s_{12\ldots n}=\sum_{1\leq i<j}^n s_{ij}$):
\begin{align}
R_{\vec{\eta}}(\ep_k) = \left\{ \begin{array}{cl} r_{\vec{\eta}}(\ep_k) &: \ k\geq 4 \\
r_{\vec{\eta}}(\ep_0) {-} 2 \zeta_2 s_{12\ldots n} {-} 2\pi i \sum_{j=2}^n \bar \eta_j \partial_{\eta_j} &: \ k=0 \end{array} \right. \, .
\label{looprev.14}
\end{align}

\subsubsection{Basic definitions of eMZVs and MGFs}
\label{sec:2.6}

We shall now review the definitions of the eMZVs and MGFs that occur as the expansion
coefficients of the above genus-one integrals. The $\eta_j$- and $\alpha'$-expansion of the open-string
integrals $Z^\tau_{\vec{\eta}}(\gamma|\rho )$ in (\ref{looprev.1}) gives rise to
A-cycle eMZVs \cite{Broedel:2014vla, Broedel:2017jdo, Broedel:2019vjc}
\beq
\omega(n_1,n_2,\ldots,n_r|\tau) = \! \! \! \! \! \! \! \! \!  \int \limits_{0<z_1<z_2<\ldots < z_r<1}
\! \! \! \! \! \! \! \! \!  \dd z_1 \, f^{(n_1)}(z_1,\tau) \,
\dd z_2 \, f^{(n_2)}(z_2,\tau) \, \ldots \, \dd z_r \, f^{(n_r)}(z_r,\tau)
\label{newdef.1}
\eeq
introduced by Enriquez \cite{Enriquez:Emzv}
which are said to carry weight $n_1+n_2+\ldots+ n_r$ and length $r$.
Endpoint divergences in case of $n_1=1$ or $n_r=1$ are shuffle-regularized
as in section 2.2.1 of \cite{Broedel:2014vla}.
The specific eMZVs at a given order of $Z^\tau_{\vec{\eta}}(\gamma|\rho )$ in $s_{ij}$ and $\eta_j$ can
be obtained from the differential equations (\ref{looprev.11}) along
with the initial values $ Z^{\tau  \rightarrow i\infty}_{\vec{\eta}}(\gamma|\rho )$ in \cite{Mafra:2019xms}
or from matrix representations of the elliptic KZB associator \cite{Broedel:2019gba, Broedel:2020tmd}.

The closed-string integrals
$Y^\tau_{\vec{\eta}}(\gamma|\rho )$ in (\ref{looprev.8}) in turn introduce multiple sums over the
momentum lattice of a torus \cite{Gerken:2018jrq, Gerken:2019cxz}
\beq
\Lambda = \ZZ + \tau \ZZ \, , \ \ \ \ \ \ \Lambda' = \Lambda \setminus \{ 0 \}
\label{newdef.2}
\eeq
that are known as MGFs \cite{DHoker:2015wxz, DHoker:2016mwo}. With the removal of $p=0$ from $\Lambda$, they
can be thought of as infrared-regulated and discretized versions of
Feynman integrals on a torus. The MGFs associated with Feynman graphs of
dihedral topology are defined by\footnote{Note that the definition of ${\cal C}[\ldots]$ in this work follows the conventions of \cite{Gerken:2019cxz, Gerken:2020yii, Gerken:2020aju} but differs from those in \cite{DHoker:2016mwo, DHoker:2016quv, Gerken:2018jrq, DHoker:2019txf} by factors of $\Im \tau$ and $\pi$.}
\beq
\cform{ a_1 &a_2 &\ldots &a_r \\ b_1 &b_2 &\ldots &b_r } = \sum_{p_1,p_2,\ldots,p_r \in \Lambda '} \frac{ \delta(p_1+p_2+\ldots+p_r) }{ p_1^{a_1}\bar p_1^{b_1}
p_2^{a_2}\bar p_2^{b_2} \ldots p_r^{a_r}\bar p_r^{b_r} }\, ,
\label{newdef.3}
\eeq
and more general topologies are for instance discussed in \cite{DHoker:2016mwo, Gerken:2020aju}.
The simplest examples of dihedral MGFs (\ref{newdef.3}) have two columns and are
associated with one-loop graphs on the world-sheet
\begin{align}
\cform{ a &0 \\ b &0 } = \sum_{p \in \Lambda'} \frac{1}{p^a \bar p^b} \, , \label{newdef.4}
\end{align}
whereas $\cform{ a_1 &a_2 &\ldots &a_r \\ b_1 &b_2 &\ldots &b_r }$ are referred to as $(r{-}1)$-loop MGFs.
As long as the entries obey $a+b >2$, the lattice sums (\ref{newdef.4}) are absolutely convergent and the one-loop MGFs are expressible in terms of non-holomorphic Eisenstein series ${\rm E}_k(\tau)$
and their Cauchy--Riemann derivatives
\begin{align}
{\rm E}_k(\tau) = \bigg( \frac{ \Im \tau }{\pi} \bigg)^k \cform{ k &0 \\ k &0 }  \, , \ \ \ \ \ \
\nabla^m {\rm E}_k(\tau) &=  \frac{ (\Im \tau)^{k+m} }{\pi^k} \frac{(k{+}m{-}1)!}{(k{-}1)!} \cform{ k+m &0 \\ k-m &0 }\,,\nn\\
\overline{\nabla}^m {\rm E}_k(\tau) &=  \frac{ (\Im \tau)^{k+m} }{\pi^k} \frac{(k{+}m{-}1)!}{(k{-}1)!} \cform{ k-m &0 \\ k+m &0 }
 \, ,
\label{newdef.5}
\end{align}
where $\nabla = 2i (\Im \tau)^2 \partial_\tau$ and $\overline{\nabla} = -2i (\Im \tau)^2 \partial_{\bar\tau}$.
As will be detailed below, both eMZVs (\ref{newdef.1}) and MGFs such as (\ref{newdef.3})
can be represented via iterated integrals of holomorphic Eisenstein series ${\rm G}_k = \cform{ k &0 \\ 0 &0 }$
defined by (\ref{looprev.12}).
Both eMZVs \cite{Broedel:2015hia} and MGFs \cite{DHoker:2015sve, DHoker:2016mwo, Basu:2016kli, DHoker:2016quv, Gerken:2020aju} exhibit a multitude of relations over rational combinations of MZVs,
all of which are automatically exposed in their iterated-Eisenstein-integral representation.\footnote{This
relies on the linear-independence result of \cite{Nilsnewarticle} on
holomorphic iterated Eisenstein integrals.} A computer
implementation for the decomposition of a large number of eMZVs and MGFs
into basis elements is available in \cite{WWWe} and \cite{Gerken:2020aju}, respectively.

\section{New types of genus-one integrals}
\label{sec:3}

The goal of this paper is to relate the $\alpha'$-expansions of suitable
generating functions of genus-one open- and closed-string integrals. The
$Z^\tau_{\vec{\eta}}(\gamma|\rho )$ and $Y^\tau_{\vec{\eta}}(\gamma|\rho )$
in (\ref{looprev.1}) and (\ref{looprev.8})
can be anticipated to not yet furnish the optimal building blocks for this purpose since
\begin{itemize}
\item [(i)] The $\tau$-dependence $\sim {\rm G}_k(\tau)$ and $\sim (\tau{-}\bar \tau)^{k-2}{\rm G}_k(\tau)$ of the
open- and closed-string differential equations (\ref{looprev.11})
does not match, even in absence of $\bar \tau$.
\item [(ii)] The contributions $\sim   \bar \eta_j \partial_{\eta_j}$ to the
closed-string derivations $R_{\vec{\eta}}(\ep_0)$ in (\ref{looprev.13}) and
(\ref{looprev.14}) do not have any open-string counterpart in $r_{\vec{\eta}}(\ep_0)$.
\end{itemize}
Both of these shortcomings will be fixed by the improved open- and closed-string
generating functions $B^\tau_{\vec{\eta}}(\gamma|\rho )$ and $J^\tau_{\vec{\eta}}(\gamma|\rho )$
to be introduced in this section.

\subsection{Genus-one open-string B-cycle integrals}
\label{sec:3.1}

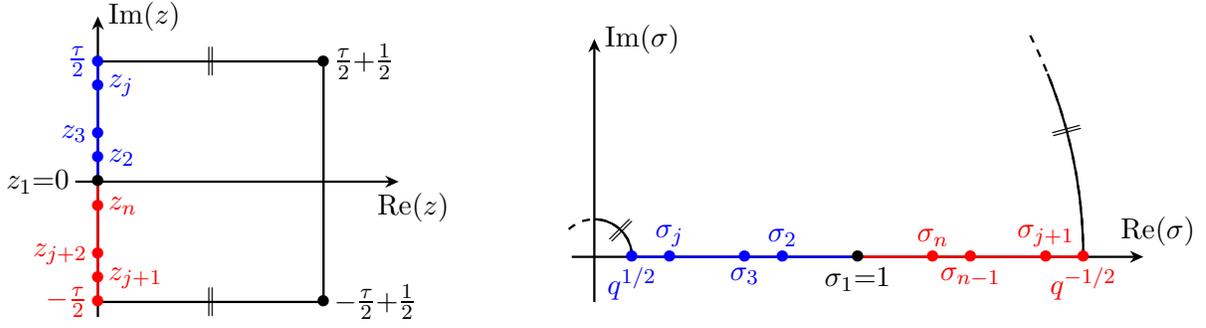
\begin{figure}
\begin{center}
\begin{tikzpicture}[line width=0.30mm]
\begin{scope}[yshift=0cm, xshift=-6.6cm]
  \draw [arrows={-Stealth[width=1.6mm, length=1.8mm]}] (-0.3,0) -- (4.0,0) node[below]{$\quad{\rm Re}(z)$};
  \draw [arrows={-Stealth[width=1.6mm, length=1.8mm]}] (0,-1.8) -- (0,2.2) node[right]{${\rm Im}(z)$};
%
\draw[red](0,-1.6) -- (0,0);
\draw[blue](0,0) -- (0,1.6);
\draw(3,-1.6) -- (3,1.6);
\draw(0,-1.6) -- (3,-1.6);
\draw(0,1.6) -- (3,1.6);
\draw(1.5,1.6)node{$| \! |$};
\draw(1.5,-1.6)node{$| \! |$};
\draw(0,0)node{$\bullet$};
\draw(-0.8,0)node{$z_1{=}0$};
\draw[blue](0,1.6)node{$\bullet$}node[left]{$\frac{\tau}{2}$};
\draw(3,1.6)node{$\bullet$}node[right]{$\frac{\tau}{2}{+}\frac{1}{2}$};
\draw(3,-1.6)node{$\bullet$}node[right]{${-}\frac{\tau}{2}{+}\frac{1}{2}$};
\draw[red](0,-1.6)node{$\bullet$}node[left]{${-}\frac{\tau}{2}$};
\draw[blue](0,0.32)node{$\bullet$}node[right]{$z_2$};
\draw[blue](0,0.64)node{$\bullet$}node[left]{$z_3$};
\draw[blue](0,1.28)node{$\bullet$}node[right]{$z_j$};
\draw[red](-0,-0.32)node{$\bullet$}node[right]{$z_{n}$};
\draw[red](-0,-0.96)node{$\bullet$}node[left]{$z_{j+2}$};
\draw[red](-0,-1.28)node{$\bullet$}node[right]{$z_{j+1}$};
\end{scope}
\begin{scope}[yshift=-1.0cm]
  \draw [arrows={-Stealth[width=1.6mm, length=1.8mm]}] (-0.3,0) -- (7.3,0) node[above]{$\quad{\rm Re}(\sigma)$};
  \draw [arrows={-Stealth[width=1.6mm, length=1.8mm]}] (0,-0.6) -- (0,2.9) node[right]{${\rm Im}(\sigma)$};
\draw[red] (6.5,0) -- (3.5,0);
\draw[blue] (0.5,0) -- (3.5,0);
\draw (0.5,0) arc (0:90:0.5cm);
\draw[dashed] (0.5,0) arc (0:135:0.5cm);
\draw (6.5,0) arc (0:22:6.5cm);
\draw[dashed] (6.5,0) arc (0:27:6.5cm);
\draw[rotate=45] (0.5,0)node[rotate=-45]{ $| \! |$};
\draw[rotate=15] (6.5,0)node[rotate=-75]{ $| \! |$};
  \draw[blue] (0.5,0)node{$\bullet$}  node[below]{$q^{1/2}$};
  \draw[blue] (1,0)node{$\bullet$}  node[above]{$\sigma_j$};
  \draw[blue] (2.0,0)node{$\bullet$}  node[below]{$\sigma_3$};
  \draw[blue] (2.5,0)node{$\bullet$}  node[above]{$\sigma_2$};
   \draw (3.5,0)node{$\bullet$}  node[below]{$\sigma_1{=}1$};
  \draw[red] (4.5,0)node{$\bullet$}  node[above]{$\sigma_n$};
  \draw[red] (5,0)node{$\bullet$}  node[below]{$\sigma_{n-1}$};
  \draw[red] (6,0)node{$\bullet$}  node[above]{$\sigma_{j+1}$};
  \draw[red] (6.5,0)node{$\bullet$}  node[below]{$q^{-1/2}$};
\end{scope}
\end{tikzpicture}
\end{center}
\caption{\textit{The parametrization (\ref{newBB.2}) of the B-cycle is mapped to the positive real
axis in the $\sigma_j= e^{2\pi i z_j}$ variables which exhausts all of $\RR^+$ as $\tau \rightarrow i\infty$
and $q = e^{2\pi i \tau} \rightarrow 0$. The line segments in the $z$-coordinate and the semicircles in the
$\sigma$-coordinate marked by $| \! |$ are identified by the periodic direction of the cylinder, i.e.\ the
B-cycle of the parental torus.}}
\label{BtoJ2}
\end{figure}

Instead of parametrizing the cylinder boundary
through the A-cycle of a torus as in (\ref{looprev.1}), one can perform a modular $S$ transformation
\begin{align}
\label{newBB.1}
B^\tau_{\vec{\eta}}(\gamma|\rho ) =
Z^{-1/\tau}_{\vec{\eta}}(\gamma|\rho )
=\int\limits_{\mathfrak B(\gamma)} \! \! \Big( \prod_{j=2}^n \dd z_j \Big)  \vph^\tau_{\tau\vec{\eta}}(1,\rho(2,\ldots,n)) \! \! \prod_{1\leq i <j }^n \! \!  e^{s_{ij} {\cal G}_{\mathfrak B}(z_{ij},\tau)}
\end{align}
to attain a parametrization through the B-cycle (recalling that $z_1=0$ and $\tau\in i\mathbb{R}^+$)
\begin{align}
\label{newBB.2}
\mathfrak B(\gamma) &= \bigoplus_{j=1}^n \mathfrak B_j(\gamma)   \\
\mathfrak B_j(\gamma)&= \{ z_{i}=\tau u_i ,  \ \ -\tfrac{1}{2}<u_{\gamma(j+1)}<u_{\gamma(j+2)}< \! \ldots \! < u_{\gamma(n)} <0< u_{\gamma(2)}<u_{\gamma(3)}<\! \ldots \!<u_{\gamma(j)}<\tfrac{1}{2}\} \, , \notag
\end{align}
where $u_i \in \mathbb R$, and the B-cycle Green function ${\cal G}_{\mathfrak B}(z,\tau)$ is constructed in two steps: First, we define ${\cal G}_{\mathfrak B}(z,\tau)$ for $z$ on the line $(0,\tau)$ by \cite{Zerbini:2018hgs}
\begin{align}
{\cal G}_{\mathfrak B}(z,\tau) &=
{\cal G}_{\mathfrak A}\Big(\frac{z}{\tau},-\frac{1}{\tau} \Big)
= - \frac{ i \pi z^2 }{\tau}
- \log \Big( \frac{ \theta_1(z,\tau) }{ \eta(\tau) } \Big) - \frac{ i \pi }{6 \tau} + i \pi\,,
\qquad z\in(0,\tau)
\label{newBB.3z}
\,.
\end{align}
Then, we extend this to $z\in(-\tau,0)$ by imposing $\mathcal{G}_{\mathfrak{B}}(z,\tau)=\mathcal{G}_{\mathfrak{B}}(-z,\tau)$ for compatibility with \eqref{looprev.3} under modular $S$ transformations, leading to the combined expression
\begin{align}
{\cal G}_{\mathfrak B}(u\tau,\tau) &=
-  i \pi u^{2}\tau
- \log \Big( \frac{ \theta_1(|u|\tau,\tau) }{ \eta(\tau) } \Big) - \frac{ i \pi }{6 \tau} + i \pi\,,
\qquad u\in(-1,1)
 \label{newBB.3}
\,.
\end{align}
Instead of integrating over $z_i =\tau u_i$ with $u_i \in (0,1)$, we have chosen the
representative $u_i \in (-\frac{1}{2},\frac{1}{2})$ of the B-cycle in order to facilitate
the comparison with genus-zero integration cycles as $\tau \rightarrow i \infty$. Figure
\ref{BtoJ2} illustrates the integration cycle (\ref{newBB.2}) in both the $z_j$ and $\sigma_j = e^{2\pi i z_j}$
variables (the latter becoming the coordinates on the sphere as $\tau \rightarrow i \infty$), where $z_{j}\in i\mathbb{R}$ and $\sigma_{j}\in \mathbb{R}^{+}$ for purely imaginary choices of $\tau$. Note that non-planar versions of the B-cycle integrals involve additional punctures at $z_j \in \frac{1}{2} + i \mathbb R$
or negative $\sigma_j \in (-q^{-1/2},-q^{1/2})$.

The modular transformation $\Omega(z,\eta,-\frac{1}{\tau}) = \tau \Omega(\tau z, \tau \eta,\tau)$
of the doubly-periodic Kronecker--Eisenstein series (\ref{looprev.4})
leads to the rescaling $\eta_j \rightarrow \tau \eta_j$ in the subscript of the
$\rho$-dependent integrand $\varphi^\tau_{\tau \vec{\eta}}$ of (\ref{newBB.1}).

\subsection{Dual closed-string integrals}
\label{sec:3.2}

The doubly-periodic integrands $\vph^\tau_{\vec{\eta}}$
in (\ref{looprev.5}) are non-holomorphic, so their complex conjugates in (\ref{looprev.8}) obey
\beq
\partial_{z_j} \overline{ \vph^\tau_{\vec{\eta}}(1,\gamma(2,\ldots,n)) }=
\frac{2\pi i \bar \eta_j}{\tau{-}\bar \tau}   \overline{ \vph^\tau_{\vec{\eta}}(1,\gamma(2,\ldots,n))}
 \label{newBB.5}
\eeq
which leads to the terms $\sim  \bar \eta_j \partial_{\eta_j}$ in the closed-string derivations $R_{\vec{\eta}}(\ep_0)$ in (\ref{looprev.14}).
This introduces a tension between the open- and closed-string differential equations (\ref{looprev.11}) such that the
$\vph^\tau_{\vec{\eta}}$ do not qualify as Betti--deRham duals of open-string integration cycles.
In order to generalize the interplay of Parke--Taylor factors (\ref{treerev.4}) with single-valued integration \cite{Schnetz:2013hqa, Brown:2018omk} to genus one, the factor of $\overline{ \vph^\tau_{\vec{\eta}}(\ldots) }$ in the $Y^\tau_{\vec{\eta}}$ integrals (\ref{looprev.8}) needs to be replaced by an antimeromorphic function that is still well-defined on the torus, i.e.\ the complex conjugate of an elliptic function in all of $z_1,z_2,\ldots, z_n$.

Such elliptic functions of $n$ punctures can be generated by cycles of Kronecker--Eisenstein series \cite{Dolan:2007eh}
\begin{align}
 \Omega(z_{12},\xi,\tau) \Omega(z_{23},\xi,\tau) \ldots \Omega(z_{n1},\xi,\tau)
=: \xi^{-n} \sum_{w=0}^\infty \xi^w V_w(1,2,\ldots,n|\tau)  \ , \label{newBB.0}
\end{align}
where $V_w$ has holomorphic modular weight $w$.
Even though the individual Kronecker--Eisenstein series $\Omega$ are not meromorphic in the $z_j$,
the $V_w$ are elliptic functions since the non-holomorphic phase factors in (\ref{looprev.4})
cancel from the cyclic product in~\eqref{newBB.0}. The simplest examples are
\begin{align}
V_0(1,2,\ldots,n|\tau) &= 1 \notag 
\\
V_1(1,2,\ldots,n|\tau) &= \sum_{j=1}^n f^{(1)}(z_j{-}z_{j+1},\tau) 
\label{V012ex} \\
V_2(1,2,\ldots,n|\tau) &= \sum_{j=1}^n f^{(2)}(z_j{-}z_{j+1},\tau)
+ \sum_{1\leq j<k}^{n}  f^{(1)}(z_j{-}z_{j+1},\tau)  f^{(1)}(z_k{-}z_{k+1},\tau)
\notag
\end{align}
with $z_{n+1}=z_1$ and Kronecker--Eisenstein coefficients
$f^{(w)}$ defined by (\ref{looprev.6}), also see (\ref{shortv}) for the analogous expressions at general $w$.

As will be detailed below, these elliptic functions degenerate to suitable combinations of Parke--Taylor factors
when forming the linear combinations
\beq
V(1,2,\ldots,n|\tau) = \sum_{w=0}^{n-2} \frac{ V_w(1,2,\ldots,n|\tau) }{(2\pi i)^w \, (n{-}w{-}1)!}
 \label{newBB.6}
\eeq
such as (see sections \ref{sec:5.4} and \ref{sec:5.5} for detailed discussions of the three-
and four-point examples)
\begin{align}
\label{newBB.7}
V(1,2|\tau)&=1\, , \ \ \ \ \ \  \ \ \ \ \ \  \ \ \ \ \ \  V(1,2,3|\tau) = \frac{1}{2} + \frac{V_1(1,2,3|\tau) }{2\pi i}
\notag \\
V(1,2,3,4|\tau)&=\frac{1}{6} + \frac{1}{2} \frac{V_1(1,2,3,4|\tau) }{2\pi i}  + \frac{V_2(1,2,3,4|\tau) }{(2\pi i)^2}
\\
V(1,2,3,4,5|\tau)&=\frac{1}{24} + \frac{1}{6} \frac{V_1(1,2,3,4,5|\tau) }{2\pi i}     + \frac{1}{2} \frac{V_2(1,2,3,4,5|\tau) }{(2\pi i)^2}
+ \frac{V_3(1,2,3,4,5|\tau) }{(2\pi i)^3}  \, . \notag
\end{align}
To lend credence to this definition of the $V$-function, let us see how their properties parallel those of the genus-zero case:
The Betti--deRham duality at genus zero relies on the simple-pole residues
\beq
    {\rm Res}_{z_j=z_{j\pm 1}} {\rm PT}(1,2,\ldots,j,\ldots,n)
= \pm {\rm PT}(1,2,\ldots,j{-}1,j{+}1,\ldots,n)
\eeq
of the Parke--Taylor factors (\ref{treerev.4}).
These residues correspond to the situation when two neighboring points of the disk ordering (\ref{treerev.2}) at $z_j = z_{j\pm 1}$ come together, which is crucial for sphere integrals being
single-valued disk integrals~\cite{Brown:2019wna, Schlotterer:2018abc}.

Similarly, at genus one, the generating function (\ref{newBB.0}) of the elliptic $V_w$ functions
exposes the recursive structure of their simple-pole residues
\begin{align}
{\rm Res}_{z_j=z_{j\pm 1}}V_w(1,2,\ldots,j,\ldots,n)
= \pm V_{w-1}(1,2,\ldots,j{-}1,j{+}1,\ldots,n)
\label{resid1}
\end{align}
and the absence of higher poles in $z_j{-}z_{j\pm1}$. Consequently, the pole structure of
the elliptic combinations (\ref{newBB.6})
\begin{align}
{\rm Res}_{z_j=z_{j\pm 1}}V(1,2,\ldots,j,\ldots,n)
= \pm \frac{1}{2\pi i} V(1,2,\ldots,j{-}1,j{+}1,\ldots,n)
\label{resid2}
\end{align}
mirrors the boundaries of the open-string integration cycles as $z_j=z_{j\pm 1}$,
i.e.\ one recovers mutually consistent $V$-functions and cycles at multiplicity $n{-}1$ in both cases.\footnote{On the closed-string side of the `genus-one Betti--deRham duality' we note that, by double-periodicity of the $V$-functions, additional poles with identical residues occur as
$z_j\rightarrow z_{j\pm 1} + m\tau +n$ ($m,n,\in\ZZ$).
On the open-string side in turn, the delimiters of the integration cycles in the B-cycle parametrization of figure~\ref{BtoJ2} are separated by $\tau$.}

The absence of $V_w$ with $w\geq n{-}1$ in (\ref{newBB.6}) can be understood from
\begin{itemize}
\item the vanishing of $V_{n-1}(1,2,\ldots,n|\tau)$ since (\ref{newBB.0}) would otherwise be an elliptic function of $\xi$ with a simple pole at the origin \cite{Dolan:2007eh}
\item the breakdown of uniform transcendentality when expanding Koba--Nielsen integrals
involving $V_{n}(1,2,\ldots,n|\tau)$ \cite{Gerken:2018jrq} (which is in tension
with the transcendentality properties of open-string integrals \cite{Mafra:2019xms})
%
\item the fact that $V_{w\geq n+1}(1,2,\ldots,n|\tau)$ is expressible in terms of ${\rm G}_{w-k}
V_{k}(1,2,\ldots,n|\tau)$ with $k \leq n{-}2$ \cite{Dolan:2007eh}
\end{itemize}
Similar to the closed-string integrals $Y^\tau_{\vec{\eta}}$, we define
an $(n{-}1)! \times (n{-}1)!$ matrix of torus integrals
\begin{align}
\label{newBB.11}
J^\tau_{\vec{\eta}}(\gamma|\rho ) &= (2i)^{n-1} \! \!  \int_{\mathfrak T^{n-1}} \! \! \Big( \prod_{j=2}^n \dd^2 z_j \Big)  \! \!  \prod_{1\leq i <j }^n \! \!  e^{s_{ij} {\cal G}_{\mathfrak T}(z_{ij},\tau)}
 \overline{ V(1,\gamma(2,\ldots,n)|\tau) } \vph^\tau_{(\tau-\bar \tau)\vec{\eta}}(1,\rho(2,\ldots,n))
\end{align}
indexed by permutations $\gamma,\rho\in S_{n-1}$ of (\ref{newBB.6}) and (\ref{looprev.5}).
Note that the cyclic symmetry
\beq
V_w(2,\ldots,n,1|\tau)=V_w(1,2,\ldots,n|\tau) \, , \ \ \ \ \ \
V(2,\ldots,n,1|\tau)=V(1,2,\ldots,n|\tau)
\label{cycVw}
\eeq
exposed by the generating function (\ref{newBB.0}) has been used to bring
the integrand of (\ref{newBB.11}) into the form of $\overline{V(1,\ldots|\tau)}$.

\subsection{Asymptotics at the cusp}
\label{sec:3.3}

The modular $S$ transformation in (\ref{newBB.1}) maps the A-cycle eMZVs (\ref{newdef.1}) in the $\eta_j$- and $\alpha'$-expansion of $Z^\tau_{\vec{\eta}}$ to B-cycle eMZVs \cite{Enriquez:Emzv} in the analogous expansion of $B^\tau_{\vec{\eta}}$.
As detailed in \cite{Broedel:2018izr, Zerbini:2018sox, Zerbini:2018hgs}, the asymptotic expansion of B-cycle eMZVs
as $\tau \rightarrow i\infty$ is governed by Laurent polynomials in $T = \pi \tau\in i \RR^+$ whose coefficients
are $\mathbb Q$-linear combinations of MZVs, for instance
\begin{align}
\omega(0,0,2|{-}\tfrac{1}{\tau}) &= -\frac{T^2}{180} -\frac{ \zeta_2}{2} + \frac{i \zeta_3}{T} + \frac{3 \zeta_4}{2 T^2}
+ {\cal O}(e^{2 i T})
\notag \\
\omega(0,0,1,0|{-}\tfrac{1}{\tau}) &= \frac{i T}{120} - \frac{i \zeta_2}{4 T} - \frac{3 \zeta_3}{4 T^2} + \frac{3 i \zeta_4}{4 T^3}+ {\cal O}(e^{2 i T})
\label{asbemzv} \\
\omega(0,0,3,0|{-}\tfrac{1}{\tau}) &= \frac{i T^3}{1260} - \frac{3 i \zeta_4}{4 T} - \frac{9 \zeta_5}{2 T^2}
+ \frac{15 i \zeta_6}{ 2 T^3} + {\cal O}(e^{2 i T})\, .
\notag
\end{align}
The suppressed terms $ {\cal O}(e^{2 i T})$ are series in $q=e^{2\pi i \tau}=e^{2iT}$ with Laurent polynomials
in $T$ as their coefficients.

The MGFs (\ref{newdef.3}) in the $\eta_j$- and $\alpha'$-expansion of (\ref{looprev.8}) admit similar expansions around the cusp, where the leading term is a Laurent polynomial in $y = \pi \Im \tau$ instead of $T$. The
coefficients in the Laurent polynomials of MGFs were shown to be $\mathbb Q$-linear combinations of
MZVs\footnote{See \cite{Zerbini:2015rss} for an earlier proof of the weaker statement that the
Laurent polynomials of modular graph functions are $\mathbb Q$-linear combinations of
cyclotomic MZVs.} \cite{Panzertalk} and are conjectured to be single-valued MZVs \cite{Zerbini:2015rss, DHoker:2015wxz}. Simple examples of the asymptotics of MGFs include
\begin{align}
{\rm E}_2(\tau) &= \frac{y^2 }{45} + \frac{ \zeta_3}{y} + {\cal O}(e^{-2 y})
\notag \\
\frac{\pi \overline \nabla {\rm E}_2(\tau) }{y^2}&= \frac{2y}{45} - \frac{\zeta_3}{y^2} + {\cal O}(e^{-2 y})
\label{asmgfs} \\
{\rm E}_3(\tau) &=  \frac{2 y^3 }{945}+ \frac{ 3 \zeta_5}{4y^2} + {\cal O}(e^{-2 y})\, ,
\notag
\end{align}
see (\ref{newdef.5}) for the lattice-sum representations of the non-holomorphic Eisenstein series.

In a variety of examples, the Laurent polynomials of MGFs and B-cycle eMZVs
have been related by an extension of the single-valued map (\ref{treerev.6}) to \cite{Broedel:2018izr, Gerken:2018jrq, Zagier:2019eus}
\beq
{\rm sv}\, T = 2i y \ \ \ \Leftrightarrow \ \ \  {\rm sv} \, \log(q) = \log|q|^2\, .
\label{onerev.6}
\eeq
By (\ref{asbemzv}) and (\ref{asmgfs}), for instance, the Laurent polynomials of $\omega(0,0,2|{-}\tfrac{1}{\tau})  \longrightarrow {\rm E}_2(\tau)$ as well as $\omega(0,0,1,0|{-}\tfrac{1}{\tau}) \longrightarrow -\frac{3}{8} \frac{\pi \overline \nabla {\rm E}_2(\tau) }{y^2}$ and $\omega(0,0,3,0|{-}\tfrac{1}{\tau})  \longrightarrow 3 {\rm E}_3(\tau)$ are related by (\ref{onerev.6}).

The A-cycle eMZVs in $Z^\tau_{\vec{\eta}}$, by contrast,
enjoy a Fourier expansion in $q=e^{2\pi i \tau}$ whose coefficients are
$\mathbb Q[(2\pi i)^{-1}]$ combinations of MZVs \cite{Enriquez:Emzv, Broedel:2015hia} and do not feature
any analogues of the Laurent polynomials in the expansion of $B^\tau_{\vec{\eta}}$.
This is yet another indication besides their differential equations that the B-cycle integrals (\ref{newBB.1}) are a more suitable starting point for comparison with closed-string integrals than their A-cycle counterparts (\ref{looprev.1}).

\subsection{Single-valued correspondence of the Laurent polynomials}
\label{sec:3.4}

As a particular convenience of the elliptic combinations (\ref{newBB.6}) in the integrands of $J^\tau_{\vec{\eta}}(\gamma|\rho )$, their degeneration
at the cusp gives rise to Parke--Taylor factors in $n{+}2$ punctures ($\sigma_1 = 1$ by $z_1=0$)
\beq
\sigma_j = e^{2\pi i z_j} \, , \ \ \ \ \ \ \sigma_+ = 0 \, , \ \ \ \ \ \ \sigma_- \rightarrow \infty \, .
\label{newBB.12}
\eeq
Since the non-holomorphic exponentials of $\Omega(z,\xi,\tau)=\exp( 2\pi i \xi \frac{ \Im z }{\Im \tau})F(z,\xi,\tau)$ cancel from the cyclic products in (\ref{newBB.0}),
one can determine the asymptotics of $V(\ldots|\tau)$ as $\tau \rightarrow i \infty$ by using the degeneration
of the holomorphic Kronecker--Eisenstein series
\beq
F(z_{ij},\xi,\tau) = \pi \cot( \pi \xi) + i \pi \frac{ \sigma_i + \sigma_j }{\sigma_i - \sigma_j} + {\cal O}(q) \, .
\label{newBB.13}
\eeq
The relative factors of the $V_w$ in (\ref{newBB.6}) have been engineered to obtain the
following cyclic combinations of Parke--Taylor factors at the cusp,
\begin{align}
\label{newBB.14}
&\lim_{\tau \rightarrow i\infty} \frac{ V(1,2,\ldots,n|\tau) }{\sigma_1 \sigma_2 \ldots \sigma_n }
= (-1)^{n-1}  \lim_{\sigma_- \rightarrow \infty} |\sigma_-|^2
 \big[ {\rm PT}(+,n,n{-}1,\ldots,2,1,-) + {\rm cyc}(1,2,\ldots,n) \big]\, ,
\end{align}
which have featured in the context of one-loop gauge-theory amplitudes in ambitwistor
string theories \cite{Geyer:2015bja}. The denominators on the left-hand side of (\ref{newBB.14})
arise from $\dd z_j = \frac{\dd \sigma_j }{2\pi i \sigma_j}$,
and the factor of $|\sigma_-|^2$ on the right-hand side identifies functions on a degenerate torus
with ${\rm SL}_2$-fixed expressions at genus zero~\cite{Zfunctions}. Given that Parke--Taylor factors are Betti--deRham dual
to disk orderings $\mathfrak D(\ldots)$ in (\ref{treerev.2}), the $\tau \rightarrow i\infty$ asymptotics of $J^\tau_{\vec{\eta}}(\gamma|\rho )$
should yield the single-valued map of suitably chosen disk integrals.
In fact, upon rewriting the B-cycle ordering in terms of the $\sigma_j$ variables (\ref{newBB.12}), each of the
contributions $\mathfrak B_j(\gamma)$
in (\ref{newBB.2}) and figure \ref{BtoJ2} degenerates to a single disk ordering
\begin{align}
\mathfrak B_j(\gamma) \, \big|_{\tau \rightarrow i \infty} = (-1)^{n-1} \mathfrak D(+,\gamma(j,j{-}1,\ldots,3,2),1,\gamma(n,n{-}1,\ldots,j{+}1),-)\, ,
\end{align}
such that the overall B-cycle ordering $\mathfrak B(\gamma) = \bigoplus_{j=1}^n \mathfrak B_j(\gamma)$ at the cusp becomes the Betti--deRham dual to the cyclic combination of Parke--Taylor factors in
(\ref{newBB.14}),
\begin{align}
\mathfrak B(2,3,\ldots,n) \, \big|_{\tau \rightarrow i \infty} =  (-1)^{n-1} \bigoplus_{j=1}^n \mathfrak D(+,j,j{-}1,\ldots,3,2,1,n,n{-}1,\ldots,j{+}1,-)\, .
\end{align}
Hence, the tree-level
result (\ref{treerev.7}) provides evidence for our central conjecture
\beq
J^\tau_{\vec{\eta}}(\gamma|\rho ) \, \big|_{\rm LP} = {\rm sv} \, B^\tau_{\vec{\eta}}(\gamma|\rho ) \, \big|_{\rm LP} \, ,
 \label{newBB.15}
\eeq
where the notation $ \big|_{\rm LP}$
instructs to only keep the Laurent polynomials in $\tau$ and $\Im \tau$ while discarding any contribution
$\sim q,\bar q$. The conjectural part of (\ref{newBB.15}) concerns the non-constant terms in the Laurent
polynomials, i.e.\ corrections $\sim (\log q)^{\pm 1}$ to the expansion
around the cusp $q=0$, so it is not implied by the Betti--deRham duality of (\ref{newBB.2}) and (\ref{newBB.14})
which only holds at the cusp.
That is why we support (\ref{newBB.15}) by extensive tests at low orders in $\eta_j,\alpha'$ as detailed below,
and by the fact that the asymptotic expansions of the Green functions are related by the single-valued
map with ${\rm sv} \, \log \sigma_{ij} = \log |\sigma_{ij}|^2$,
\begin{align}
{\cal G}_{\mathfrak{B}}(z_{ij},\tau) \, \big|_{\rm LP}&=  -  \frac{  i T}{6} - \frac{ i \zeta_2 }{T }
 + \frac{1}{2} (\log \sigma_i + \log \sigma_j)  - \log |\sigma_{ij}| +  \frac{i (\log \sigma_i - \log \sigma_j)^2 }{4 T}
 \label{newBB.71} \\
{\cal G}_{\mathfrak T}(z_{ij},\tau) \, \big|_{\rm LP}&=  \frac{y}{3}
+ \log |\sigma_i| + \log |\sigma_j| -2 \log |\sigma_{ij}|
+  \frac{\big(\!  \log |\sigma_i| - \log |\sigma_j|  \big)^2}{2y}  \notag \\
&= {\rm sv} \, G_{\mathfrak{B}}(z_{ij},\tau) \, \big|_{\rm LP} \, .
 \label{newBB.72}
\end{align}
Note that the absolute value in \eqref{newBB.71} is due to the argument $|u|\tau$ of $\theta_{1}$ in \eqref{newBB.3}. For the two-point instances $B^\tau_{\vec{\eta}}$ and $J^\tau_{\vec{\eta}}$
of the open- and closed-string integrals (\ref{newBB.1}) and (\ref{newBB.11}), the Laurent
polynomials in the asymptotics at the cusp can be determined \cite{yinprogress} by mild generalizations
of the techniques in \cite{DHoker:2019xef, Zagier:2019eus} (also see \cite{Vanhove:2020qtt} for an
alternative approach to the closed-string case):
\begin{align}
B^\tau_{\eta}(2|2 ) \, \big|_{\rm LP} &= \exp\Big(  {-} \frac{i s_{12} T }{6} - \frac{i s_{12} \zeta_2}{T} \Big)  \label{apeq.1} \\
&\hspace{-2.1cm}\times \bigg\{
 \big[ i \cot( \eta_2 T ) {+} 1 \big] \exp \Big( \frac{ i s_{12} }{4 T}  \partial_{\eta_2}^2 \Big) \frac{1}{s_{12}{+}2\eta_2} \bigg[
 \frac{ \Gamma(1{+}\tfrac{ s_{12}}{2} {+} \eta_2) \Gamma(1{-}s_{12} ) }{\Gamma(1 {-} \tfrac{ s_{12}}{2} {+} \eta_2)}
  - e^{i T(\tfrac{ s_{12} }{2} + \eta_2) }\bigg] \notag \\
 &\hspace{-2.1cm} \ \ +  \big[ i \cot( \eta_2 T ) {-} 1 \big] \exp \Big( \frac{ i s_{12} }{4 T}  \partial_{\eta_2}^2 \Big)
 \frac{1}{s_{12}{-}2\eta_2} \bigg[
 \frac{ \Gamma(1{+}\tfrac{ s_{12}}{2} {-} \eta_2) \Gamma(1{-}s_{12} ) }{\Gamma(1 {-} \tfrac{ s_{12}}{2} {-} \eta_2)}
  - e^{i T(\tfrac{ s_{12} }{2} - \eta_2) }\bigg]
   \notag \\
 &\hspace{-2.1cm} \ \ + \frac{1}{s_{12}} \exp \Big( \frac{ i s_{12} }{4 T}  \partial_{\eta_2}^2 \Big) \bigg[
  \frac{ \Gamma(1{+}\tfrac{ s_{12}}{2} {-} \eta_2) \Gamma(1{-}s_{12} ) }{\Gamma(1 {-} \tfrac{ s_{12}}{2} {-} \eta_2)}
  -  \frac{ \Gamma(1{+}\tfrac{ s_{12}}{2} {+} \eta_2) \Gamma(1{-}s_{12} ) }{\Gamma(1 {-} \tfrac{ s_{12}}{2} {+} \eta_2)}  \bigg]
\bigg\}
\notag \\
J^\tau_{\eta}(2|2 ) \, \big|_{\rm LP} &= \exp\Big(  \frac{ s_{12} y }{3} \Big) \label{apeq.2} \\
&\hspace{-2.1cm}\times \bigg\{
 \big[ i \cot( 2i\eta_2 y ) {+} 1 \big] \exp \Big( \frac{ s_{12} }{8y}  \partial_{\eta_2}^2 \Big) \frac{1}{s_{12}{+}2\eta_2} \bigg[
 \frac{ \Gamma(1{+}\tfrac{ s_{12}}{2} {+} \eta_2) \Gamma(1{-}s_{12}) \Gamma(1{+}\tfrac{ s_{12}}{2} {-} \eta_2)   }{\Gamma(1 {-} \tfrac{ s_{12}}{2} {+} \eta_2)  \Gamma(1{+}s_{12}) \Gamma(1 {-} \tfrac{ s_{12}}{2} {-} \eta_2)}
  - e^{-y( s_{12} +2 \eta_2) }\bigg] \notag \\
 &\hspace{-2.1cm} \ \ +  \big[ i \cot( 2i\eta_2 y ) {-} 1 \big] \exp \Big( \frac{ s_{12} }{8y}  \partial_{\eta_2}^2 \Big)
 \frac{1}{s_{12}{-}2\eta_2} \bigg[
 \frac{
\Gamma(1{+}\tfrac{ s_{12}}{2} {+} \eta_2) \Gamma(1{-}s_{12}) \Gamma(1{+}\tfrac{ s_{12}}{2} {-} \eta_2)   }{\Gamma(1 {-} \tfrac{ s_{12}}{2} {+} \eta_2)  \Gamma(1{+}s_{12}) \Gamma(1 {-} \tfrac{ s_{12}}{2} {-} \eta_2)}
  - e^{-y(s_{12} -2 \eta_2) }\bigg]
\bigg\}
\notag
\end{align}
These two-point expressions are easily seen to line up with the all-multiplicity claim (\ref{newBB.15}) since
\beq
{\rm sv} \,\bigg[ \frac{\Gamma(1{-}a)\Gamma(1{-}b)}{\Gamma(1{-}a{-}b)}\bigg]
= \frac{\Gamma(1{-}a)\Gamma(1{-}b)\Gamma(1{+}a{+}b)}{\Gamma(1{+}a)\Gamma(1{+}b)\Gamma(1{-}a{-}b)} \, ,
\label{apeq.3}
\eeq
and the last line of (\ref{apeq.1}) therefore vanishes under sv.
Moreover, we have checked the three-point Laurent polynomials
to obey (\ref{newBB.15}) to the orders in the $s_{ij}$- and $\eta_j$-expansions
where MGFs such as (\ref{newdef.3}) of total modular weight 10  occur\footnote{This amounts to performing the $\alpha'$- and $\eta$-expansion to order 10 in the terminology of section 3.4.2 of \cite{Gerken:2020yii}.}.
Finally, we have checked (\ref{newBB.15}) to hold at four points to the orders where MGFs of
total modular weight 8 occur, at least for contributions from $\varphi^\tau_{\vec{\eta}}$ in (\ref{looprev.5})
without any singular factors of $f^{(1)}(z_{ij},\tau)$.\footnote{We have excluded the singular functions
$f^{(1)}(z_{ij},\tau) = \frac{1}{z_{ij}} + {\cal O}(z_{ij})$ in the integrand from our checks to avoid the tedious
treatment of the resulting kinematic poles in the $\alpha'$-expansion. For the contributions of $V_0$ and $V_1$
to the integrand $\overline{V(1,2,3,4|\tau)}$ in (\ref{newBB.7}), we have checked the Laurent polynomials from
up to one factor of $f^{(1)}(z_{ij},\tau)$ in the integrand to obey (\ref{newBB.15}), see section
\ref{sec:3.6} for the disentanglement of different $V_w$ entering $\overline{V(1,2,\ldots,n|\tau)}$.}
These checks are based on Enriquez' methods \cite{Enriquez:Emzv} (also see appendix B of \cite{Broedel:2018izr}) to determine the Laurent polynomials of B-cycle eMZVs. The Laurent polynomials for all B-cycle eMZVs with (length$+$weight) $\leq 16$ obtained from a {\tt FORM} implementation \cite{Vermaseren:2000nd} of these methods are available for download \cite{WWWbcyc}.

While the two-point Laurent-polynomials generated by (\ref{apeq.1}) and (\ref{apeq.2})
only involve Riemann zeta values, higher-point examples also introduce irreducible
MZVs of depth $\geq 2$. The appearance of $\zeta_{3,5}$ in B-cycle Laurent polynomials
is later on exemplified in (\ref{appbb.4}) and (\ref{cusps.5}). Moreover, the
appearance of $\zeta_{3,5,3}$ in open- and closed-string calculations
at three points in agreement with (\ref{newBB.15}) was observed in
section 3.3.5 of \cite{Broedel:2018izr},
based on earlier closed-string computations \cite{Zerbini:2015rss}.

\subsection{Single-valued correspondence of the differential equations}
\label{sec:3.5}

The holomorphic derivatives of the $B^\tau_{\vec{\eta}}$ and $J^\tau_{\vec{\eta}}$-integrals
(\ref{newBB.1}) and (\ref{newBB.11}) can be easily deduced from (\ref{looprev.11}):
In the open-string case, the modular $S$ transformation relating
$B^\tau_{\vec{\eta}}=Z^{-1/\tau}_{\vec{\eta}}$ and the modular weight $(k,0)$ of the holomorphic Eisenstein series ${\rm G}_k$ give rise to
\begin{align}
 \label{newBB.16}
\frac{ \partial}{\partial \log q} B^\tau_{\vec{\eta}}(\gamma|\rho ) &= \frac{1}{(2\pi i)^2} \sum_{k=0}^\infty (1{-}k) \tau^{k-2}{\rm G}_k(\tau)
 \sum_{\alpha \in S_{n-1}} r_{\vec{\eta}}(\ep_k)_\rho{}^\alpha B^\tau_{\vec{\eta}}(\gamma|\alpha )\, ,
\end{align}
see \cite{Mafra:2019xms} for the $n$-point derivations $r_{\vec{\eta}}(\ep_k)$ and (\ref{looprev.13}) for their
two-point examples.
Since the single-valued map at genus zero acts on transcendental constants, we have passed to the differential operator $\frac{ \partial}{\partial \log q}=(2\pi i)^{-1} \partial_\tau$ in comparison to (\ref{looprev.11}) and in preparation for the extended single-valued map to be introduced below around (\ref{newBB.19}).


In the closed-string case, the $\overline{V(\ldots|\tau)}$ in (\ref{newBB.11}) are not affected by holomorphic
derivatives, and one can import a simplified version of the differential equations in \cite{Gerken:2019cxz} where contributions $\sim \bar \eta_j \partial_{\eta_j}$ are absent,
\begin{align}
 \label{newBB.17}
  \frac{\partial}{\partial \log q} J^\tau_{\vec{\eta}}(\gamma|\rho ) &= \frac{1}{(2\pi i)^2} \sum_{k=0}^\infty (1{-}k) (\tau{-}\bar \tau)^{k-2}{\rm G}_k(\tau)
 \sum_{\alpha \in S_{n-1}} {\rm sv} \, r_{\vec{\eta}}(\ep_k)_\rho{}^\alpha J^\tau_{\vec{\eta}}(\gamma|\alpha )\, .
\end{align}
By the differential equations (\ref{looprev.10}) of the Green functions, also the term $\sim \zeta_2$ in
$ r_{\vec{\eta}}(\ep_0)$ is absent which we have indicated through the sv notation,
\begin{align}
{\rm sv} \, r_{\vec{\eta}}(\ep_k) = \left\{ \begin{array}{cl} r_{\vec{\eta}}(\ep_k) &: \ k\geq 4 \\
r_{\vec{\eta}}(\ep_0) - 2 \zeta_2 s_{12\ldots n} &: \ k=0 \end{array} \right.\, ,
\label{newBB.18}
\end{align}
where $r_{\vec{\eta}}(\ep_2)=0$. The building blocks of the closed-string
differential operator in (\ref{newBB.17}) are related to those in the open-string analogue (\ref{newBB.16})
through an extension SV of the single-valued map (recall that $T= \pi \tau$ and $y= \pi \Im \tau$)
\begin{align}
{\rm SV} \,  \bigg[ (2\pi i\tau)^{k-2} \frac{ {\rm G}_k(\tau) }{(2\pi i)^k} r_{\vec{\eta}}(\ep_k) \bigg]
&= {\rm SV} \,  \big[(2 i T)^{k-2} \big] {\rm SV} \,  \bigg[  \frac{ {\rm G}_k(\tau) }{(2\pi i)^k}  \bigg]
{\rm SV} \,  \big[r_{\vec{\eta}}(\ep_k) \big]\notag \\
&= (-4 y)^{k-2} \frac{ {\rm G}_k(\tau) }{(2\pi i)^k} \,  {\rm sv} \, r_{\vec{\eta}}(\ep_k) \label{newBB.19}\\
&= \frac{1}{(2\pi i)^2}  (\tau{-}\bar\tau)^{k-2} {\rm G}_k(\tau)  \,  {\rm sv} \, r_{\vec{\eta}}(\ep_k)
\notag
\end{align}
%
which is taken to preserve the properties of sv,
\beq
{\rm SV} \,  \zeta_{n_1,n_2,\ldots,n_r} = {\rm sv} \,  \zeta_{n_1,n_2,\ldots,n_r} \, , \ \ \ \ \ \
{\rm SV} \,  T = 2i y
\label{newBB.19b}
\eeq
and to furthermore preserve $(2\pi i)^{-k} {\rm G}_k(\tau)$
(the inverse powers of $\pi$ ensuring rational coefficients in the $q$-expansion) and the $\eta_j$-variables, cf.\ (\ref{newBB.19}).
In other words, the differential operator ${\cal O}^\tau_{\vec\eta}=(2\pi i)^{-2}
\sum_{k=0}^\infty (1{-}k) \tau^{k-2}{\rm G}_k(\tau)  r_{\vec{\eta}}(\ep_k)$
appearing in (\ref{newBB.16}) and its closed-string
analogue in (\ref{newBB.17}) are related by
\beq
\frac{\partial  B^\tau_{\vec{\eta}}(\gamma|\rho )}{\partial \log q}
 =  \! \! \! \sum_{\alpha \in S_{n-1}}  \! \! \! {\cal O}^\tau_{\vec\eta}(\rho|\alpha)B^\tau_{\vec{\eta}}(\gamma|\alpha )
\ \ \ \ \leftrightarrow \ \ \ \
\frac{\partial  J^\tau_{\vec{\eta}}(\gamma|\rho )}{\partial \log q}
 =  \! \! \!  \sum_{\alpha \in S_{n-1}}  \! \! \! \big[{\rm SV} \, {\cal O}^\tau_{\vec\eta}(\rho|\alpha) \big]J^\tau_{\vec{\eta}}(\gamma|\alpha ) \, .
\label{newBB.19a}
\eeq
From the above discussion, both the $\tau \rightarrow i \infty$ asymptotics and the differential operators
of the open- and closed-string integrals $B^\tau_{\vec{\eta}}$ and $J^\tau_{\vec{\eta}}$ are related by the SV map
(\ref{newBB.19b}). Hence, we propose that the solutions
of (\ref{newBB.19a}) yield an appropriate extension of the SV map
\beq
J^\tau_{\vec{\eta}}(\gamma|\rho )  = {\rm SV} \, B^\tau_{\vec{\eta}}(\gamma|\rho ) \, .
\label{newBB.20}
\eeq
This proposal is key the result of this work, relating the open-string integrals $B^\tau_{\vec{\eta}}(\gamma|\rho )$ in (\ref{newBB.1}) with integration ordering $\gamma$ 
 to the closed-string integrals $J^\tau_{\vec{\eta}}(\gamma|\rho )$, where the
ordering $\gamma$ governs the singularity structure of the antielliptic integrand $\overline{V}$ in (\ref{newBB.11}). By construction, this SV map commutes with the 
holomorphic $\tau$-derivative and, under the assumption (\ref{newBB.15}),
it is consistent at the level of the Laurent polynomials at the cusp. Compatibility
of a single-valued map at genus one with $\tau \rightarrow i \infty$ generalizes the fact that the single-valued
map of multiple polylogarithms commutes with evaluation \cite{Brown:2013gia}. Moreover,
by the evidence to be discussed in section \ref{sec:4.3},
the SV map is expected to be compatible with the shuffle product.
As we will see in the next section, the $\ap$-expansion of (\ref{newBB.20}) induces an elliptic single-valued
map for the eMZVs generated by $B^\tau_{\vec{\eta}}$ which yields the MGFs generated by $J^\tau_{\vec{\eta}}$.

Let us consider the scope of our definition~\eqref{newBB.20}.
Firstly, not all the holomorphic iterated Eisenstein integrals appear in the $\alpha'$-expansion of
$B^\tau_{\vec{\eta}}$. As was discussed in~\cite{Broedel:2015hia, Gerken:2020yii} and will become clearer when we discuss the $\alpha'$-expansion of the solution of~\eqref{newBB.19a}, relations among the $r_{\vec{\eta}}(\epsilon_k)$
such as (\ref{derrels}) lead to dropouts of certain iterated Eisenstein integrals from eMZVs and $Y^\tau_{\vec{\eta}}$ and thereby from
$B^\tau_{\vec{\eta}}$ and $J^\tau_{\vec{\eta}}$. Hence, (\ref{newBB.20}) does not comprise the SV map for the
combinations of iterated Eisenstein integrals affected by these dropouts, starting with double integrals involving ${\rm G}_4$ and ${\rm G}_{10}$.

By contrast, the SV map of arbitrary convergent eMZVs can be extracted from
(\ref{newBB.20}) at sufficiently high multiplicity:
As will be detailed in section \ref{sec:5.8}, for $\omega(n_1,\ldots,n_r|\tau)$ in
(\ref{newdef.1}) with given entries $n_j$ (where $n_1,n_r \neq 1$), one can engineer a combination of genus-one
open-string integrals, where the desired eMZV occurs at the zeroth order in $s_{ij}$.\footnote{By a similar argument, each MGF can be realized in the $s_{ij}^0$-order of $Y^\tau$-integrals at sufficiently high multiplicity, see section 2.5 of \cite{Gerken:2019cxz}.}

Finally, one could wonder whether holomorphic cusp forms lead to ambiguities in the definition of the
$V(1,\ldots,n|\tau)$ in (\ref{newBB.6}):\footnote{We are grateful to Nils Matthes for valuable discussions
on this point.}
Starting from $n=14$ points, their defining properties including simple-pole residues,
the modularity of their constituents
and their behavior (\ref{newBB.14}) at the cusp are unchanged when adding
combinations of holomorphic cusp forms and lower-weight $V_w(1,\ldots,n|\tau)$.
However, adding a cusp form without any $z_j$-dependent coefficient to $V(1,\ldots,n|\tau)$
leads to a contradiction with the requirement that the $\tau$-independent $\eta^{1-n}$ order
of $B^\tau_{\vec{\eta}}$ is mapped to the same term $\sim \eta^{1-n}$ in the corresponding $J^\tau_{\vec{\eta}}$ integral.
Products of $V_w(1,\ldots,n|\tau)$ with cusp forms in turn would violate the pole structure
(\ref{resid2}) that reflects the boundary structure of the dual cycles. Hence, the above requirements do not
leave any room to modify $V(1,\ldots,n|\tau)$ by holomorphic cusp forms.

\subsection{Dual modular weights for cycles}
\label{sec:3.6}

Given that the antielliptic $\overline{V_w(\ldots|\tau)}$-functions (\ref{newBB.0}) carry
modular weight $(0,w)$, their combinations $\overline{V(\ldots|\tau)}$ (\ref{newBB.6})
mix different modular weights. Hence, the $\alpha'$-expansion of the generating function
(\ref{newBB.11}) with $\overline{V(\ldots|\tau)}$ in the integrand mixes modular
forms of different weight, even at fixed order in $\eta_j$. One may wish to
isolate the contributions at fixed modular weights and study
\begin{align}
\label{mix.1}
\! \! \! \! J^\tau_{w,\vec{\eta}}(\gamma|\rho ) &= \frac{ (2i)^{n-1} }{(-2\pi i)^w} \! \!  \int_{\mathfrak T^{n-1}} \! \! \Big( \prod_{j=2}^n \dd^2 z_j \Big)  \! \!  \prod_{1\leq i <j }^n \! \!  e^{s_{ij} {\cal G}_{\mathfrak T}(z_{ij},\tau)}
 \overline{ V_w(1,\gamma(2,\ldots,n)|\tau) } \vph^\tau_{(\tau-\bar \tau)\vec{\eta}}(1,\rho(2,\ldots,n)) \, ,
\end{align}
with $0\leq w\leq n{-}2$,
where the terms at homogeneity degree $m$ in the $\eta_j$ are modular forms of weight $(0,1{-}n{-}m{+}w)$.
One can still identify combinations of integration cycles (\ref{newBB.2}) to write (\ref{mix.1})
at fixed modular weight $w$ and ordering $\gamma$ as the single-valued version of known open-string integrals:
Each $V_w(1,2,\ldots,n|\tau)$ with $w\leq n{-}2$ is expressible via permutation sums
\beq
V_w(1,2,\ldots,n|\tau) =  (2\pi i)^w \sum_{\gamma \in S_{n-1}} c_{w,\gamma} V(1,\gamma(2,\ldots,n)|\tau)
\label{mix.71}
\eeq
with coefficients $c_{w,\gamma} \in \QQ$, e.g.\
\begin{align}
V_0(1,\ldots,n|\tau) &= 1 = \sum_{\gamma \in S_{n-1}}V(1,\gamma(2,\ldots,n)|\tau)
\label{mix.2} \\
V_1(1,2,3|\tau) &= i \pi \big[ V(1,2,3|\tau)- V(1,3,2|\tau) \big]
\label{mix.3} \\
V_1(1,2,3,4|\tau) &=2 \pi i \big[ V(1,2,3,4|\tau)- V(1,4,3,2|\tau) \big]
\label{mix.3a} \\
V_2(1,2,3,4|\tau) &= \frac{(2\pi i)^2}{6}\big[  2V(1,2,3,4|\tau)+2V(1,4,3,2|\tau)  -V(1,2,4,3|\tau)\notag \\
& \ \ \ \
-V(1,3,4,2|\tau) -V(1,3,2,4|\tau)-V(1,4,2,3|\tau)\big]\,.
\label{mix.4}
\end{align}
These relations and coefficients $ c_{w,\gamma}$ can be traced back to the symmetries
of the $V_w$-functions including the cyclicity (\ref{cycVw}), the reflection property
\beq
V_w(1,2,\ldots,n|\tau) = (-1)^{w} V_w(n,\ldots,2,1|\tau)
\eeq
and corollaries of the Fay identity \cite{Mumford:Tata1} which have
been discussed in \cite{Dolan:2007eh, Mafra:2018pll}. An independent method
based on the degeneration (\ref{newBB.14}) to determine the $c_{w,\gamma}$
is described in appendix \ref{appcws}.
As a result, there are
less than $(n{-}1)!$ independent permutations $V_w(1,\gamma(2,\ldots,n)|\tau)$
at fixed $0\leq w\leq n{-}2$ and $n \geq 3$. Their counting is governed by the unsigned Stirling
number $S_{n-1,n-w-1}$ of the first kind (where $S_{a,b}$ counts the number of permutations
of $a$ elements with $b$ disjoint cycles) as exemplified in table \ref{tabonvw}.
%
\begin{table}[h]
\begin{center}
\begin{tabular}{c||c|c|c|c|c|c}
\diagbox[]{$n \! \!$}{$\! \! w$} &0&1&2&3&4&5 \\\hline\hline
2 &1 &0&0&0&0&0 \\\hline
3 &1 &1&0&0&0&0 \\\hline
4 &1 &3&2&0&0&0 \\\hline
5 &1 &6&11&6&0&0 \\\hline
6 &1 &10&35&50&24&0 \\\hline
7 &1 &15&85&225&274&120
\end{tabular}
\end{center}
\caption{\textit{Examples of the unsigned Stirling numbers $S_{n-1,n-w-1}$ which count the
number of independent permutations $\gamma \in S_{n-1}$ of $V_w(1,\gamma(2,\ldots,n)|\tau)$.}}
\label{tabonvw}
\end{table}
In particular, permutations of $V_{w=n-2}( 1,\ldots,n|\tau)$ are related by Kleiss--Kuijf relations \cite{Kleiss:1988ne, Mafra:2018pll}
\beq
V_{n-2}(1,(a_2,\ldots,a_j) \shuffle (a_{j+1},\ldots,a_n)|\tau) = 0 \, , \ \ \ \ \ \ j=2,3,\ldots,n{-}1
\label{mix.72}
\eeq
such as
\beq
V_1(1,2,3|\tau) = - V_{1}(1,3,2|\tau)  \, , \ \ \  \ \ \ V_2(1,2,3,4|\tau) + {\rm cyc}(2,3,4)  = 0\, ,
\label{mix.73}
\eeq
consistent with the counting $S_{n-1,1}=(n{-}2)!$ of independent permutations.

Given the decomposition (\ref{mix.71}) of a given $V_w$ function with rational coefficients $c_{w,\gamma}$,
one can by (\ref{newBB.20}) write each $J^\tau_{w,\vec{\eta}}$ integral (\ref{mix.1}) as a combination
of single-valued B-cycle integrals
\beq
J^\tau_{w,\vec{\eta}}(2,\ldots,n|\rho )  = {\rm SV} \sum_{\gamma \in S_{n-1}} c_{w,\gamma}
B^\tau_{\vec{\eta}}(\gamma |\rho ) \, .
\label{mix.74}
\eeq
For instance, the equivalent
\begin{align}
J^\tau_{0,\eta_2,\eta_3}(2{,}3|\rho ) &=  {\rm SV} \big[ B^\tau_{\eta_2,\eta_3}(2{,}3|\rho ) + B^\tau_{\eta_2,\eta_3}(3{,}2|\rho ) \big]   \label{mix.5}
\\
J^\tau_{1,\eta_2,\eta_3}(2{,}3|\rho ) &= \frac{1}{2} {\rm SV} \big[ B^\tau_{\eta_2,\eta_3}(2{,}3|\rho ) - B^\tau_{\eta_2,\eta_3}(3{,}2|\rho ) \big] \notag
\end{align}
of (\ref{newBB.20}) together with (\ref{mix.2}) and (\ref{mix.3})
suggests to assign a formal ``dual modular weight'' $0$ and $1$
to the symmetric and antisymmetric three-point
cycles, respectively,
\begin{align}
\mathfrak B(2,3) + \mathfrak B(3,2)  \ \ &\leftrightarrow \ \ {\rm dual}\ {\rm modular} \ {\rm weight} \ 0
\label{mix.75.1} \\
\mathfrak B(2,3) - \mathfrak B(3,2)  \ \ &\leftrightarrow \ \ {\rm dual}\ {\rm modular} \ {\rm weight} \ 1\, .
\notag
\end{align}
Similarly, combining (\ref{newBB.20}) with (\ref{mix.2}), (\ref{mix.3a}) and (\ref{mix.4})
leads to the following dual modular weights (d.m.w.) for four-point cycles
\begin{align}
\mathfrak B(2,3,4){+}\mathfrak B(4,3,2)  {+}\mathfrak B(2,4,3) {+}\mathfrak B(3,4,2) {+}\mathfrak B(3,2,4){+}\mathfrak B(4,2,3)
 \ \ &\leftrightarrow \ \ {\rm d.m.w.} \ 0
 \notag \\
 \mathfrak B(2,3,4)-\mathfrak B(4,3,2)
  \ \ &\leftrightarrow \ \ {\rm d.m.w.} \ 1
  \label{mix.75.2} \\
2\mathfrak B(2,3,4){+}2\mathfrak B(4,3,2)  {-}\mathfrak B(2,4,3) {-}\mathfrak B(3,4,2) {-}\mathfrak B(3,2,4){-}\mathfrak B(4,2,3)
 \ \ &\leftrightarrow \ \ {\rm d.m.w.} \ 2\, , \notag
\end{align}
see section \ref{sec:5.5} for a more detailed discussion of the weight-two case. Finally, the all-multiplicity formula (\ref{mix.71}) translates into
\beq
 \sum_{\gamma \in S_{n-1}} c_{w,\gamma}
\mathfrak B(\gamma(2,3,\ldots,n)) \ \ \leftrightarrow \ \ {\rm dual}\ {\rm modular} \ {\rm weight} \ w \, ,
\label{mix.75}
\eeq
see appendix \ref{appcws} for the rational coefficients $c_{w,\gamma}$ and table \ref{tabonvw} for
the counting of independent $n$-point cycles with dual modular weight $w$.

\section{Single-valued iterated Eisenstein integrals from $\alpha'$-expansions}
\label{sec:4}

The goal of this section is to provide the explicit form of the single-valued map SV for the iterated-Eisenstein-integral representation of eMZVs \cite{Broedel:2015hia}
by reading (\ref{newBB.20}) at the level of the $\alpha'$- and $\eta_j$-expansions of
$B^\tau_{\vec{\eta}}$ and $J^\tau_{\vec{\eta}}$.
We will employ the formulation of iterated Eisenstein integrals
with integration kernels $\tau^j {\rm G}_k, \ k \geq 4$ \cite{Brown:mmv},
\begin{align}
\EBR{j_1 &j_2 &\ldots &j_\ell}{k_1 &k_2 &\ldots &k_\ell}{\tau}
&=   (-1)^\ell \! \! \! \! \!
\int \limits_{0<q_1<q_2<\ldots < q_\ell<q}\! \! \!   \frac{ \dd q_1 }{q_1}\frac{ \dd q_2 }{q_2}\ldots \frac{ \dd q_\ell }{q_\ell}  \prod_{r=1}^\ell
\frac{ (2\pi i \tau_r)^{j_r}  {\rm G}_{k_r}(\tau_r)}{(2\pi i)^{k_r} }\, .
\label{svEMZV.1}
\end{align}
The entries are taken to obey $k_r\geq 4$ and $0\leq j_r \leq k_r{-}2$, and we use tangential-base-point regularization for the divergences as $q_r \rightarrow 0$ \cite{Brown:mmv}, which implies that the iterated Eisenstein integrals $\mathcal{E}[\ldots;\tau]$ vanish in the regularized limit $\tau \to i \infty$.

\subsection{Improving the differential equations}
\label{sec:4.1}

We shall now derive the structure
of the $\alpha'$-expansion of $B^\tau_{\vec{\eta}}$
and $J^\tau_{\vec{\eta}}$ by repeating the key steps of \cite{Gerken:2020yii}
in solving the differential equation (\ref{looprev.11}) of $Y^\tau_{\vec{\eta}} $.
The first step is to introduce redefined generating series $\Bhat^\tau_{\vec{\eta}}$ and $\Jhat^\tau_{\vec{\eta}}$ by
\begin{align}
\Bhat^\tau_{\vec{\eta}}  &= \exp \Big( {-} \frac{  r_{\vec{\eta}}(\epsilon_0) }{2\pi i \tau} \Big)  B^\tau_{\vec{\eta}}
\, , \ \ \ \ \ \
\Jhat^\tau_{\vec{\eta}}  = \exp \Big( {-} \frac{ {\rm sv} \, r_{\vec{\eta}}(\epsilon_0) }{2\pi i (\tau{-}\bar \tau)} \Big)  J^\tau_{\vec{\eta}}\, .
\label{svEMZV.2}
\end{align}
After this redefinition, the $k=0$ terms involving $\frac{{\rm G}_0}{\tau^2}$ and $\frac{{\rm G}_0}{(\tau{-}\bar \tau)^2}$ with ${\rm G}_0=-1$ are absent from the analogues of the differential equations
(\ref{newBB.16}) and (\ref{newBB.17}),  see \eqref{svEMZV.4} and \eqref{svEMZV.5} below.
Throughout this section, we suppress the permutations $\gamma,\rho$ labeling $B^\tau_{\vec{\eta}}(\gamma|\rho )$
and $J^\tau_{\vec{\eta}}(\gamma|\rho )$, and all matrix representations $r_{\vec{\eta}}(\epsilon_k) $ are
understood to act matrix-multiplicatively on the $\rho$-entry.

Since the $r_{\vec{\eta}}(\epsilon_k)$ are expected
(as tested for a wide range of $k$ and $n$) to inherit the ad-nilpotency
relations of the derivation algebra,
\beq
{\rm ad}_{ r_{\vec{\eta}}(\epsilon_0) }^{k-1}  r_{\vec{\eta}}(\epsilon_k) = 0 \, ,  \ \ \ \ \ \ k \geq 4 \,
\label{svEMZV.3}
\eeq
the combinations $\exp ( {-} \frac{  r_{\vec{\eta}}(\epsilon_0) }{2\pi i \tau} )r_{\vec{\eta}}(\epsilon_{k\geq 4})
\exp ( \frac{  r_{\vec{\eta}}(\epsilon_0) }{2\pi i \tau} )$
in the differential equations of the redefined integrals (\ref{svEMZV.2})
truncate to a finite number of terms and we obtain
\begin{align}
\label{svEMZV.4}
2\pi i \partial_\tau \Bhat^\tau_{\vec{\eta}} &= \sum_{k=4}^{\infty} (1{-}k)
\sum_{j=0}^{k-2} \frac{1}{j!} \Big( \frac{ -1}{2\pi i} \Big)^j  \tau^{k-2-j}  {\rm G}_k(\tau)
r_{\vec{\eta}}\big( {\rm ad}_{ \epsilon_0}^j (\epsilon_k) \big)
 \Bhat^\tau_{\vec{\eta}}
\\
\label{svEMZV.5}
2\pi i \partial_\tau \Jhat^\tau_{\vec{\eta}} &= \sum_{k=4}^{\infty} (1{-}k)
\sum_{j=0}^{k-2} \frac{1}{j!} \Big( \frac{ -1}{2\pi i} \Big)^j  (\tau{-}\bar\tau)^{k-2-j}  {\rm G}_k(\tau)
r_{\vec{\eta}}\big( {\rm ad}_{ \epsilon_0}^j (\epsilon_k) \big)
 \Jhat^\tau_{\vec{\eta}} \, .
\end{align}
We have used that ${\rm ad}_{ r_{\vec{\eta}}(\epsilon_0) }(\cdot)=[ r_{\vec{\eta}}(\epsilon_0) , \cdot ]={\rm ad}_{{\rm sv}\, r_{\vec{\eta}}(\epsilon_0) }(\cdot)$ (since the term $\sim \zeta_2$ in (\ref{newBB.18}) suppressed by sv is commutative) and employ the shorthands
\beq
r_{\vec{\eta}}\big( {\rm ad}_{ \epsilon_0}^j (\epsilon_k) \big)
= {\rm ad}_{ r_{\vec{\eta}}(\epsilon_0) }^j r_{\vec{\eta}}(\epsilon_k)\, , \ \ \ \
r_{\vec{\eta}}(\epsilon_{k_1} \epsilon_{k_2})=r_{\vec{\eta}}(\epsilon_{k_1})r_{\vec{\eta}}(\epsilon_{k_2})\, .
\label{svEMZV.6}
\eeq
In equations (\ref{svEMZV.4}), (\ref{svEMZV.5}) and below, we use the derivative with respect to $\tau$ instead of $\log q$ as compared to section \ref{sec:3.5} above.


\subsection{The $\alpha'$-expansion of $B^\tau_{\vec{\eta}}$}
\label{sec:4.2}

By the differential equation
\begin{align}
2\pi i \partial_\tau \EBR{j_1 &j_2 &\ldots &j_\ell}{k_1 &k_2 &\ldots &k_\ell}{\tau}
&= - (2\pi i)^{2-k_\ell} (2\pi i \tau)^{j_\ell} {\rm G}_{k_\ell}(\tau)
\EBR{j_1 &j_2 &\ldots &j_{\ell-1}}{k_1 &k_2 &\ldots &k_{\ell-1}}{\tau}
 \label{svEMZV.31}
 \end{align}
of the iterated Eisenstein integrals (\ref{svEMZV.1}), one can solve the
differential equation (\ref{svEMZV.4}) of the generating series
through the path-ordered exponential
\begin{align}
\label{svEMZV.7}
\Bhat^{\tau}_{\vec{\eta}} &=
\sum_{\ell=0}^\infty \sum_{k_1,k_2,\ldots,k_\ell \atop{=4,6,8,\ldots}}
\sum_{j_1=0}^{k_1-2} \sum_{j_2=0}^{k_2-2} \ldots \sum_{j_\ell=0}^{k_\ell-2}
\bigg(  \prod_{i=1}^\ell \frac{ (-1)^{j_i} (k_i{-}1)}{(k_i{-}j_i{-}2)!} \bigg)  \EBR{j_1 &j_2 &\ldots &j_\ell}{k_1 &k_2 &\ldots &k_\ell}{\tau}  \\
&\quad \times    r_{\vec{\eta}}\big( {\rm ad}_{ \epsilon_0}^{k_\ell-j_\ell-2} (\epsilon_{k_{\ell}})\ldots
 {\rm ad}_{ \epsilon_0}^{k_2-j_2-2} (\epsilon_{k_2}) {\rm ad}_{ \epsilon_0}^{k_1-j_1-2} (\epsilon_{k_1}) \big)
\Bhat^{i\infty}_{\vec{\eta}} \notag
\end{align}
for some initial value $\Bhat^{i\infty}_{\vec{\eta}}$ to be discussed below. By inverting
the redefinition (\ref{svEMZV.2}) and moving the exponential to act directly
on the initial value, we obtain the open-string analogue
\begin{align}
  \label{svEMZV.8}
B^{\tau}_{\vec{\eta}} &=
\sum_{\ell=0}^\infty \sum_{k_1,k_2,\ldots,k_\ell \atop{=4,6,8,\ldots}}
\sum_{j_1=0}^{k_1-2} \sum_{j_2=0}^{k_2-2} \ldots \sum_{j_\ell=0}^{k_\ell-2}
\bigg(  \prod_{i=1}^\ell \frac{ (-1)^{j_i} (k_i{-}1)}{(k_i{-}j_i{-}2)!} \bigg) \betaBR{j_1 &j_2 &\ldots &j_\ell}{k_1 &k_2 &\ldots &k_\ell}{\tau}  \\
&\quad \times
 r_{\vec{\eta}}\big( {\rm ad}_{ \epsilon_0}^{k_\ell-j_\ell-2} (\epsilon_{k_{\ell}})\ldots
 {\rm ad}_{ \epsilon_0}^{k_2-j_2-2} (\epsilon_{k_2}) {\rm ad}_{ \epsilon_0}^{k_1-j_1-2} (\epsilon_{k_1}) \big) \exp \Big( \frac{  r_{\vec{\eta}}(\epsilon_0) }{2\pi i \tau} \Big)
\Bhat^{i\infty}_{\vec{\eta}}   \notag
\end{align}
of the key result for the $\alpha'$-expansion of $Y^\tau_{\vec{\eta}}$ in
(3.11) of \cite{Gerken:2020yii}. In commuting $\exp ( \frac{  r_{\vec{\eta}}(\epsilon_0) }{2\pi i \tau}) $
past the $\ep_{k_j}$, the iterated Eisenstein integrals are rearranged into the combinations
\begin{align}
\betaBR{j_1 }{k_1 }{\tau} &= \! \! \sum_{p_1=0}^{k_1- j_1-2} \! \!  { k_1{-}j_1{-}2 \choose p_1} \Big( \frac{i}{2T} \Big)^{p_1}
\EBR{j_1+p_1}{k_1}{\tau}
  \label{svEMZV.9}
\end{align}
and more generally
\begin{align}
\betaBR{j_1 &j_2 &\ldots &j_\ell}{k_1 &k_2 &\ldots &k_\ell}{\tau} &= \sum_{p_1=0}^{k_1- j_1-2}\sum_{p_2=0}^{k_2- j_2-2} \ldots \sum_{p_\ell=0}^{k_\ell- j_\ell-2}
 \binom{k_1{-}j_1{-}2}{p_1} \binom{k_2{-}j_2{-}2}{p_2}\cdots \binom{k_\ell{-}j_\ell{-}2}{p_\ell}   \notag\\
 &\ \ \ \ \ \times  \Big( \frac{i}{2T} \Big)^{p_1+p_2+\ldots +p_\ell} \EBR{j_1+p_1 & j_2 + p_2 &\ldots & j_\ell + p_\ell}{k_1 & k_2 &\ldots &k_\ell}{\tau}\, .
   \label{svEMZV.10}
\end{align}
Note that \eqref{svEMZV.8} is an alternative\footnote{
On top of the modular $S$ transformation relating $Z^{\tau}_{\vec{\eta}}$ and $B^{\tau}_{\vec{\eta}}$, the ${\cal E}[\ldots]$ in (\ref{svEMZV.7}) involve integration kernels $\tau^j {\rm G}_k$ with $0 \leq j \leq k{-}2$ instead of the ${\rm G}^0_0=-1$ in \cite{Mafra:2019xms,Mafra:2019ddf}. In other words, the relations (\ref{svEMZV.3}) in the derivation algebra are built into (\ref{svEMZV.7}), whereas the results in the references may require the use of shuffle relations to manifest the absence of ${\cal E}[\ldots \smallmatrix j \geq k-1 \\ k \endsmallmatrix \ldots]$.
}
organization of open-string $\alpha'$-expansions at genus one as compared to \cite{Mafra:2019xms,Mafra:2019ddf}.
Non-planar B-cycle integrals obey the same differential equation \eqref{newBB.16} as the planar ones and therefore have an $\alpha'$-expansion of the same form (\ref{svEMZV.8}), only their initial values $\Bhat^{i\infty}_{\vec{\eta}} $ need to be adapted to the non-planar integration cycle.

The modified iterated Eisenstein integrals $\beta[\ldots]$ in (\ref{svEMZV.10})
satisfy the differential equations
\begin{align}
2\pi i \tau^2 \partial_\tau
\betaBR{j_1 &j_2&\ldots &j_\ell}{k_1 &k_2 &\ldots &k_\ell}{\tau} & =
\sum_{i=1}^\ell
 (k_i{-}j_i{-}2)  \betaBR{j_1 &j_2 &\ldots  &j_{i-1} &j_i +1 &j_{i+1} &\ldots &j_\ell}{k_1 &k_2 &\ldots &k_{i-1} &k_i  &k_{i+1} &\ldots &k_\ell}{\tau}  \label{svEMZV.11} \\
 &\quad  - \delta_{j_\ell,k_\ell-2} \tau^{k_\ell} {\rm G}_{k_\ell}(\tau) \betaBR{j_1 &j_2 &\ldots &j_{\ell-1}}{k_1 &k_2 &\ldots &k_{\ell-1}}{\tau} \,,
\notag
\end{align}
which allows us to directly check that (\ref{svEMZV.8}) obeys (\ref{newBB.16}). The integrals $\beta[\ldots]$ inherit the property that they vanish for $\tau\to i \infty$ from the $\mathcal{E}[\ldots]$. Note that the definition (\ref{svEMZV.10})
is equivalent to integral representations such as
\begin{align}
 \betaBR{j_1 }{k_1 }{\tau} &= - \frac{ (2\pi i)^{1+j_1-k_1} }{\tau^{k_1-j_1-2} }
  \int\limits^\tau_{i\infty} \dd \tau_1 \, {\rm G}_{k_1}(\tau_1) (\tau {-} \tau_1)^{k_1-j_1-2} \tau_1^{j_1}
\label{oa9.1} \\
 \betaBR{j_1 &j_2}{k_1 &k_2}{\tau} &= \frac{ (2\pi i)^{2+j_1+j_2-k_1-k_2} }{\tau^{k_1+k_2-j_1-j_2-4} }
  \int\limits^\tau_{i\infty} \dd \tau_2 \, {\rm G}_{k_2}(\tau_2) (\tau {-} \tau_2)^{k_2-j_2-2} \tau_2^{j_2}
 \int\limits^{\tau_2}_{i\infty} \dd \tau_1 \, {\rm G}_{k_1}(\tau_1) (\tau {-} \tau_1)^{k_1-j_1-2} \tau_1^{j_1} \, .
 \notag
\end{align}
The definition (\ref{svEMZV.10}) of the $\beta[\ldots]$ preserves the shuffle relations of the
iterated Eisenstein integrals (\ref{svEMZV.1}), for instance
\beq
 \EBR{j_1 }{k_1 }{\tau}   \EBR{j_2}{k_2}{\tau} =
 \EBR{j_1 &j_2}{k_1 &k_2}{\tau}  +  \EBR{j_2 &j_1}{k_2 &k_1}{\tau}
\ \ \Rightarrow \ \
 \betaBR{j_1 }{k_1 }{\tau}   \betaBR{j_2}{k_2}{\tau} =
 \betaBR{j_1 &j_2}{k_1 &k_2}{\tau}  +  \betaBR{j_2 &j_1}{k_2 &k_1}{\tau} \, .
 \label{holshuffle}
\eeq

\subsection{The $\alpha'$-expansion of $J^\tau_{\vec{\eta}}$}
\label{sec:4.3}

One can
extend the above strategy to expand $B^\tau_{\vec{\eta}}$ via (\ref{svEMZV.4}) to
the $J^\tau_{\vec{\eta}}$ integrals. The idea is to solve their differential equation
(\ref{svEMZV.5}) order by order in $\alpha'$ via
\begin{align}
\label{svEMZV.33}
2\pi i \partial_\tau \EsvBR{j_1 &j_2 &\ldots &j_\ell}{k_1 &k_2 &\ldots &k_\ell}{\tau}
&=
 - (2\pi i)^{2-k_\ell+j_\ell} (\tau{-}\bar \tau)^{j_\ell} {\rm G}_{k_\ell}(\tau)\EsvBR{j_1 &j_2 &\ldots &j_{\ell-1}}{k_1 &k_2 &\ldots &k_{\ell-1}}{\tau}\, ,
 \end{align}
using the combinations ${\cal E}^{\rm sv}$ of holomorphic iterated Eisenstein integrals
(\ref{svEMZV.1}) and their complex conjugates introduced in \cite{Gerken:2020yii}.
Their depth $\ell{=}1$ instances are completely known from the reference
\begin{align}
 \label{svEMZV.34}
\EsvBR{j_1}{k_1}{\tau} = \sum_{r_1=0}^{j_1} (-2\pi i \bar\tau)^{r_1} \binom{j_1}{r_1} \Big( \EBR{j_1-r_1}{k_1}{\tau}
+(-1)^{j_1-r_1} \overline{  \EBR{j_1-r_1}{k_1}{\tau} } \Big)\, ,
\end{align}
and their generalizations to depth $\ell \geq 2$
involve antiholomorphic integration
constants $\overline{\alpha[\ldots]}$,
\begin{align}
&\EsvBR{j_1 &j_2}{k_1 &k_2}{\tau} =  \sum_{r_1=0}^{j_1}\sum_{r_2=0}^{j_2} (-2\pi i \bar\tau)^{r_1+r_2}
\binom{j_1}{r_1}\binom{j_2}{r_2} \label{svEMZV.35} \\
&   \times
\Big\{ \EBR{j_1 -r_1&j_2-r_2}{k_1 &k_2}{\tau} {+}(-1)^{j_1-r_1} \overline{ \EBR{j_1-r_1}{k_1}{\tau} }  \EBR{j_2-r_2}{k_2}{\tau}  \notag \\
&\ \ \ \ +(-1)^{j_1+j_2-r_1-r_2} \overline{ \EBR{j_2-r_2 &j_1-r_1}{k_2 &k_1}{\tau} }
\Big\}  +\overline{\alphaBR{j_1 &j_2}{k_1 &k_2}{\tau}} \, .
\notag
\end{align}
The integration constants $\overline{\alpha[\ldots]}$ are invariant under $\tau \rightarrow \tau+1$
since the ${\cal E}^{\rm sv}[\ldots]$ and the contributions from the ${\cal E}[\ldots], \overline{{\cal E}[\ldots]}$
to (\ref{svEMZV.35}) are. They are known on a case-by-case basis, for~instance
\begin{align}
 \alphaBRno{1& 0}{4& 4} &=  \alphaBRno{0& 1}{4& 4} = 0  \,,\notag \\
  \alphaBRno{2& 0}{4& 4} &=  \frac{2 \zeta_3}{3}  \Big(   \EBRno{0}{4}
 + \frac{ i \pi   \tau}{360} \Big)
 = -  \alphaBRno{0& 2}{4& 4}  \,, \label{eq4.24} \\
  \alphaBRno{2& 1}{4& 4}  &=  \frac{2 \zeta_3}{3}  \Big(  2 \pi i \tau   \EBRno{0}{4}  -    \EBRno{1}{4} - \frac{ \pi^2  \tau^2}{360} \Big)
= -    \alphaBRno{1& 2}{4& 4} \, , \notag
\end{align}
and the complete list of $\overline{\alpha[\smallmatrix j_1 &j_2 \\ k_1 &k_2 \endsmallmatrix]}$ at
 $k_1+k_2\leq 12$ can be found in an ancillary file within the arXiv submission of this work.
The integration constants at arbitrary depth can be determined from the reality properties of the $Y^\tau_{\vec{\eta}}$
integrals \cite{Gerken:2020yii}. The method in the reference to fix the $\overline{\alpha[\ldots]}$
hinges on the fact that the coefficients in the $\eta_j$- and $\bar \eta_j$-expansion of $Y^\tau_{\vec{\eta}}$
are closed under complex conjugation. For the $n$-point $J^\tau_{\vec{\eta}}$-series in turn
the antiholomorphic modular weights $\bar w$ of the integrands $\overline{V(\ldots)}$ in (\ref{newBB.11})
are bounded by $\bar w \leq n{-}2$, so the complex conjugates of higher orders in the $\eta_j$-expansion
are not part of the series. Hence, in the present
formulation, the expansion
of the $Y^\tau_{\vec{\eta}}$ in \cite{Gerken:2020yii} is
a necessary input to obtain well-defined
${\cal E}^{\rm sv}$. This expansion depends on the knowledge of the initial values of $Y^{\tau}_{\vec{\eta}}$ which is currently available from sphere integrals to arbitrary weight only for two points and is under investigation for higher multiplicity \cite{yinprogress}. Still, the torus-integral- and lattice-sum representations of single-valued
eMZVs in section \ref{sec:5.8} do not require any knowledge of $Y^{i\infty}_{\vec{\eta}}$ and $\overline{\alpha[\ldots]}$.

By repeating the steps towards (\ref{svEMZV.7}) and (\ref{svEMZV.8}),
we arrive at the structure of the $\alpha'$-expansion
\begin{align}
  \label{svEMZV.80}
J^{\tau}_{\vec{\eta}} &=
\sum_{\ell=0}^\infty \sum_{k_1,k_2,\ldots,k_\ell \atop{=4,6,8,\ldots}}
\sum_{j_1=0}^{k_1-2} \sum_{j_2=0}^{k_2-2} \ldots \sum_{j_\ell=0}^{k_\ell-2}
\bigg(  \prod_{i=1}^\ell \frac{ (-1)^{j_i} (k_i{-}1)}{(k_i{-}j_i{-}2)!} \bigg) \bsvBR{j_1 &j_2 &\ldots &j_\ell}{k_1 &k_2 &\ldots &k_\ell}{\tau}  \\
&\quad \times
 r_{\vec{\eta}}\big( {\rm ad}_{ \epsilon_0}^{k_\ell-j_\ell-2} (\epsilon_{k_{\ell}})\ldots
 {\rm ad}_{ \epsilon_0}^{k_2-j_2-2} (\epsilon_{k_2}) {\rm ad}_{ \epsilon_0}^{k_1-j_1-2} (\epsilon_{k_1}) \big) \exp \Big({-} \frac{ {\rm sv} \,  r_{\vec{\eta}}(\epsilon_0) }{4y} \Big)
\Jhat^{i\infty}_{\vec{\eta}}
\notag
\end{align}
with an initial value $\Jhat^{i\infty}_{\vec{\eta}}$ to be discussed below and the
combinations analogous to (\ref{svEMZV.10}) \cite{Gerken:2020yii}
\begin{align}
\bsvBR{j_1 &j_2 &\ldots &j_\ell}{k_1 &k_2 &\ldots &k_\ell}{\tau} &= \sum_{p_1=0}^{k_1- j_1-2}\sum_{p_2=0}^{k_2- j_2-2} \ldots \sum_{p_\ell=0}^{k_\ell- j_\ell-2}
 \binom{k_1{-}j_1{-}2}{p_1}\binom{k_2{-}j_2{-}2}{p_2}\cdots \binom{k_\ell{-}j_\ell{-}2}{p_\ell}  \notag \\
 &\ \ \ \ \ \times \Big( \frac{1}{4y} \Big)^{p_1+p_2+\ldots +p_\ell}
\EsvBR{j_1+p_1 & j_2 + p_2 &\ldots & j_\ell + p_\ell}{k_1 & k_2 &\ldots &k_\ell}{\tau}\, .
  \label{svEMZV.90}
\end{align}
The expansion (\ref{svEMZV.80}) solves (\ref{newBB.17}) since the $\beta^{\rm sv}$ inherit their
differential equation from (\ref{svEMZV.33}),
\begin{align}
2\pi i (\tau{-}\bar \tau)^2 \partial_\tau
\bsvBR{j_1 &j_2&\ldots &j_\ell}{k_1 &k_2 &\ldots &k_\ell}{\tau} &=
\sum_{i=1}^\ell
 (k_i{-}j_i{-}2) \bsvBR{j_1 &j_2 &\ldots  &j_{i-1} &j_i +1 &j_{i+1} &\ldots &j_\ell}{k_1 &k_2 &\ldots &k_{i-1} &k_i  &k_{i+1} &\ldots &k_\ell}{\tau}  \label{svEMZV.sv} \\
 &\quad - \delta_{j_\ell,k_\ell-2} (\tau{-}\bar \tau)^{k_\ell} {\rm G}_{k_\ell}(\tau) \bsvBR{j_1 &j_2 &\ldots &j_{\ell-1}}{k_1 &k_2 &\ldots &k_{\ell-1}}{\tau} \,,
\notag
\end{align}
see (\ref{svEMZV.11}) for the holomorphic counterpart for $\partial_\tau \beta[\ldots]$.
Both the ${\cal E}^{\rm sv}[\ldots]$ and the $\beta^{\rm sv}[\ldots]$ are expected to preserve the shuffle
multiplication of their holomorphic counterparts (\ref{svEMZV.1}) and (\ref{svEMZV.10}):
The differential equations (\ref{svEMZV.33}) and (\ref{svEMZV.90}) recursively imply that shuffle
relations among ${\cal E}^{\rm sv}[\ldots]$ and the $\beta^{\rm sv}[\ldots]$ can at most
be violated by antiholomorphic functions such as the integration constants
$\overline{\alpha[\ldots]} $ in (\ref{svEMZV.35}).\footnote{Moreover, any such
violation of shuffle relations would need to be a combination of antiholomorphic iterated Eisenstein integrals (by the differential equation \cite[(2.37)]{Gerken:2020yii} for $\partial_{\bar \tau} Y^\tau_{\vec{\eta}}$)
but at the same time line up with the modular weights in the $\eta_j$-expansion of
$J^\tau_{\vec{\eta}}$, $Y^\tau_{\vec{\eta}}$ and the reality properties of the latter. It would be interesting to find a rigorous argument to rule out the existence of such antiholomorphic functions.}
All examples of $\alpha [ \begin{smallmatrix} j_1& j_2\\k_1& k_2\end{smallmatrix}]$
up to including $k_1+k_2=12$ were checked to preserve the shuffle relations, and their explicit form can also be
found in an ancillary file to the arXiv submission of this work. Note that these checks
cover the more intricate cases with $(k_1,k_2)=(4,6)$ and $(k_1,k_2)=(4,8)$
where imaginary cusp forms occur among the MGFs~\cite{Gerken:2020yii,Gerken:2020aju}.

The ${\cal E}^{\rm sv}$ and $\beta^{\rm sv}$ are expected to occur in Brown's generating
series of single-valued iterated Eisenstein integrals \cite{Brown:mmv, Brown:2017qwo, Brown:2017qwo2}. The construction of non-holomorphic
modular forms in the references -- so-called equivariant iterated Eisenstein integrals -- are obtained
by augmenting their single-valued counterparts by combinations of MZVs and objects of lower depth.
At depth one, the equivariant iterated Eisenstein integrals are non-holomorphic Eisenstein series along
with their Cauchy--Riemann derivatives \cite{Brown:mmv, Brown:2017qwo, Brown:2017qwo2}. From their representation \cite{Gerken:2020yii}
\begin{align}
(\pi \nabla)^m {\rm E}_k &= \Big( {-}\frac{1}{4} \Big)^{m} \frac{ (2k{-}1)! }{(k{-}1)! (k{-}1{-}m)!}
\bigg\{
{-} \betasv{ k-1+m\\ 2k} + \frac{ 2 \zeta_{2k-1} }{(2k{-}1) (4y)^{k-1-m} }
\bigg\}
\notag \\
{\rm E}_k &= \frac{ (2k{-}1)! }{ [(k{-}1)!]^2} \bigg\{
{-} \betasv{ k-1\\ 2k} + \frac{ 2 \zeta_{2k-1} }{(2k{-}1) (4y)^{k-1} }
\bigg\}
\label{eq:EE}
\\
\frac{(\pi \overline{\nabla})^m {\rm E}_k}{y^{2m}} &=  \frac{({-}4)^{m} (2k{-}1)! }{(k{-}1)! (k{-}1{-}m)!}
\bigg\{
{-} \betasv{ k-1-m\\ 2k} + \frac{ 2 \zeta_{2k-1} }{(2k{-}1) (4y)^{k-1+m} }
\bigg\} \, ,
\notag
\end{align}
the $\beta^{\rm sv}$ are seen to take the role of the single-valued rather than equivariant iterated Eisenstein
integrals at depth one. At higher depth, the precise relation of the $\beta^{\rm sv}$ to Brown's construction is
an open question at the time of writing.

\subsection{Initial values}
\label{sec:4.4}

It remains to specify the initial values $\Bhat^{i\infty}_{\vec{\eta}}$ and $\Jhat^{i\infty}_{\vec{\eta}}$ in the $\alpha'$-expansions (\ref{svEMZV.8}) and (\ref{svEMZV.80}). The Laurent-polynomial contributions from the asymptotics (\ref{newBB.71}) and (\ref{newBB.72}) of the Green functions are still functions of $\tau$ and need to be translated into series that solely depend on $\eta_j$ and $s_{ij}$. Following the construction of a similar initial value for $Y^\tau_{\vec{\eta}}$ in section 3.4 of \cite{Gerken:2020yii}, we import the constant parts $\sim \tau^0$ and $\sim (\Im \tau)^0$ of the respective Laurent polynomials
\begin{align}
\Bhat^{i\infty}_{\vec{\eta}} &= \exp \Big( \frac{ i r_{\vec{\eta}}(\epsilon_0) }{2T} \Big)  B^{\tau}_{\vec{\eta}} \, \big|_{\rm LP} \, \big|_{\tau^0}
\label{svEMZV.51} \\
\Jhat^{i\infty}_{\vec{\eta}} &= \exp \Big(\frac{ {\rm sv} \, r_{\vec{\eta}}(\epsilon_0) }{4y} \Big)   J^{\tau}_{\vec{\eta}} \, \big|_{\rm LP} \, \big|_{(\Im \tau)^0} \, .
\label{svEMZV.52}
\end{align}
In both cases, the exponentials ensure that the negative powers of $T$ in
$B^{\tau}_{\vec{\eta}} \, \big|_{\rm LP}$ and $y$ in $J^{\tau}_{\vec{\eta}} \, \big|_{\rm LP}$
disappear order by order in $\alpha'$. Hence, (\ref{svEMZV.51}) and
(\ref{svEMZV.52}) pick up the lowest powers of $T,y$ present in
$ \exp ( \frac{ i r_{\vec{\eta}}(\epsilon_0) }{2T} )  B^{\tau}_{\vec{\eta}} \, \big|_{\rm LP}$
and  $\exp (\frac{ {\rm sv} \, r_{\vec{\eta}}(\epsilon_0) }{4y})   J^{\tau}_{\vec{\eta}} \, \big|_{\rm LP}$.
The leading $\alpha'$- and $\eta_2$-orders of the two-point initial values following from the expressions in (\ref{apeq.1}) and (\ref{apeq.2}) are
\begin{align}
\Bhat^{i\infty}_{\eta_2}& =   \frac{1}{\eta_2} \Big[ 1 + \frac{1}{6} s_{12}^2 \zeta_2+\frac{1}{12} s_{12}^3 \zeta_3  +
     \frac{131}{720} s_{12}^4   \zeta_4  +
      s_{12}^5 \Big( \frac{17}{360} \zeta_2 \zeta_3 + \frac{43 \zeta_5}{720} \Big) + {\cal O}(s_{12}^6) \Big]
      \notag \\
& \ \ \ \ + \eta_2 \Big[ {-} 2 \zeta_2 - s_{12} \zeta_3 - \frac{29}{12} s_{12}^2   \zeta_4
 -  s_{12}^3 \Big( \frac{1}{3} \zeta_2 \zeta_3 + \frac{5 \zeta_5}{6} \Big)
  -  s_{12}^4 \Big( \frac{1}{12} \zeta_3^2 + \frac{ 87 \zeta_6}{40} \Big)  + {\cal O}(s_{12}^5) \Big]
        \notag \\
& \ \ \ \  +   \eta_2^3 \Big[ {-}2 \zeta_4 + s_{12} (2 \zeta_2 \zeta_3 - \zeta_5) +
      s_{12}^2 \Big( \frac{ \zeta_3^2}{2} - \frac{33 \zeta_6}{8} \Big) +
      s_{12}^3 \Big( \frac{9}{4} \zeta_3 \zeta_4 + \frac{ 3}{2} \zeta_2 \zeta_5
       - \frac{   7 \zeta_7}{4} \Big)   +{\cal O}(s_{12}^4) \Big]
    \notag \\
& \ \ \ \ + \eta_2^5 \Big[ {-}2 \zeta_6 +
      s_{12} ( 2 \zeta_4 \zeta_3 + 2 \zeta_2 \zeta_5 - \zeta_7) +
      s_{12}^2 \Big({-}\zeta_2 \zeta_3^2 + \zeta_3 \zeta_5 - \frac{43 \zeta_8}{6} \Big)  +{\cal O}(s_{12}^3) \Big]
    \label{svEMZV.54} \\
& \ \ \ \ + \eta_2^7 \big[ {-}2 \zeta_8 + s_{12} ( 2 \zeta_3 \zeta_6 + 2 \zeta_4 \zeta_5 +
         2 \zeta_2 \zeta_7 - \zeta_9)  +{\cal O}(s_{12}^2) \big]
 + \eta_2^9 \big[ {-}2 \zeta_{10} +{\cal O}(s_{12}) \big] + {\cal O}(\eta_2^{11})
\notag
\end{align}
as well as
\begin{align}
\Jhat^{i\infty}_{\eta_2}& =  \frac{1}{\eta_2} \Big[1+
\frac{1}{6} s_{12}^3 \zeta_3  + \frac{43}{360} s_{12}^5 \zeta_5 +{\cal O}(s_{12}^6) \Big]
\notag \\
 & \ \  + \eta_2 \Big[   {-} 2  s_{12} \zeta_3  - \frac{5}{3}   s_{12}^3 \zeta_5
  - \frac{ 1}{3}  s_{12}^4 \zeta_3^2 +{\cal O}(s_{12}^5) \Big] \notag \\
 & \ \   + \eta^3_2 \Big[ {-}  2   s_{12} \zeta_5 +  2   s_{12}^2 \zeta_3^2   - \frac{ 7}{2}   s_{12}^3 \zeta_7 +{\cal O}(s_{12}^4) \Big]
\label{svEMZV.55} \\
 & \ \  + \eta^5_2 \big[{-} 2   s_{12} \zeta_7 + 4   s_{12}^2 \zeta_3 \zeta_5  +{\cal O}(s_{12}^3) \big] \notag \\
 & \ \  + \eta^7_2 \big[ {-} 2  s_{12} \zeta_9 +{\cal O}(s_{12}^2) \big]
  + \eta_2^9 {\cal O}(s_{12}) + {\cal O}(\eta^{11}_2) \, . \notag
\end{align}
Higher orders in $s_{12}$ and $\eta_2$ are readily available through the straightforward
expansion of the exponentials and $\Gamma$-functions in (\ref{apeq.1}) and (\ref{apeq.2}).
In particular, these two-point expressions imply that all the coefficients in the $s_{12}$- and
$\eta_2$-expansions are combinations of Riemann zeta values for $\Bhat^{i\infty}_{\eta_2}$ and
odd Riemann zeta values for $\Jhat^{i\infty}_{\eta_2}$.

Starting from $n=3$ points, the initial values $\Bhat^{i\infty}_{\vec{\eta}}$ will also feature
irreducible MZVs of higher depth. Based on Enriquez' method to generate the
Laurent polynomial of B-cycle eMZVs \cite{Enriquez:Emzv} (also see appendix B of \cite{Broedel:2018izr})
we have determined the three-point initial values to certain orders, and the results are included in an ancillary file to the arXiv submission of this article.
To the orders under consideration, we find the following coefficients of $\zeta_{3,5}$
\begin{align}
&\widehat{B}^{i\infty}_{\eta_2,\eta_3}(2,3|2,3) \, \big|_{\zeta_{3,5}} =
\frac{1}{10} (\eta_{23} - 2 \eta_{3}) (2 \eta_{23} - \eta_{3}) (\eta_{23} +
   \eta_{3})  \notag \\
   &\ \ \times \Big[ 2 \eta_{23}^2 s_{13} -2 \eta_{23} \eta_{3} s_{12}  -
   4 \eta_{23} \eta_{3} s_{13} + 2 \eta_{3}^2 s_{13}
   - 2 \eta_{23} \eta_{3} s_{23}   + \eta_{23}  s_{12} s_{13} + \eta_{3}  s_{13} s_{23}+ {\cal O}(s_{ij}^3) \Big]
\notag \\
&\widehat{B}^{i\infty}_{\eta_2,\eta_3}(2,3|3,2)  \, \big|_{\zeta_{3,5}} =
  -\frac{1}{10} (\eta_{2} - 2 \eta_{23}) (2 \eta_{2} - \eta_{23}) (\eta_{2} +
   \eta_{23}) \label{appbb.4}   \\
   &\ \ \times \Big[2 \eta_{2}^2 s_{12} - 4 \eta_{2} \eta_{23} s_{12} + 2 \eta_{23}^2 s_{12} -
   2 \eta_{2} \eta_{23} s_{13}  - 2 \eta_{2} \eta_{23} s_{23}
   - \eta_{23}   s_{12} s_{13} -  \eta_{2}   s_{12} s_{23} + {\cal O}(s_{ij}^3) \Big]  \, , \notag
\end{align}
which by the single-valued maps ${\rm sv}\, \zeta_{3,5}=-10\zeta_3 \zeta_5$ and
${\rm sv}\, \zeta_{3} \zeta_{5}=4\zeta_3 \zeta_5$
enter the closed-string initial values via
\beq
\widehat{J}^{i\infty}_{\eta_2,\eta_3}(2,3|\rho(2,3)) \, \big|_{\zeta_{3} \zeta_{5}}=
-10 \widehat{B}^{i\infty}_{\eta_2,\eta_3}(2,3|\rho(2,3)) \, \big|_{\zeta_{3,5}} + 4 \widehat{B}^{i\infty}_{\eta_2,\eta_3}(2,3|\rho(2,3)) \, \big|_{\zeta_{3} \zeta_{5}}\, .
\label{appbb.4extra}
\eeq
The coefficients of $\zeta_{3,5}$ and $\zeta_{3} \zeta_{5}$ in (\ref{appbb.4}) and (\ref{appbb.4extra})
are extracted after reducing all MZVs at weight 8 to $\mathbb Q$-linear combinations of $\{ \zeta_8, \zeta_2\zeta_3^2, \zeta_{3} \zeta_{5}, \zeta_{3,5} \}$ \cite{Blumlein:2009cf}.
Similarly, the MZV $\zeta_{3,5,3}$ seen in Laurent polynomials of both B-cycle integrals
\cite{Broedel:2018izr} and modular graph functions \cite{Zerbini:2015rss} will occur in both $\Bhat^{i\infty}_{\eta_2,\eta_3}$
and $\Jhat^{i\infty}_{\eta_2,\eta_3}$. The contributions to $\Bhat^{i\infty}_{\eta_2,\eta_3}$ involving MZVs of
weight up to and including four can be found in appendix \ref{app:bhat}.

Since the initial values are obtained from the Laurent polynomials and
the exponents in (\ref{svEMZV.51}) and (\ref{svEMZV.52}) are related
by the single-valued map, the conjecture (\ref{newBB.15}) supported
by tree-level results and extensive genus-one tests is equivalent to
\beq
\Jhat^{i\infty}_{\vec{\eta}}(\gamma|\rho )  = {\rm sv} \, \Bhat^{i\infty}_{\vec{\eta}}(\gamma|\rho )  \, ,
 \label{svEMZV.56}
\eeq
in agreement with (\ref{svEMZV.54}) and (\ref{svEMZV.55}).

The $\tau \rightarrow i \infty$ asymptotics of $n$-point A-cycle
integrals (\ref{looprev.1}) has been expressed in terms of $(n{+}2)$-point
disk integrals (\ref{treerev.1}) in suitable kinematic limits \cite{Mafra:2019xms}.
Similarly, the Laurent polynomials of $n$-point genus-one integrals $B^\tau_{\vec{\eta}},
J^\tau_{\vec{\eta}}$ are determined by genus-zero integrals at multiplicity $n{+}2$
and below, see (\ref{apeq.1}) and (\ref{apeq.2}) for the explicit two-point result.
As will be further investigated in \cite{yinprogress}, the main challenge is to
determine the admixture of lower-point genus-zero integrals that
generalize the subtraction of $e^{iT(\frac{s_{12}}{2} \pm \eta_2)}$
and $e^{-y(s_{12} \pm 2\eta_2)}$ from the $\Gamma$-functions in (\ref{apeq.1}) and (\ref{apeq.2}).

\subsection{The single-valued map on iterated Eisenstein integrals}
\label{sec:4.5}

The proposed single-valued map~\eqref{newBB.20} can now also be studied at the level of the $\alpha'$-expansions.
Using~\eqref{svEMZV.56} and~\eqref{newBB.18}, we find that
one obtains $J^{\tau}_{\vec{\eta}}  $ as the single-valued
version of $B^{\tau}_{\vec{\eta}}$ if the coefficients obey
\beq
\bsvBR{j_1 &j_2&\ldots &j_\ell}{k_1 &k_2 &\ldots &k_\ell}{\tau} = {\rm SV} \, \betaBR{j_1 &j_2&\ldots &j_\ell}{k_1 &k_2 &\ldots &k_\ell}{\tau}\,.
 \label{eesv.1}
\eeq
This follows from the relation (\ref{svEMZV.56}) among the initial values and
the form of the $r_{\vec{\eta}}$ operators in the respective $\alpha'$-expansions, recalling
that $r_{\vec{\eta}}({\rm ad}_{\ep_0}^j(\ep_k))
= {\rm sv}\, r_{\vec{\eta}}({\rm ad}_{\ep_0}^j(\ep_k))$.

On the one hand, (\ref{eesv.1}) fixes the single-valued map of the eMZVs in the expansion of
$B^{\tau}_{\vec{\eta}}$ that enter through the iterated Eisenstein integrals $\beta[\ldots]$.
On the other hand, (\ref{eesv.1}) only
applies to the combinations $\beta[\ldots]$ and $\beta^{\rm sv}[\ldots]$ that occur in the path-ordered exponentials
(\ref{svEMZV.8}) and (\ref{svEMZV.80}).
The SV map of individual $ \betaBR{j_1  &\ldots &j_\ell}{k_1   &\ldots &k_\ell}{\tau}$ remains undetermined
whenever the relations in the derivation algebra such as (\ref{derrels}) lead to dropouts of
certain $\beta[\ldots]$ and $\beta^{\rm sv}[\ldots]$ from $B^{\tau}_{\vec{\eta}}$ and $J^{\tau}_{\vec{\eta}}$ (starting
with cases at $(k_1,k_2)=(10,4)$ at depth $\ell=2$ and any instance where $j_i > k_i{-}2$).

Since the factors of $\frac{i}{2T}$
and $\frac{1}{4y}$ in (\ref{svEMZV.10}) and (\ref{svEMZV.90}) are furthermore related
by SV, (\ref{eesv.1}) is equivalent to
\beq
\EsvBR{j_1 &j_2&\ldots &j_\ell}{k_1 &k_2 &\ldots &k_\ell}{\tau} = {\rm SV} \, \EBR{j_1 &j_2&\ldots &j_\ell}{k_1 &k_2 &\ldots &k_\ell}{\tau}\, ,
 \label{eesv.2}
\eeq
again up to cases where the relations in the derivation algebra cause dropouts.
For instance, (\ref{svEMZV.34}) implies that the single-valued
version of holomorphic Eisenstein integrals (\ref{svEMZV.1}) at depth one is given by
\begin{align}
\label{eesv.2a}
&{\rm SV} \, \EBR{j}{k}{\tau} = \sum_{r=0}^{j} (-2\pi i \bar\tau)^{r} \binom{j}{r}  \Big( \EBR{j-r}{k}{\tau}
+(-1)^{j-r} \overline{  \EBR{j-r}{k}{\tau} } \Big) \,,
\end{align}
where the contributions on the right-hand side can be recognized as
\begin{align}
\sum_{r=0}^{j} (-2\pi i \bar\tau)^{r} \binom{j}{r} 	\EBR{j-r}{k}{\tau} &= (2\pi i)^{1-k+j} \int^{i\infty}_\tau \dd \tau_1 \, (\tau_1{-}\bar \tau)^{j} {\rm G}_k(\tau_1)\,,\\
(-1)^j\sum_{r=0}^{j} (2\pi i \bar\tau)^{r} \binom{j}{r} 	\overline{\EBR{j-r}{k}{\tau}} &= - (2\pi i)^{1-k+j} \int^{-i\infty}_{\bar \tau} \dd \bar \tau_1 \, (\bar \tau_1{-}\bar \tau)^{j} \overline{{\rm G}_k(\tau_1)}\,, \notag
\end{align}
respectively. Note that (\ref{eesv.1}) fixes the SV map of all the $\EBR{j}{k}{\tau}$ at depth one with $k\geq 4$
and $0\leq j \leq k{-}2$ since the caveats related to relations in the derivation algebra beyond (\ref{svEMZV.3}) only affect iterated Eisenstein integrals of depth $\ell \geq2$.

The iterated Eisenstein integrals defined in (\ref{svEMZV.1}) may be reorganized in terms of~\cite{Broedel:2015hia}
\begin{equation}
	\mathcal{E}_0(k_1,k_2,\dots,k_\ell;\tau)= 2\pi i \int_{\tau}^{i\infty}\dd \tau_\ell  \, \frac{{\rm G}^0_{k_\ell}(\tau_\ell)}{(2\pi i)^{k_\ell}} \, \mathcal{E}_0(k_1,k_2,\dots,k_{\ell-1};\tau)
	\label{ezero.1}
\end{equation}
with $k_i \in 2 \mathbb N_0$ and $\mathcal{E}_0(;\tau)=1$. By subtracting the zero mode of the holomorphic Eisenstein series
\begin{align}
	{\rm G}^0_k(\tau)&={\rm G}_k(\tau) - 2 \zeta_k \quad \text{for} \quad  k > 0 \ \text{even} \, , \notag\\ {\rm G}_0^0(\tau) &= {\rm G}_0(\tau) = -1\,,
	\label{ezero.2}
\end{align}
the integrals (\ref{ezero.1}) are made to converge if $k_1>0$, and all other cases are shuffle-regularized based on $\mathcal{E}_0(0;\tau) = 2\pi i \tau$.

At depth one, they are related to the holomorphic iterated Eisenstein integrals via \cite{Broedel:2018izr}
\begin{equation}\label{eq:ExtractConvergentDepthOne}
	\EBR{j}{k}{\tau} = j! \, \mathcal{E}_0 (0^j,k\,;\tau) + \frac{B_k (2\pi i \tau)^{j+1} }{k!(j{+}1)}
\end{equation}
with Bernoulli numbers $B_k$.
From this, we see that the implicit action of SV on these functions at depth one is given by
\begin{equation}
{\rm SV}\, \mathcal{E}_0 (0^j,k\,;\tau) = \frac{1}{j!}\EsvBR{j}{k}{\tau}  - \frac{B_k(-4y)^{j+1} }{k!(j{+}1)!}\,.
\label{ezero.4}
\end{equation}
By (\ref{eesv.2a}), (\ref{eq:ExtractConvergentDepthOne}) and the shuffle relation
\begin{equation}
	\mathcal{E}_0 (0^j,k;\tau) = \sum_{r=0}^j \frac{(-1)^{j-r}}{r!} (2\pi i \tau)^r \mathcal{E}_0 (k,0^{j-r};\tau)\,,
	\label{ezero.5}
\end{equation}
two equivalent formulations of (\ref{ezero.4}) are
\begin{align}
{\rm SV}\, 	\mathcal{E}_0 (0^j,k\,;\tau) &= \mathcal{E} _0 (0^j,k\,;\tau) + \sum_{r=1}^{j}  \overline{\mathcal{E}_0 (0^r\,;\tau)}\mathcal{E} _0 (0^{j-r},k\,;\tau) + \overline{\mathcal{E}_0 (k,0^j\,;\tau)}
	\label{ezero.6}
		\\
{\rm SV}\, 	\mathcal{E}_0 (k,0^j\,;\tau) &= \mathcal{E} _0 (k,0^j\,;\tau) + \sum_{r=1}^{j}
\overline{\mathcal{E}_0 (0^{j-r},k\,;\tau)}\mathcal{E} _0 (0^{r}\,;\tau) + \overline{\mathcal{E}_0 (0^j,k\,;\tau)}\,,
\notag
\end{align}
which match the expectation from~\cite{Panzerknewit}.

\section{Examples}
\label{sec:5}

We shall now spell out a variety of examples that illustrate both
the pairing of cycles with dual antielliptic integrands $\overline{V(\ldots|\tau)}$
and the action of the single-valued map on eMZVs. Both eMZVs and MGFs
occur in the simultaneous expansion of the generating series $B^\tau_{\vec{\eta}},J^\tau_{\vec{\eta}}$
in $s_{ij}$ and $\eta_j$ which results in a lattice-sum representation of all convergent
${\rm SV} \, \omega(n_1,\ldots,n_r|{-}\tfrac{1}{\tau})$ in section \ref{sec:5.8}. The coefficients in
the $\eta_j$-expansions will be referred to as {\it component integrals}, and we will use the shorthand
\beq
f^{(a)}_{ij} = f^{(a)}(z_{i}{-}z_j,\tau) \, , \ \ \ \ \ \ \overline{f^{(b)}_{ij}} = \overline{ f^{(b)}(z_{i}{-}z_j,\tau) }
\label{shortf}
\eeq
for the Kronecker--Eisenstein coefficients defined by (\ref{looprev.6}) that occur in the integrands.
More precisely, the building blocks (\ref{looprev.5}) and (\ref{newBB.0}) of the integrands
of $B^\tau_{\vec{\eta}},J^\tau_{\vec{\eta}}$ involve the following combinations of (\ref{shortf})
with $\eta_{j,j+1 \ldots n} = \eta_j{+}\eta_{j+1}{+}\ldots{+}\eta_n$,
\begin{align}
\varphi^\tau_{\vec{\eta}}(1,2,\ldots,n) &=
 \sum_{a_2,\ldots,a_n \geq 0} \eta_{23\ldots n}^{a_2-1} \eta_{3\ldots n}^{a_3-1}
 \ldots \eta_{n}^{a_n-1}  f^{(a_2)}_{12} f^{(a_3)}_{23} \ldots f^{(a_n)}_{n-1,n} \notag \\
V_w(1,2,\ldots,n|\tau) &= \sum_{a_1,a_2,\ldots,a_n \geq 0 \atop{a_1+a_2+\ldots+a_n = w}}  f^{(a_1)}_{12}
f^{(a_2)}_{23} \ldots f^{(a_{n-1})}_{n-1,n} f^{(a_n)}_{n,1} \, .
\label{shortv}
\end{align}

\subsection{Two-point $\alpha'$-expansions}
\label{sec:5.1}

At two points, the general definitions (\ref{newBB.1}) and (\ref{newBB.11}) only admit a single
permutation of the integrands and cycle in $B^\tau_{\eta_2} = B^\tau_{\eta_2}(2|2) $
and $J^\tau_{\eta_2} = J^\tau_{\eta_2}(2|2) $,
\begin{align}
B^\tau_{\eta_2} &=  \int_{-\tau/2}^{\tau/2} \frac{\dd z_2 }{ \tau \eta_2} \sum_{a=0}^\infty \tau^a \eta_2^{a} f^{(a)}_{12} e^{s_{12} {\cal G}_{\mathfrak B}(z_{12},\tau)}
\label{comp2pA}
\\
J^\tau_{\eta_2} &=  \int_{\mathfrak T} \frac{ \dd^2 z_2}{\Im \tau} \,   \frac{1}{\eta_2} \sum_{a=0}^\infty (\tau{-}\bar \tau)^a \eta_2^{a} f^{(a)}_{12} e^{s_{12} {\cal G}_{\mathfrak T}(z_{12},\tau)}   \, ,
\notag
\end{align}
and we introduce the following notation for component integrals
\begin{align}
 B^\tau_{(a)}  &= B^\tau_{\eta_2} \, \big|_{\eta_2^{a-1}}
 =  \tau^a  \int_{-\tau/2}^{\tau/2} \frac{ \dd z_2}{ \tau} \,
f^{(a)}_{12} e^{s_{12} {\cal G}_{\mathfrak B}(z_{12},\tau)}
\label{comp2p}
\\
 J^\tau_{(a)}  &= J^\tau_{\eta_2} \, \big|_{\eta_2^{a-1}}
 =  (\tau{-}\bar \tau)^a  \int_{\mathfrak T} \frac{ \dd^2 z_2}{\Im \tau} \,
f^{(a)}_{12} e^{s_{12} {\cal G}_{\mathfrak T}(z_{12},\tau)} \, .
\notag
 \end{align}
Then, combining the initial values (\ref{svEMZV.54}) and (\ref{svEMZV.55})
with the $\alpha'$-expansions (\ref{svEMZV.8}) and (\ref{svEMZV.80}) yields
expressions like
 \begin{align}
B^\tau_{(0)} &=  1 + s_{12}^2 \Big({-}3 \betaBR{1}{4}{\tau} + \frac{ \zeta_2}{6} + \frac{i \zeta_3}{2 T}
+ \frac{    3 \zeta_4}{4 T^2} \Big) \notag \\
&\ \ +
 s_{12}^3 \Big({-}5 \betaBR{2}{6}{\tau} + 12 \zeta_2  \betaBR{0}{4}{\tau} + \frac{ \zeta_3}{12}
+ \frac{19 i \zeta_4}{24 T}    - \frac{ \zeta_5}{    4 T^2} + \frac{   \zeta_2 \zeta_3}{T^2}
 - \frac{4 i \zeta_6}{3 T^3} \Big)   + {\cal O}(s_{12}^4)
\notag \\
B^\tau_{(2)} &=  -2 \zeta_2 + s_{12} \Big(3 \betaBR{2}{4}{\tau} - \zeta_3 + \frac{3 i \zeta_4}{T} \Big)  \label{expls.1} \\
&\ \ +  s_{12}^2 \Big(10 \betaBR{3}{6}{\tau} - 18 \zeta_2 \betaBR{1}{4}{\tau}
 - \frac{29 \zeta_4}{12}  - \frac{i \zeta_5}{T} + \frac{    3 i \zeta_2 \zeta_3}{T} + \frac{    43 \zeta_6}{8 T^2} \Big)  + {\cal O}(s_{12}^3)
 \notag \\
B^\tau_{(4)} &=  -2 \zeta_4 +
 s_{12} \Big(5 \betaBR{4}{6}{\tau} - 6 \zeta_2 \betaBR{2}{4}{\tau}   +  2 \zeta_2 \zeta_3 - \zeta_5 - \frac{11 i \zeta_6}{2 T} \Big)+ {\cal O}(s_{12}^2)
\notag
  \end{align}
as well as
\begin{align}
J^{\tau}_{(0)}&=
1 + s_{12}^2 \Big( {-}3 \bsvBR{1}{4}{\tau} + \frac{ \zeta_3}{2 y}  \Big) +s_{12}^3 \Big( {-}5 \bsvBR{2}{6}{\tau} + \frac{ \zeta_3}{6} + \frac{  \zeta_5}{8 y^2}  \Big)  + {\cal O}(s_{12}^4)
 \notag \\
J^{\tau}_{(2)}&=
s_{12} ( 3 \bsvBR{2}{4}{\tau} - 2 \zeta_3 ) + s_{12}^2 \Big( 10 \bsvBR{3}{6}{\tau} - \frac{\zeta_5}{y}  \Big)
 + {\cal O}(s_{12}^3) \label{expls.2}   \\
J^{\tau}_{(4)}&=
s_{12} ( 5 \bsvBR{4}{6}{\tau} - 2 \zeta_5 )
  + {\cal O}(s_{12}^2) \notag
\end{align}
upon extracting suitable powers of $\eta_2$. The action of SV on
the $\zeta_n, T, \beta[\ldots]$ as in (\ref{newBB.19b}) and (\ref{eesv.1})
relates $J^\tau_{(a)}= {\rm SV} \, B^\tau_{(a)}$ as expected from (\ref{newBB.20}).
Examples of $\beta[\ldots]$
beyond depth one occur at the next orders in $s_{ij}$, e.g.\
\begin{align}
B^\tau_{(0)} \, \big|_{s_{12}^4} &=
{-}21 \betaBR{3}{8}{\tau} + 9 \betaBR{1& 1}{4& 4}{\tau} -
    18 \betaBR{2& 0}{4& 4}{\tau}  - \frac{\zeta_2}{2} \betaBR{1}{4}{\tau}  +
    40  \zeta_2  \betaBR{1}{6}{\tau}  \notag \\
    &\ \   + 6 \zeta_3 \betaBR{0}{4}{\tau}     - \frac{    3 i  \zeta_3  }{2 T} \betaBR{1}{4}{\tau} - \frac{    18 i  \zeta_4}{T}  \betaBR{0}{4}{\tau}
    - \frac{9 \zeta_4 }{ 4 T^2} \betaBR{1}{4}{\tau}  \label{expls.3} \\
    &\ \ + \frac{131 \zeta_4}{720} + \frac{5 i \zeta_5}{12 T}
    + \frac{i \zeta_2 \zeta_3}{12 T}  + \frac{23 \zeta_6}{32 T^2}
    + \frac{\zeta_3^2}{8 T^2}   - \frac{9 i \zeta_3 \zeta_4}{8 T^3}
    + \frac{i \zeta_2 \zeta_5}{T^3}
    - \frac{ 3 i \zeta_7}{8 T^3} + \frac{85 \zeta_8}{192 T^4}
   \notag
\end{align}
as well as
\begin{align}
J^\tau_{(0)} \, \big|_{s_{12}^4} &=
 {-}21 \bsvBR{3}{8}{\tau} + 9  \bsvBR{1& 1}{4& 4}{\tau} - 18 \bsvBR{2& 0}{4& 4}{\tau} \notag \\
&\ \ +
 12  \zeta_3 \bsvBR{0}{4}{\tau}  - \frac{ 3 \zeta_3}{2 y}  \bsvBR{1}{4}{\tau}
  + \frac{ 5 \zeta_5}{12 y} - \frac{  \zeta_3^2}{8 y^2} + \frac{ 3 \zeta_7}{ 32 y^3}    \notag \\
J^\tau_{(0)} \, \big|_{s_{12}^5} &=    {-}135 \bsvBR{4}{10}{\tau} - 60 \bsvBR{3& 0}{6& 4}{\tau} +
 15 \bsvBR{1& 2}{4& 6}{\tau} + 15 \bsvBR{2& 1}{6& 4}{\tau} -
 60 \bsvBR{2& 1}{4& 6}{\tau}  \label{expls.3a} \\
 &\ \  - \frac{ 1}{2} \zeta_3 \bsvBR{1}{4}{\tau}
 + \frac{ 6  \zeta_5}{y}  \bsvBR{0}{4}{\tau} - \frac{3  \zeta_5}{8 y^2} \bsvBR{1}{4}{\tau}
  + 40 \zeta_3 \bsvBR{1}{6}{\tau}  - \frac{5  \zeta_3}{2 y}  \bsvBR{2}{6}{\tau}  \notag \\
  &\ \
 + \frac{43 \zeta_5}{360}  + \frac{ \zeta_3^2}{12 y} + \frac{7 \zeta_7}{32 y^2}  - \frac{3 \zeta_3 \zeta_5}{ 16 y^3}
 + \frac{15 \zeta_9}{128 y^4} \, . \notag
\end{align}

\subsection{Extracting single-valued eMZVs}
\label{sec:5.2}

The above $\alpha'$-expansions at two
points have been generated in earlier work in terms of eMZVs \cite{Broedel:2018izr, Mafra:2019ddf} and
MGFs \cite{DHoker:2015wxz,DHoker:2016mwo, Gerken:2019cxz}, respectively. The results in the references include
\begin{align}
B^\tau_{(0)} &=1+ s_{12}^2 \Big( \frac{1}{2} \omega(0,0,2|{-}\tfrac{1}{\tau}) + \frac{ 5 \zeta_2}{12} \Big)
+ s_{12}^3  \Big( \frac{1}{18} \omega(0, 0, 3, 0|{-}\tfrac{1}{\tau}) - \frac{4}{3} \zeta_2 \omega(0, 0, 1, 0|{-}\tfrac{1}{\tau}) + \frac{ \zeta_3}{12} \Big) \notag \\
& \ \ \ \ + s_{12}^4 \Big(
{-} \omega(0, 0, 0, 2, 2|{-}\tfrac{1}{\tau})- \frac{5}{4} \omega(0, 0, 0, 0, 4|{-}\tfrac{1}{\tau})  + \frac{1}{8} \omega(0, 0, 4|{-}\tfrac{1}{\tau}) \label{expls.4} \\
&\ \ \ \ \ \ \ \
+ \frac{5}{8} \omega(0, 0, 2|{-}\tfrac{1}{\tau})^2  + \frac{ 13}{24}  \zeta_2 \omega(0, 0, 2|{-}\tfrac{1}{\tau})  -
 2 \zeta_2 \omega(0, 0, 0, 0, 2|{-}\tfrac{1}{\tau}) + \frac{343 \zeta_4}{576}
\Big)
+ {\cal O}(s_{12}^5)
\notag \\
J^\tau_{(0)} &= 1 + \frac{1}{2} s_{12}^2 {\rm E}_2(\tau) + \frac{1}{6} s_{12}^3 \big( {\rm E}_3(\tau) + \zeta_3\big)
+s_{12}^4\Big( {\rm E}_{2,2}(\tau){+}\frac{3}{20} {\rm E}_{4}(\tau){+}\frac{1}{8}{\rm E}_{2}^{2}(\tau) \Big)
+ {\cal O}(s_{12}^5)\, ,
\notag
\end{align}
where for instance (by comparison with (\ref{expls.1}) and (\ref{expls.2}))
\beq
 \omega(0,0,2|{-}\tfrac{1}{\tau}) = - 6 \betaBR{1}{4}{\tau}  - \frac{ \zeta_2 }{2} + \frac{i \zeta_3}{T}
 + \frac{3 \zeta_4}{2 T^2}
 \, , \ \ \ \ \ \
{\rm E}_2(\tau)= - 6 \bsvBR{1}{4}{\tau} + \frac{ \zeta_3}{y}
\, ,
\eeq
and the modular transformation may be evaluated to yield \cite{Broedel:2018izr, Zerbini:2018sox}
\begin{align}
  \label{expls.5}
 \omega(0,0,2|{-}\tfrac{1}{\tau}) &= - \frac{ T^2 }{180} - \frac{ \zeta_2 }{6} + \frac{ 3 \zeta_4 }{2T^2}+ \omega(0,0,2|\tau) + \frac{8i}{T} \zeta_2 \omega(0,1,0,0|\tau) \, .
\end{align}
For the $s_{12}^2$-order in (\ref{expls.4}), the component version $J^\tau_{(a)}= {\rm SV} \, B^\tau_{(a)}$
of $J^\tau_{\eta_2}= {\rm SV} \, B^\tau_{\eta_2}$ implies that
\begin{align}
 {\rm SV} \,   \omega(0,0,2|{-}\tfrac{1}{\tau}) = {\rm E}_2(\tau)
\label{expls.6}
\end{align}
and a similar analysis for higher orders in $s_{12}$ and at $a\neq 0$ yields for instance
\begin{align}
{\rm SV} \, \omega(0,3|{-}\tfrac{1}{\tau})  &= 2 \pi \nabla {\rm E}_2(\tau) \, , \ \ \ \ \ \ &
{\rm SV} \, \omega(0,0,4|{-}\tfrac{1}{\tau})  &= -\frac{ 4}{3} \pi \nabla {\rm E}_3(\tau)  \label{SVex04}  \\
{\rm SV} \,\omega(0,0,3,0|{-}\tfrac{1}{\tau}) &= 3 {\rm E}_3(\tau) \, , \ \ \ \ \ \ &
{\rm SV} \,\omega(0, 5|{-}\tfrac{1}{\tau}) &= - \frac{4}{3} (\pi \nabla)^2 {\rm E}_3(\tau)
\, . \notag
\end{align}
At depth two, relating $B^\tau_{(0)} \big|_{s_{12}^4}  \leftrightarrow J^\tau_{(0)} \big|_{s_{12}^4} $
 (see (\ref{expls.3}) and (\ref{expls.3a})) or $B^\tau_{(2)} \big|_{s_{12}^3}  \leftrightarrow J^\tau_{(2)} \big|_{s_{12}^3} $
 yields
\begin{align}
{\rm E}_{2,2}(\tau) &=
  {\rm SV}\,  \Big(
{-}\frac{7}{5} \omega(0,0,0,0,4 | {-}\tfrac{1}{\tau}) - \omega(0,0,0,2,2 | {-}\tfrac{1}{\tau})  + \frac{1}{2}\omega(0,0,2 | {-}\tfrac{1}{\tau})^2 + \frac{3}{20} \omega(0,0,4 | {-}\tfrac{1}{\tau}) \Big)  \notag \\
 \pi \nabla {\rm E}_{2,2} (\tau) &=
{\rm SV} \, \Big( {-}\frac{1}{60} \omega(0, 5|{-}\tfrac{1}{\tau}) + \frac{3}{5} \omega(0, 0, 0, 5|{-}\tfrac{1}{\tau}) -\frac{1}{2} \omega(0, 0, 2, 3|{-}\tfrac{1}{\tau}) \Big)
\, ,
\label{expls.7}
\end{align}
where the combinations
\begin{align}
{\rm E}_{2,2}&=  \bigg(\frac{\Im \tau}{\pi} \bigg)^4 \Big( \cform{ 2 &1 &1 \\  2 &1 &1 }  - \frac{9}{10} \cform{ 4 &0 \\ 4 &0 }  \Big) \notag \\
&= -18 \bsvBRno{2& 0}{4& 4} + 12 \zeta_3 \bsvBRno{0}{ 4} + \frac{5 \zeta_5}{12y} - \frac{ \zeta_3^2}{4y^2}
\label{expls.7a}  \\
\pi \nabla {\rm E}_{2,2}&= \frac{(\Im \tau)^5}{\pi^3}  \Big( \cform{ 3 &1 &1 \\  1 &1 &1 }  - \frac{8}{5} \cform{ 5 &0 \\ 3 &0 }  \Big) \notag \\
&= 9 \bsvBRno{2& 1}{4& 4} - 6 \zeta_3 \bsvBRno{1}{ 4} - \frac{5 \zeta_5}{12} + \frac{ \zeta_3^2}{2y} \notag
\end{align}
of MGFs (\ref{newdef.3}) are engineered to avoid ${\rm G}_8$ in the differential equations \cite{Broedel:2018izr}.
The systematics of depth-one relations between eMZVs and non-holomorphic Eisenstein series
including higher-weight generalizations of (\ref{SVex04}) is detailed in appendix \ref{app:2pt.1}
and leads to the closed-form results
\begin{align}
{\rm SV}\, \sum_{j=0}^{\ell-1} \frac{ B_j }{j!} \omega(0^{\ell-j},2k{+}\ell | {-}\tfrac{1}{\tau})
&= (-1)^\ell
\frac{ (k{+}\ell{-}1) ! }{ (2k{+}\ell{-}1) ! }  (-4 \pi \nabla)^k {\rm E}_{k+\ell}
 \, , \ \ \ \; k{\geq} 0\, , \ \ \ell {\geq} 1 \, , \ \ k{+}\ell {\geq} 2 \label{alldepth} \\
{\rm SV}\, \sum_{j=0}^{\ell-1} \frac{ B_j }{j!} \omega(0^{\ell-j},\ell{-}2k | {-}\tfrac{1}{\tau})
&= (-1)^{\ell+k} \frac{ (\ell{-}k{-}1) ! }{(\ell{-}1)!} \frac{ (\pi \overline \nabla)^k {\rm E}_{\ell-k} }{(2y)^{2k}} \, ,
\ \ \ \ \ \, k{\geq}0 \, , \ \ \ell {-} 2k {\geq}1
\, , \ \ \ell {-} k {\geq}2  \notag
 \end{align}
for combinations of eMZVs of different length that are weighted by Bernoulli numbers
$B_j$ \cite{privNils}. Similarly, the analogue of (\ref{expls.7}) for the MGF ${\rm E}_{2,3 } =
(\frac{\Im \tau}{\pi} )^5 \big( \cform{ 3 &1 &1 \\  3 &1 &1 }  - \frac{43}{35} \cform{ 5 &0 \\ 5 &0 }  \big)$
and its holomorphic derivatives is spelled out in appendix \ref{app:2pt.2}.

Note that the single-valued map of A-cycle eMZVs at argument $\tau$ rather than $-\frac{1}{\tau}$
generically leads to combinations of MGFs of different modular weights. For instance, changing
the argument $-\frac{1}{\tau}$ to $\tau$ in (\ref{expls.6}) gives rise to
\beq
{\rm SV} \, \omega(0,0,2|\tau) = - \frac{y^2}{15} + {\rm E}_2(\tau) + \frac{ \pi \overline \nabla {\rm E}_2(\tau)}{y}
\label{expls.7h}
\eeq
instead of a single modular invariant $ {\rm E}_2(\tau)$. This can be seen by
expressing all of $ \omega(0,0,2|\tau)$, $ {\rm E}_2(\tau)$ and $\overline \nabla {\rm E}_2(\tau)$
in terms of convergent iterated Eisenstein
integrals (\ref{ezero.1}) and applying their single-valued map (\ref{ezero.6})\footnote{The representations in terms
of convergent iterated Eisenstein integrals needed to verify (\ref{expls.7h}) are
\begin{align*}
\omega(0,0,2|\tau) &= - 6 {\cal E}_0(4,0;\tau) -\frac{1}{3} \zeta_2
\\
{\rm E}_2(\tau) &= \frac{y^2}{45} + \frac{ \zeta_3}{y} - 12 {\rm Re}[ {\cal E}_0(4,0;\tau)] - \frac{6}{y}  {\rm Re}[ {\cal E}_0(4,0,0;\tau)]
\\
\pi \overline \nabla {\rm E}_2(\tau) &=
 \frac{2y^3}{45} -  \zeta_3 + 24 y^2 \overline{ {\cal E}_0(4;\tau) } + 12 y  \overline{ {\cal E}_0(4,0;\tau) }
 + 6 {\rm Re}[ {\cal E}_0(4,0,0;\tau)]
\end{align*}
}.
Alternatively, (\ref{expls.7h}) can be deduced by
setting $\tau \rightarrow - \frac{1}{\tau}$ in (\ref{expls.5}) and exploiting the result
$ {\rm SV} \, \omega(0,1,0,0|{-}\tfrac{1}{\tau}) = \frac{  3 \pi\overline \nabla {\rm E}_2(\tau) }{8y^2}$
that will be extracted from a four-point example in section \ref{sec:5.5}. The much cleaner result (\ref{expls.6}) for
${\rm SV} \, \omega(0,0,2|{-}\frac{1}{\tau})$ as compared to ${\rm SV} \, \omega(0,0,2|\tau)$
is another manifestation of the fact that the differential equations (\ref{newBB.16}) of
B-cycle integrals are more closely related to the closed-string counterparts (\ref{newBB.17}) than the
A-cycle differential equations in (\ref{looprev.11}).


\subsection{Symmetrized cycles and graph functions}
\label{sec:5.3}

At $n\geq 3$ points, most of the antielliptic functions in (\ref{newBB.7})
introduce non-constant $\overline{f^{(a)}_{ij}}$ into the closed-string integrands,
except for the simplest case $V_0(1,2,\ldots,n|\tau)=1$ dual to a permutation
sum over B-cycles, see (\ref{mix.2}),
\begin{align}
J^\tau_{0,\vec{\eta}}(\ast|\rho) &= \int_{\mathfrak T^{n-1}} \bigg( \prod_{j=2}^n \frac{ \dd^2 z_j }{\Im \tau}\bigg)
\! \!  \prod_{1\leq i <j }^n \! \!  e^{s_{ij} {\cal G}_{\mathfrak T}(z_{ij},\tau)} \rho \bigg\{ \prod_{k=2}^n \sum_{a_k=0}^\infty (\tau{-}\bar \tau)^{a_k} (\eta_{k,k+1\ldots n})^{a_k-1} f_{k-1,k}^{(a_k)}  \bigg\}  \label{expls.8}\\
&= {\rm SV} \sum_{\gamma \in S_{n-1}} B^\tau_{\vec{\eta}}(\gamma|\rho) \notag \, .
\end{align}
As indicated by the $\ast$-notation, the $J^\tau_{0,\vec{\eta}}(\ast|\rho) $ integral (\ref{mix.1})
on the left-hand side is independent of the ordering $\ast$ since its integrand
$V_0$ is. The symmetrized open-string integrals on the right-hand side were studied
in \cite{Broedel:2018izr, Zerbini:2018hgs, Zagier:2019eus} as the generating series of
holomorphic graph functions,
\begin{align}
 \sum_{\gamma \in S_{n-1}} &B^\tau_{\vec{\eta}}(\gamma|\rho) = \bigg(  \int \limits_{-\tau/2}^{\tau/2} \prod_{j=2}^n  \frac{\dd z_j}{\tau} \bigg)
\! \!  \prod_{1\leq i <j }^n \! \!  e^{s_{ij} {\cal G}_{\mathfrak B}(z_{ij},\tau)} \rho \bigg\{ \prod_{k=2}^n \sum_{a_k=0}^\infty \tau^{a_k}(\eta_{k,k+1\ldots n})^{a_k-1} f_{k-1,k}^{(a_k)}  \bigg\}  \, ,  \label{expls.9}
\end{align}
where each puncture is integrated independently over the entire B-cycle.
More specifically, the references considered the components $f_{k-1,k}^{(a_k)} \rightarrow f^{(0)}_{k-1,k}= 1$
at the most singular order in the $\eta_j$,
\begin{align}
M_n^{\rm open} &= \sum_{\gamma \in S_{n-1}} B^\tau_{\vec{\eta}}(\gamma|\rho)  \, \big|_{\eta_{23\ldots n}^{-1}
\eta_{3\ldots n}^{-1}\ldots \eta_{n}^{-1}} =
  \bigg(  \prod_{j=2}^n \int_{-\tau/2}^{\tau/2}  \frac{ \dd z_j }{\tau} \bigg)
  \prod_{1\leq i <j }^n \! \!  e^{s_{ij} {\cal G}_{\mathfrak B}(z_{ij},\tau)}  \label{expls.9sing}\\
&
\tikzpicture
\draw(0,0)node{$\displaystyle  = 1+ \frac{1}{2}  \sum_{1\leq i<j}^n s_{ij}^2 $};
\draw(4,0)node{$\displaystyle  + \frac{1}{6}   \sum_{1\leq i<j}^n s_{ij}^3 $};
\draw(8,0)node{$\displaystyle  + \! \! \!  \sum_{1\leq i<j<k}^n s_{ij} s_{ik} s_{jk}  \! \! \! $};
\draw(11.9,0)node{$\displaystyle  +\, {\cal O}(s_{ij}^4)\, ,$};
\scope[xshift=1.8cm]
\draw(0.37,0)node{$ {\bf B}\Big[ \ \ \ \ \ \ \Big]$};
\draw(0.3,0)node{$\bullet$};
\draw(0.8,0)node{$\bullet$};
\draw(0.3,0) .. controls (0.45,0.25) and (0.65,0.25) .. (0.8,0);
\draw(0.3,0) .. controls (0.45,-0.25) and (0.65,-0.25) .. (0.8,0);
\endscope
\scope[xshift=5.4cm]
\draw(0.37,0)node{${\bf B}\Big[ \ \ \ \ \ \ \Big]$};
\draw(0.3,0)node{$\bullet$};
\draw(0.8,0)node{$\bullet$};
\draw(0.3,0) .. controls (0.45,0.25) and (0.65,0.25) .. (0.8,0);
\draw(0.3,0) .. controls (0.45,-0.25) and (0.65,-0.25) .. (0.8,0);
\draw(0.3,0) -- (0.8,0);
\endscope
\scope[xshift=10cm]
\draw(0.37,0)node{${\bf B}\Big[ \ \ \ \ \ \ \Big]$};
\draw(0.3,-0.2)node{$\bullet$};
\draw(0.8,-0.2)node{$\bullet$};
\draw(0.55,0.22)node{$\bullet$};
\draw(0.3,-0.2)--(0.8,-0.2);
\draw(0.55,0.22)--(0.8,-0.2);
\draw(0.3,-0.2)--(0.55,0.22);
\endscope
\endtikzpicture
\notag
\end{align}
where the dependence on the permutation $\rho$ drops out, and the integrands
at fixed order in $s_{ij}$ are polynomials in B-cycle Green functions.
In passing to the second line, each monomial in ${\cal G}_{\mathfrak B}(z_{ij},\tau)$ is
mapped to a graph $\Gamma$ that labels the B-cycle graph functions ${\bf B}[\Gamma]$,
where a factor of ${\cal G}_{\mathfrak B}(z_{ij},\tau)$ is represented by an edge connecting vertices $z_i$ and $z_j$.
One-particle reducible graphs $\Gamma_{1  {\rm PR}}$ lead to vanishing ${\bf B}[\Gamma_{1 {\rm PR}}]$
since $\int^{\tau/2}_{-\tau/2} \dd z \, {\cal G}_{\mathfrak B}(z,\tau) =0$, i.e.\ higher
orders of (\ref{expls.9sing}) stem from all combinations of one-particle irreducible graphs with
four and more edges in total. Any ${\bf B}[\Gamma]$ is expressible in terms of B-cycle eMZVs
\cite{Broedel:2018izr} since the $\alpha'$-expansion of each component integral of the series $B^\tau_{\vec{\eta}}(\gamma|\rho) $ is.

Similarly, modular graph functions ${\bf D}[\Gamma]$ (as opposed to modular graph forms)
were defined \cite{DHoker:2015wxz,DHoker:2016mwo} by $n$-point
torus integrals over monomials in ${\cal G}_{\mathfrak T}(z_{ij},\tau)$,
where each torus Green function
is again visualized through an edge between vertices $z_i$ and $z_j$.
The ${\bf D}[\Gamma]$ associated with dihedral graphs $\Gamma$ are
proportional to the lattice sums (\ref{newdef.3}) with $a_j=b_j$, and also more
complicated graph topologies can be straightforwardly translated into lattice sums.

The generating series of $n$-point modular graph functions resides at the
most singular order of (\ref{expls.8}) w.r.t.\ $\eta_j$ where the insertions
of $f^{(a)}_{ij}$ are absent,
\begin{align}
M_n^{\rm closed} &= J^\tau_{0,\vec{\eta}}(\ast|\rho)  \, \big|_{\eta_{23\ldots n}^{-1}
\eta_{3\ldots n}^{-1}\ldots \eta_{n}^{-1}} =\int_{\mathfrak T^{n-1}} \bigg( \prod_{j=2}^n \frac{ \dd^2 z_j }{\Im \tau}\bigg)
\! \!  \prod_{1\leq i <j }^n \! \!  e^{s_{ij} {\cal G}_{\mathfrak T}(z_{ij},\tau)}  \label{cl.9sing}\\
&\tikzpicture
\draw(0,0)node{$\displaystyle  = 1+ \frac{1}{2}  \sum_{1\leq i<j}^n s_{ij}^2 $};
\draw(4,0)node{$\displaystyle  + \frac{1}{6}   \sum_{1\leq i<j}^n s_{ij}^3 $};
\draw(8,0)node{$\displaystyle  + \! \! \!  \sum_{1\leq i<j<k}^n s_{ij} s_{ik} s_{jk}  \! \! \! $};
\draw(11.9,0)node{$\displaystyle  +\, {\cal O}(s_{ij}^4)\, ,$};
\scope[xshift=1.8cm]
\draw(0.37,0)node{$ {\bf D}\Big[ \ \ \ \ \ \ \Big]$};
\draw(0.3,0)node{$\bullet$};
\draw(0.8,0)node{$\bullet$};
\draw(0.3,0) .. controls (0.45,0.25) and (0.65,0.25) .. (0.8,0);
\draw(0.3,0) .. controls (0.45,-0.25) and (0.65,-0.25) .. (0.8,0);
\endscope
\scope[xshift=5.4cm]
\draw(0.37,0)node{${\bf D}\Big[ \ \ \ \ \ \ \Big]$};
\draw(0.3,0)node{$\bullet$};
\draw(0.8,0)node{$\bullet$};
\draw(0.3,0) .. controls (0.45,0.25) and (0.65,0.25) .. (0.8,0);
\draw(0.3,0) .. controls (0.45,-0.25) and (0.65,-0.25) .. (0.8,0);
\draw(0.3,0) -- (0.8,0);
\endscope
\scope[xshift=10cm]
\draw(0.37,0)node{${\bf D}\Big[ \ \ \ \ \ \ \Big]$};
\draw(0.3,-0.2)node{$\bullet$};
\draw(0.8,-0.2)node{$\bullet$};
\draw(0.55,0.22)node{$\bullet$};
\draw(0.3,-0.2)--(0.8,-0.2);
\draw(0.55,0.22)--(0.8,-0.2);
\draw(0.3,-0.2)--(0.55,0.22);
\endscope
\endtikzpicture
\notag
\end{align}
and we have ${\bf D}[\Gamma_{1{\rm PR}}]=0$ by $\int_{\mathfrak T} \dd^2 z \,  {\cal G}_{\mathfrak T}(z,\tau)=0$.
As a consequence of (\ref{expls.8}) at the most singular order in the $\eta_j$, modular graph functions
are single-valued B-cycle graph functions,
\beq
M_n^{\rm closed} =  {\rm SV} \,  M_n^{\rm open} \ \ \Leftrightarrow \ \
{\bf D}[\Gamma] ={\rm SV} \, {\bf B}[\Gamma]\, ,
\label{bsvd.1}
\eeq
which ultimately follows from the `Betti--deRham duality' between $V_0=1$ and the
symmetrized cycles $ \sum_{\gamma \in S_{n-1}} \mathfrak B(\gamma(2,\ldots,n))$.

The relations in (\ref{bsvd.1}) have firstly appeared in \cite{Broedel:2018izr} with
a proposal ``esv'' for an elliptic single-valued map in the place of SV. The esv map  of
\cite{Broedel:2018izr} has the same action (\ref{newBB.19b}) on MZVs and Laurent
polynomials in $\tau$ as the SV map in this work. In particular, all
pairs of B-cycle eMZVs and modular graph functions related via ${\rm esv} \, \omega(\ldots|{-}\frac{1}{\tau})\sim
{\bf D}[\ldots]$ in the reference are also related via
${\rm SV} \, \omega(\ldots|{-}\frac{1}{\tau})\sim {\bf D}[\ldots]$ as a consequence of (\ref{bsvd.1}).
For suitable representations of the $q$-series of eMZVs via ${\cal E}_0$ defined by (\ref{ezero.1}),
the Fourier expansions of all modular graph functions up to weight six could be reproduced
from the replacement ${\cal E}_0 \rightarrow 2 \Re({\cal E}_0)$ prescribed by esv \cite{Broedel:2018izr}.
However, it was an open problem in the reference to reconcile esv
with the shuffle property of iterated Eisenstein integrals.
The SV action (\ref{eesv.1}) in turn is expected to be compatible with
the shuffle multiplication of the $\beta[\ldots]$ and $\beta^{\rm sv}[\ldots]$
by the discussion in section \ref{sec:4.3} as detailed below (\ref{svEMZV.sv}).

Note that subleading orders in the $\eta_j$-expansion of (\ref{expls.8}) generate
infinite families of additional relation between MGFs and single-valued eMZVs
beyond (\ref{bsvd.1}). The comparison of open- and closed-string integrals
with additional insertions of $f^{(a_2)}_{12}f^{(a_3)}_{23}  \ldots f^{(a_n)}_{n-1,n}$ identifies MGFs of various modular
weights as single-valued B-cycle eMZVs.

\subsection{Three-point cycles and $\overline{V_1(1,2,3|\tau)}$}
\label{sec:5.4}

The simplest instance
of $J^\tau_{\vec{\eta}}(\gamma|\rho )  = {\rm SV} \, B^\tau_{\vec{\eta}}(\gamma|\rho ) $ with
non-constant antielliptic integrands $\overline{V(\ldots|\tau)}$ occurs at three points. The single-valued map
relates an antisymmetric integration cycle on the open-string side in (\ref{mix.5})
to the closed-string integral
\begin{align}
&J^{\tau}_{1,\eta_2,\eta_3}(2,3|2,3) = {-} \frac{1}{2\pi i } \int_{\mathfrak T^2} \frac{ \dd^2 z_2  \dd^2 z_3 }{(\Im \tau)^2} \! \!  \prod_{1\leq i <j }^3 \! \!  e^{s_{ij} {\cal G}_{\mathfrak T}(z_{ij},\tau)}  \notag \\
&\ \ \times \sum_{a,b=0}^\infty (\tau{-}\bar \tau)^{a+b} \eta_{23}^{a-1} \eta_3^{b-1} f^{(a)}_{12} f^{(b)}_{23} ( \overline{ f^{(1)}_{12} }{+}\overline{ f^{(1)}_{23} }{+}\overline{ f^{(1)}_{31} }) \, .
\label{expls.10}
\end{align}
Since contributions with even $a{+}b$ integrate to zero, the simplest component
integrals involve permutations of
$f^{(1)}_{12} \overline{ f^{(1)}_{12} }$
or $f^{(1)}_{23} \overline{ f^{(1)}_{12} }$
at the orders of $\eta_{3}^{-1}$ or $\eta_{23}^{-1}$,
\begin{align}
J^{\tau}_{1,\eta_2,\eta_3}(2,3|2,3) \, \big|_{\eta_{23}^{0}\eta_3^{-1}} &= - \frac{ \Im \tau }{\pi}
\int_{\mathfrak T^2} \frac{ \dd^2 z_2  \dd^2 z_3 }{(\Im \tau)^2} \! \!  \prod_{1\leq i <j }^3 \! \!  e^{s_{ij} {\cal G}_{\mathfrak T}(z_{ij},\tau)} f^{(1)}_{12} ( \overline{ f^{(1)}_{12} }{+}\overline{ f^{(1)}_{23} }{+}\overline{ f^{(1)}_{31} }) \notag \\
&= \frac{1}{s_{12}} + \frac{s_{123}^{2}}{2s_{12}}{\rm E}_{2}+\frac{s_{123}^{3}}{6s_{12}}({\rm E}_{3}+\zeta_{3})+\frac{s_{13}s_{23}}{2}(3{\rm E}_{3}+\zeta_{3}) \label{bsvd.2}\\
&\quad +\frac{s_{123}^{4}}{s_{12}}\Big({\rm E}_{2,2}+\frac{3}{20} {\rm E}_{4}+\frac{1}{8}{\rm E}_{2}^{2}\Big)+ s_{13}s_{23}s_{123}\Big(\frac{9}{2}{\rm E}_{2,2}+\frac{21}{20}{\rm E}_{4} \Big) +\mathcal{O}(s_{ij}^{4})
 \, ,
\notag
\end{align}
which furnish the simplest examples of kinematic poles $\sim s_{ij}^{-1}$ in
a $J^\tau_{\vec{\eta}}$-series. The corresponding antisymmetrized B-cycle integral
features the same types of kinematic poles in component integrals involving $f^{(1)}_{ij}$, e.g.\footnote{By slight abuse of notation, we denote the ordering of punctures $z_i,z_j$ on the imaginary axis by $-\frac{\tau}{2}{<}z_i{<}z_j{<}\frac{\tau}{2}$.}
\begin{align}
 \frac{1}{2} &\big[ B^\tau_{\eta_2,\eta_3}(2,3|2,3) - B^\tau_{\eta_2,\eta_3}(3,2|2,3) \big] \, \big|_{\eta_{23}^{0}\eta_3^{-1}}
 = \frac{1}{\tau} \int_{-\frac{\tau}{2}<z_2<z_3<\frac{\tau}{2}} \dd z_2 \, \dd z_3 \,  f^{(1)}_{12}
  \! \!  \prod_{1\leq i <j }^3 \! \!  e^{s_{ij} {\cal G}_{\mathfrak B}(z_{ij},\tau)} \notag \\
&= \frac{1}{s_{12}}
+\frac{ s_{123}^2 }{s_{12}} \Big( \frac{1}{2} \omega(0, 0, 2|{-}\tfrac{1}{\tau}) + \frac{5 \zeta_2}{12} \Big)
+\frac{ s_{123}^3 }{s_{12}} \Big( \frac{1}{18} \omega(0, 0, 3, 0|{-}\tfrac{1}{\tau}) - \frac{4}{3} \zeta_2 \omega(0, 0, 1, 0|{-}\tfrac{1}{\tau}) + \frac{ \zeta_3}{12} \Big) \notag \\
&\ \ + s_{13}s_{23} \Big(  \frac{1}{2} \omega(0,0,3,0|{-}\tfrac{1}{\tau})  +  \frac{ \zeta_3 }{4} \Big)
+\frac{ s_{123}^4 }{s_{12}}  \Big(
{-} \omega(0, 0, 0, 2, 2|{-}\tfrac{1}{\tau})- \frac{5}{4} \omega(0, 0, 0, 0, 4|{-}\tfrac{1}{\tau})  \label{bsvd.3} \\
&\ \ \ \ \ \
+ \frac{1}{8} \omega(0, 0, 4|{-}\tfrac{1}{\tau})
+ \frac{5}{8} \omega(0, 0, 2|{-}\tfrac{1}{\tau})^2  + \frac{ 13}{24}  \zeta_2 \omega(0, 0, 2|{-}\tfrac{1}{\tau})  -
 2 \zeta_2 \omega(0, 0, 0, 0, 2|{-}\tfrac{1}{\tau}) + \frac{343 \zeta_4}{576}
\Big)
\notag \\
&\ \
- s_{13}s_{23} s_{123} \Big(  \frac{9}{2}  \omega(0, 0, 0, 2, 2|{-}\tfrac{1}{\tau})
+  \frac{21}{4}  \omega(0, 0, 0, 0, 4|{-}\tfrac{1}{\tau})
 - \frac{1}{2}  \omega(0, 0, 4|{-}\tfrac{1}{\tau})    \notag \\
 &\ \ \ \ \ \  - \frac{9}{4}  \omega(0, 0, 2|{-}\tfrac{1}{\tau})^2 +  \frac{1}{2} \zeta_2 \omega(0, 0, 2|{-}\tfrac{1}{\tau})  - 3  \zeta_2\omega(0, 0, 0, 0, 2|{-}\tfrac{1}{\tau})  + \frac{11 \zeta_4}{40}
 \Big) + {\cal O}(s_{ij}^4)\, .
\notag
\end{align}
We have used that, by the antisymmetry $ f^{(1)}_{12}=- f^{(1)}_{21}$ of the integrand, the contribution from
the ordering $-\frac{\tau}{2}{<}z_3{<}z_2{<}\frac{\tau}{2}$ is minus that
of the ordering $-\frac{\tau}{2}{<}z_2{<}z_3{<}\frac{\tau}{2}$.
Comparison of (\ref{bsvd.3}) with (\ref{bsvd.2}) confirms the relation
(\ref{mix.5}) under the SV map at the respective orders in $s_{ij}$ and $\eta_j$.
Up to the restriction of the Koba--Nielsen factor to three instead of five punctures,
(\ref{bsvd.2}) and (\ref{bsvd.3}) are the type of integrals over $f^{(1)}_{ij} \overline{ f^{(1)}_{pq} }$
seen in genus-one five-point amplitudes of type II superstrings \cite{Richards:2008jg, Green:2013bza}.

\subsection{Four-point cycles and $\overline{V_2(1,2,3,4|\tau)}$}
\label{sec:5.5}

The esv map \cite{Broedel:2018izr}
has also been applied to the four-gluon amplitude of the heterotic string \cite{Gerken:2018jrq},
where the torus integral\footnote{The quantity $J_{\rm het}^\tau$ in (\ref{expls.41}) is defined to
be $(2\pi i )^{-2}$ times the complex conjugate of the integral ${\cal I}^{(2,0)}_{1234}$ in (2.44)
and (4.35) of \cite{Gerken:2018jrq}. Similarly, $B_{\rm het}^\tau$ in (\ref{expls.43}) is obtained
from the integral $Z^{(2)}_{1234}$ in section 5.2 of \cite{Gerken:2018jrq} through modular $S$ transformation.}
\begin{align}
J_{\rm het}^\tau &= \frac{1}{(2\pi i)^2} \int_{\mathfrak T^3} \bigg( \prod_{j=2}^4 \frac{ \dd^2 z_j }{\Im \tau} \bigg)
  \overline{V_2(1,2,3,4|\tau)}
\! \!  \prod_{1\leq i <j }^4 \! \!  e^{s_{ij} {\cal G}_{\mathfrak T}(z_{ij},\tau)} \notag \\
&= J^\tau_{2,\eta_2,\eta_3,\eta_4}(2,3,4|2,3,4 ) \, \big|_{ \eta_{234}^{-1} \eta_{34}^{-1} \eta_4^{-1} }
\label{expls.41}
\end{align}
was related to the open-string integration cycle dual to (\ref{mix.3}). More specifically,
the MGFs in \cite{Gerken:2018jrq}
\begin{align}
J_{\rm het}^\tau \, \big|_{k_j^2=0} &= {-} \frac{3 s_{13} \pi \overline \nabla {\rm E}_2 }{4 y^2}- (s_{13}^{2}{+}2s_{12}s_{23}) \frac{\pi \overline\nabla {\rm E}_3}{6 y^2}  \label{expls.42}
\\
&\ \ \ \ + s_{13}(s_{12} s_{23} - s_{13}^2) \Big(  \frac{ \pi \overline\nabla {\rm E}_4 }{5y^2}
 + \frac{ 3 {\rm E}_2 \pi \overline\nabla {\rm E}_2 }{2y^2}
 + \frac{ 3 \pi \overline\nabla {\rm E}_{2,2} }{y^2}
 \Big)+{\cal O}(s_{ij}^{4}) \notag
  \end{align}
were proposed to be the single-valued versions of the eMZVs in the $\alpha'$-expansion of
\begin{align}
B_{\rm het}^\tau &=  \frac{1}{6}\big[  2 B^\tau_{234}{+}2B^\tau_{432}  {-}B^\tau_{243}{-}B^\tau_{342}  {-} B^\tau_{324}{-} B^\tau_{423} \big]  \notag\\
B^\tau_{ijk}&= B_{\eta_2,\eta_3,\eta_4}^{\tau}(i,j,k|2,3,4) \, \big|_{ \eta_{234}^{-1} \eta_{34}^{-1} \eta_4^{-1} } \, ,\label{expls.43}
\end{align}
namely \cite{Broedel:2014vla}
\begin{align}
  B_{\rm het}^\tau \, \big|_{k_j^2=0} &= -2 s_{13} \omega(0,1,0,0|{-}\tfrac{1}{\tau}) - \frac{2}{3} (s_{13}^2 +2 s_{12}s_{23})
  \big[
 \omega(0,1,0,1,0|{-}\tfrac{1}{\tau}) + \omega(0,1,1,0,0| {-}\tfrac{1}{\tau})  \big]  \notag \\
 & \ \ + \frac{4}{3} s_{13}(s_{13}^2 - s_{12}s_{23})
 \Big[
 \omega(0,0,1,0,0,2|{-}\tfrac{1}{\tau}) + \omega(0,0,0,1,0,2|{-}\tfrac{1}{\tau})   \label{expls.44} \\
 &\ \ \ \ \ \ \ \ \ \ \ \ \ \ \ \ \ \ \ \ \ \ \ \ \ \ \ \  - \omega(0,1,0,1,1,0|{-}\tfrac{1}{\tau}) - \zeta_2 \omega(0,1,0,0|{-}\tfrac{1}{\tau})
 \Big] + {\cal O}(s_{ij}^4)  \, .\notag
 \end{align}
As indicated by $\big|_{k_j^2=0}$, the $\alpha'$-expansions (\ref{expls.42}) and (\ref{expls.44})
have been obtained in the limit of four-point on-shell kinematics with two independent Mandelstam invariants instead of six. However, the relation (\ref{newBB.20}) between $n$-point closed-string and single-valued open-string
integrals is conjectured to be valid for the $\frac{1}{2}n(n{-}1)$ independent Mandelstam variables
$\{s_{ij}, \ 1{\leq} i{<}j{\leq} n\}$ with $s_{ij}=s_{ji}$.
At four points, the corollary
\begin{align}
&J^\tau_{2,\eta_2,\eta_3,\eta_4}(2,3,4|\rho ) =  \frac{1}{6}{\rm SV} \, \big[  2 B_{\eta_2,\eta_3,\eta_4}^{\tau}(2,3,4|\rho)
+2 B_{\eta_2,\eta_3,\eta_4}^{\tau}(4,3,2|\rho) \label{genhet} \\
&\ \  -  B_{\eta_2,\eta_3,\eta_4}^{\tau}(2,4,3|\rho) -
 B_{\eta_2,\eta_3,\eta_4}^{\tau}(3,4,2|\rho) -  B_{\eta_2,\eta_3,\eta_4}^{\tau}(3,2,4|\rho) -
 B_{\eta_2,\eta_3,\eta_4}^{\tau}(4,2,3|\rho) \big] \notag
\end{align}
of the relation (\ref{mix.4}) between $V_2(1,2,3,4|\tau)$ and permutations of the $V(1,2,3,4|\tau)$
functions is claimed to hold for all of $\{s_{12},s_{13},s_{23},s_{14},s_{24},s_{34}\}$ independent.
The coefficient of $\eta_{234}^{-1}\eta_{34}^{-1} \eta_4^{-1}$ in (\ref{genhet}) with $\rho=2,3,4$ then implies
\beq
J_{\rm het}^\tau={\rm SV} \, B_{\rm het}^\tau
\label{expls.45}
\eeq
and explains the relations between the $\alpha'$-expansions (\ref{expls.42}) and (\ref{expls.44})
observed in \cite{Gerken:2018jrq} in the on-shell limit $k_j^2=0$. In particular, the prescription
(\ref{eesv.1}) for the single-valued map of the iterated-Eisenstein-integral representation of
$ B_{\rm het}^\tau$ produces the complete $q,\bar q$-expansion of the MGFs in (\ref{expls.42}),
whereas certain antiholomorphic contributions could not be reproduced by esv in \cite{Gerken:2018jrq}.

By applying (\ref{expls.45}) at the level of the $\alpha'$-expansions (\ref{expls.42}) and (\ref{expls.44}),
one can infer
\begin{align}
 \frac{   \pi\overline \nabla {\rm E}_2 }{y^2}  &= \frac{8}{3} \, {\rm SV} \, \omega(0,1,0,0|{-}\tfrac{1}{\tau})
\notag \\
 \frac{  \pi\overline \nabla {\rm E}_3 }{y^2}  &= 4 \, {\rm SV} \, \big[ \omega(0,1,0,1,0|{-}\tfrac{1}{\tau}) + \omega(0,1,1,0,0| {-}\tfrac{1}{\tau}) \big] \label{expls.46} \\
 \frac{  \pi\overline \nabla {\rm E}_{2,2} }{y^2}  &= {\rm SV} \, \Big[
\frac{ 8}{5} \omega(0, 0, 0, 0, 0, 3|{-}\tfrac{1}{\tau})
 + \frac{ 2}{5} \omega(0, 0, 0, 3|{-}\tfrac{1}{\tau})
      \notag \\
  &\ \ \ \ \ - \frac{ 11}{75} \omega(0, 3|{-}\tfrac{1}{\tau})
 - 8 \omega(0, 0, 0, 0, 1, 2|{-}\tfrac{1}{\tau})  + \frac{ \zeta_3}{3}\Big] \, .
 \notag
\end{align}
Moreover, higher orders in the $\eta_j$-expansion of (\ref{genhet})
yield infinite families of relations between the $\alpha'$-expansions of open- and closed-string integrals over
additional factors $f^{(a)}_{1i} f^{(b)}_{ij} f^{(c)}_{jk}$.

\subsection{Imaginary cusp forms and double zetas}
\label{sec:5.6}

We shall finally exemplify the appearance of cuspidal MGFs from
single-valued open-string integrals whose Laurent polynomial at the order of $q^0 \bar q^0$
vanishes. A systematic study of imaginary cusp forms among the two-loop MGFs can
be found in \cite{DHoker:2019txf}, also see \cite{Gerken:2020aju} for examples of real cusp forms. The simplest
imaginary cusp forms occur among the lattice sums (\ref{newdef.3}) at modular weights $(5,5)$
whose basis can be chosen\footnote{The choice of basis in \cite{Gerken:2020yii} is tailored to delay
the appearance of holomorphic Eisenstein to higher Cauchy-Riemann derivatives as far as possible. That is why
the real MGFs $- \frac{21}{4}\mathrm{E}_{2,3}- \frac{1}{2} \zeta_{3}\mathrm{E}_{2}$ have been
added to the imaginary cusp forms $ {\rm B}_{2,3}+\frac{1}{2}\left(\frac{\Im\tau}{\pi}\right)^5(\cform{0&2&3\\3&0&2} - \overline{\cform{0&2&3\\3&0&2}} \, )$ in (\ref{cusps.1}).} to include \cite{Gerken:2020yii}
\begin{align}
  {\rm B}_{2,3}&=\left(\frac{\Im\tau}{\pi}\right)^5 ( \cform{0&1&2&2\\1&1&0&3} -  \overline{\cform{0&1&2&2\\1&1&0&3}} \, )
  +\frac{(\nabla\mathrm{E}_{2})\overline{\nabla}\mathrm{E}_{3} -(\overline{\nabla}\mathrm{E}_{2})\nabla\mathrm{E}_{3}}{6(\Im \tau)^2} \notag \\
{\rm B}'_{2,3}&= {\rm B}_{2,3}+\frac{1}{2}\left(\frac{\Im\tau}{\pi}\right)^5(\cform{0&2&3\\3&0&2} - \overline{\cform{0&2&3\\3&0&2}} \, )- \frac{21}{4}\mathrm{E}_{2,3}- \frac{1}{2} \zeta_{3}\mathrm{E}_{2}\,.
\label{cusps.1}
\end{align}
The $\beta^{\rm sv}$-representations involve
double-integrals over ${\rm G}_4 {\rm G}_6$ \cite{Gerken:2020yii},
\begin{align}
{\rm B}_{2,3}  &=
 450 \betasv{2&1\\4&6}- 450 \betasv{3&0\\6&4}
 +270 \betasv{2&1\\6&4} -270 \betasv{1&2\\4&6}    \notag\\
 &\ \ \ \ -
 3 \zeta_3\betasv{1\\4} - 300 \zeta_3 \betasv{1\\6}
 + \frac{ 45\zeta_3 \betasv{2\\6} }{y}
 + \frac{ 45 \zeta_5 \betasv{0\\4} }{y} - \frac{27\zeta_5 \betasv{1\\4}}{ 4 y^2}
 - \frac{13 \zeta_5}{120} \,,
\notag \\
{\rm B}'_{2,3} &= 1260 \betasv{2&1\\4&6} - 840 \zeta_3\betasv{1\\6}
+ \frac{7 \zeta_5}{240}  - \frac{ \zeta_3^2}{2 y}
- \frac{147 \zeta_7}{64 y^2} + \frac{21 \zeta_3 \zeta_5}{ 8 y^3}  \,,
\label{cusps.2}
\end{align}
and the associated integration constants $\overline{\alpha[\ldots]}$ can be found in the reference and in
an ancillary file within the arXiv submission of this article.
Both ${\rm B}_{2,3} ,{\rm B}_{2,3}'$ and their Cauchy--Riemann derivatives drop out from
$J^\tau_{\eta_2}$ and $Y^\tau_{\eta_2}$ at two points.
At three points, one can identify their derivatives as single-valued eMZVs,
\begin{align}
\pi \nabla {\rm B}_{2,3} &=  {\rm SV} \, \big[  {-}\tfrac{1}{2} \omega(0, 0, 2, 2, 2|{-}\tfrac{1}{\tau})  - 2 \omega(0, 0, 0, 1, 5|{-}\tfrac{1}{\tau})  -
\tfrac{ 3}{8} \omega(0, 0, 0, 2, 4|{-}\tfrac{1}{\tau}) \notag \\
&\ \ \ \  + \tfrac{ \zeta_3 }{8} \omega(0, 3|{-}\tfrac{1}{\tau})  +
\tfrac{ 17 }{48 }\omega(0, 3|{-}\tfrac{1}{\tau}) ^2 +\tfrac{ 7}{8 } \omega(0, 0, 4|{-}\tfrac{1}{\tau})  \omega(0, 0, 2|{-}\tfrac{1}{\tau})  \notag \\
&\ \ \ \ -
\tfrac{ 7}{2 } \omega(0, 0, 0, 3|{-}\tfrac{1}{\tau})  \omega(0, 3|{-}\tfrac{1}{\tau})  +\tfrac{ 137}{16 }\omega(0, 0, 0, 0, 6|{-}\tfrac{1}{\tau})  -
 \tfrac{15}{32} \omega(0, 0, 6|{-}\tfrac{1}{\tau})  \big] \notag
\\
\pi \nabla {\rm B}_{2,3}' &=  {\rm SV} \, \big[ {-}\tfrac{1}{2} \omega(0, 0, 2, 2, 2|{-}\tfrac{1}{\tau})  - 11 \omega(0, 0, 0, 1, 5|{-}\tfrac{1}{\tau})  +
\tfrac{ 295}{8 }\omega(0, 0, 0, 0, 6|{-}\tfrac{1}{\tau}) \notag \\
&\ \ \ \ - \tfrac{25}{16} \omega(0, 0, 6|{-}\tfrac{1}{\tau})  -
 11 \omega(0, 0, 0, 3|{-}\tfrac{1}{\tau})  \omega(0, 3|{-}\tfrac{1}{\tau})  + \tfrac{ 11}{12 } \omega(0, 3|{-}\tfrac{1}{\tau}) ^2  \label{cusps.3}\\
 &\ \ \ \ -
 \tfrac{1}{4 }\omega(0, 0, 4|{-}\tfrac{1}{\tau})  \omega(0, 0, 2|{-}\tfrac{1}{\tau})   \big]
\notag \\
 (\pi \nabla)^2 {\rm B}_{2,3}' &=  {\rm SV} \, \big[
 {-}\tfrac{189}{80} \omega(0, 0, 2, 5|{-}\tfrac{1}{\tau})  +\tfrac{ 63}{160} \omega(0, 0, 4, 3|{-}\tfrac{1}{\tau})  +
 \tfrac{603}{40} \omega(0, 0, 0, 7|{-}\tfrac{1}{\tau}) \notag \\
 &\ \ \ \  - \tfrac{699}{320} \omega(0, 7|{-}\tfrac{1}{\tau})  -
 \tfrac{1323}{160}  \omega(0, 3|{-}\tfrac{1}{\tau})  \omega(0, 0, 4|{-}\tfrac{1}{\tau})  \big]
 \notag
 \\
 (\pi \nabla)^3 {\rm B}_{2,3}' &=    {\rm SV} \, \big[
 {-}\tfrac{63}{40} \omega(0, 3, 5|{-}\tfrac{1}{\tau})  + \tfrac{ 567}{64} \omega(0, 0, 8|{-}\tfrac{1}{\tau})  -
\tfrac{ 63}{8 }\omega(0, 3|{-}\tfrac{1}{\tau})  \omega(0, 5|{-}\tfrac{1}{\tau})  \big]
\notag
\end{align}
by inspecting the contributions of $f^{(3)}_{12} f^{(3)}_{23}$ or $f^{(4)}_{12}$ to $J^{\tau}_{0,\eta_2,\eta_3}$
and $f^{(3)}_{12}$ to $J^{\tau}_{1,\eta_2,\eta_3}$. The appearance of the undifferentiated
${\rm B}_{2,3}$ and ${\rm B}_{2,3}'$ is relegated to the $J^\tau_{\vec{\eta}}$-series at four points
(or the $Y^\tau_{\vec{\eta}}$-series at three points \cite{Gerken:2020yii}), and comparison with the
B-cycle integrals yields
\begin{align}
{\rm B}_{2,3} &= {\rm SV} \, \big[  \tfrac{143}{20} \omega(0, 0, 0, 0, 0, 5|{-}\tfrac{1}{\tau}) - \tfrac{ 11}{2} \omega(0, 0, 0, 0, 1, 4|{-}\tfrac{1}{\tau}) +
 2 \omega(0, 0, 0, 0, 2, 3|{-}\tfrac{1}{\tau}) \notag \\
 & \! \!  - 2 \omega(0, 0, 0, 1, 2, 2|{-}\tfrac{1}{\tau})  - \tfrac{ 91}{40 } \omega(0, 0, 0, 5|{-}\tfrac{1}{\tau})
  -\tfrac{ 449}{7200} \omega(0, 5|{-}\tfrac{1}{\tau})
  - \tfrac{ 5}{12 } \omega(0, 3|{-}\tfrac{1}{\tau}) \omega(0, 0, 2|{-}\tfrac{1}{\tau})
    \notag \\
 & \! \!   + \tfrac{ 5}{2 } \omega(0, 0, 2|{-}\tfrac{1}{\tau}) \omega(0, 0, 0, 3|{-}\tfrac{1}{\tau})
   + \tfrac{ 23}{12 } \omega(0, 0, 2, 3|{-}\tfrac{1}{\tau}) - \tfrac{ 15}{2 } \omega(0, 3|{-}\tfrac{1}{\tau}) \omega(0, 0, 0, 0, 2|{-}\tfrac{1}{\tau}) \notag\\
 & \! \! +   \tfrac{\zeta_3}{4}  \omega(0, 0, 2|{-}\tfrac{1}{\tau})  - \tfrac{ 43  \zeta_5}{480} \big]
 \label{cusps.4} \\
{\rm B}'_{2,3} &= {\rm SV} \, \big[  \tfrac{463}{10} \omega(0, 0, 0, 0, 0, 5|{-}\tfrac{1}{\tau}) - 22 \omega(0, 0, 0, 0, 1, 4|{-}\tfrac{1}{\tau}) +
 5 \omega(0, 0, 0, 0, 2, 3|{-}\tfrac{1}{\tau})  \notag \\
 & \! \! - 2 \omega(0, 0, 0, 1, 2, 2|{-}\tfrac{1}{\tau})  - \tfrac{ 121}{20 } \omega(0, 0, 0, 5|{-}\tfrac{1}{\tau})  -\tfrac{1069}{3600} \omega(0, 5|{-}\tfrac{1}{\tau})
  - \tfrac{ 1}{6 } \omega(0, 3|{-}\tfrac{1}{\tau}) \omega(0, 0, 2|{-}\tfrac{1}{\tau})
  \notag \\
 & \! \!
  +  \omega(0, 0, 2|{-}\tfrac{1}{\tau}) \omega(0, 0, 0, 3|{-}\tfrac{1}{\tau})
+  \tfrac{25}{6 }\omega(0, 0, 2, 3|{-}\tfrac{1}{\tau}) - 24 \omega(0, 3|{-}\tfrac{1}{\tau}) \omega(0, 0, 0, 0, 2|{-}\tfrac{1}{\tau})  - \tfrac{ 11\zeta_5 }{96} \big] \, . \notag
\end{align}
The open-string counterparts of ${\rm B}_{2,3},{\rm B}_{2,3}'$ and their Cauchy--Riemann derivatives
involve the simplest combinations of B-cycle eMZVs with an irreducible $\zeta_{3,5}$ in their Laurent polynomials:
The methods of \cite{Enriquez:Emzv} (also see appendix B of \cite{Broedel:2018izr}) yield the following
examples of $\tau \rightarrow i\infty$ degenerations in (\ref{cusps.3}) and (\ref{cusps.4}),
\begin{align}
{\rm B}_{2,3} \, \big|_{\rm LP}&= {\rm SV} \, \Big[
\frac{ i T^3 \zeta_{2}}{5040} + \frac{29 i T \zeta_{4} }{630}  + \frac{\zeta_3 \zeta_{2}}{8}  - \frac{8153 i \zeta_{6}}{2880 T}
- \frac{ 9 \zeta_5 \zeta_{2}}{ 8 T^2}   - \frac{ 73 \zeta_3 \zeta_{4}}{16 T^2}
+ \frac{1837 i \zeta_{8}}{240 T^3} \notag \\
&\ \ \ \ \ \
+ \frac{ 3 i}{10 T^3} (2 \zeta_{3,5}  + 5 \zeta_3 \zeta_5)  - \frac{39 \zeta_5 \zeta_{4}}{8 T^4}
+ \frac{ 45 \zeta_3 \zeta_{6}}{8 T^4}
  + \frac{ 33 i \zeta_{10}}{20 T^5}  \Big]\notag \\
  &=0
\label{cusps.5}  \\
 (\pi \nabla)^3 {\rm B}_{2,3}' \, \big|_{\rm LP}&={\rm SV} \, \Big[
 \frac{T^8}{17280} + \frac{ i T^5 \zeta_3}{120}  + \frac{ T^4 \zeta_{4} }{40} - \frac{7 T^2 \zeta_{6} }{80}
  - \frac{8211 \zeta_{8}}{640}  \notag \\
  &\ \ \ \ \ \ + \frac{ 63 \zeta_{3,5}}{40}
  + \frac{ 189 i \zeta_5 \zeta_{4}}{8 T} + \frac{ 63 i \zeta_3 \zeta_{6}}{4 T} + \frac{ 14553 \zeta_{10}}{160 T^2}  \Big]\notag \\
  &= \frac{ 2 y^8}{135} - \frac{8 y^5 \zeta_3}{15} - \frac{ 63 \zeta_3 \zeta_5}{4}\, .
  \notag
\end{align}
One can see from the order of $T^{-3}$ or $y^{-3}$ that the cuspidal nature of ${\rm B}_{2,3}$ hinges on
the depth-two result ${\rm sv} \, \zeta_{3,5} = -10 \zeta_3 \zeta_5$.
The non-vanishing Laurent polynomial of $(\pi \nabla)^3 {\rm B}_{2,3}' $ is due to
the real MGFs $- \frac{21}{4}\mathrm{E}_{2,3}- \frac{1}{2} \zeta_{3}\mathrm{E}_{2}$ in (\ref{cusps.1}).

Note that the simplest instances of $\zeta_{3,7}$ and $\zeta_{3,5,3}$ arise in the Laurent polynomials of
B-cycle eMZVs with MGFs (\ref{newdef.3}) of weights $\sum_{j=1}^r(a_j+b_j)=12$ and $14$ in their SV mage.
The appearance of $\zeta_{3,5,3}$ in modular graph functions and eMZVs can be found in \cite{Zerbini:2015rss} and \cite{Broedel:2018izr}, respectively. While $\zeta_{3,7}$ drops out from MGFs under the single-valued
map, it enters for instance the $T^0$-order of the Laurent polynomial of $\omega(0,3,7|{-}\tfrac{1}{\tau})$ whose
SV image contributes to the quantity $(\pi \nabla)^4 {\rm B}'_{2,4}$ in section 9.2 of \cite{Gerken:2020aju}
\begin{align}
&\omega(0,3,7|{-}\tfrac{1}{\tau}) \, \big|_{\rm LP} =
-\frac{T^{10}}{1261260} +
 \frac{ 2 i T^5 \zeta_5}{315} +
\frac{ 2 T^4 \zeta_6 }{63}  + \frac{7 i T^3 \zeta_7}{45}   + \frac{7 T^2 \zeta_8}{6}   \\
&\ \ \ \
- \zeta_{3,7} - 14 \zeta_3 \zeta_7  - 6 \zeta_5^2 + \frac{27 \zeta_{10}}{2} + \frac{84 i \zeta_{11}}{T} + \frac{30 i \zeta_5 \zeta_6}{T}
+ \frac{ 84 i \zeta_3 \zeta_8}{T}+ \frac{1353 \zeta_{12}}{2 T^2} \, .   \notag
\end{align}
One can eventually find all $\mathbb Q$-independent MZVs\footnote{See \cite{Blumlein:2009cf} for a computer
implementation of $\mathbb Q$-relations among MZVs.} in the Laurent polynomials of
B-cycle eMZVs. This follows from both the degeneration limits of the elliptic
KZB associator \cite{Enriquez:Emzv} and from the fact that any MZV is expressible via
$\mathbb Q[2\pi i]$-linear combinations of multiple modular values \cite{Saad:2020mzv}.

Note that the Laurent polynomials of all B-cycle eMZVs with $\text{length} + \text{weight} \leq 16$ obtained from a
{\tt FORM} implementation \cite{Vermaseren:2000nd} of the methods of
\cite{Enriquez:Emzv, Broedel:2018izr} are available for download from \cite{WWWbcyc}.


\subsection{Single-valued map of individual eMZVs}
\label{sec:5.8}

\begin{figure}
\begin{center}
\begin{tikzpicture}[line width=0.30mm]
\draw[dashed](0,0) circle (2cm);
\draw(-2,0)node{$\bullet$}node[left]{$1$};
\draw(-1,1.73)node{$\bullet$}node[above]{$2$};
\draw(1,1.73)node{$\bullet$}node[above]{$3$};
\draw(-1,-1.73)node{$\bullet$}node[below]{$r{+}1$};
\draw(1,-1.73)node{$\bullet$}node[below]{$r$};
\draw(-2,0) -- (-1,1.73);
\draw(-2,0) -- (1,1.73);
\draw(-2,0) -- (-1,-1.73);
\draw(-2,0) -- (1,-1.73);
\draw(1,1)node[rotate=45]{$\vdots$};
\draw(1.5,0)node{$\vdots$};
\draw(1,-1)node[rotate=-45]{$\vdots$};
\end{tikzpicture}
\end{center}
\caption{\textit{Graphical representation of the integrand of single-valued eMZVs (\ref{nsvemzv.1}):
Solid lines represent factors of $ f^{(w)}_{jk}$ while dashed lines stand for the $\overline{f^{(w)}_{12}},
\overline{f^{(w)}_{23}},\ldots ,\overline{f^{(w)}_{r,r+1}},\overline{f^{(w)}_{r+1,1}}$ that are compatible with the cyclic arrangement of the arguments of $  \overline{ V(1,2,\ldots,r{+}1|\tau) } $.}}
\label{figsvemzv}
\end{figure}
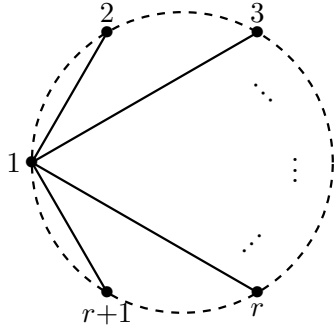

While the above combinations of single-valued eMZVs were tailored to obtaining a single MGF
in the bases of \cite{Broedel:2018izr, Gerken:2020yii}, we shall now give a closed formula
for the single-valued map of individual eMZVs. The integrands of convergent A-cycle eMZVs
(\ref{newdef.1}) with length $r$ and $n_1,n_r\neq 1$ arise at the $s_{ij}^0$-order of the series
$Z^\tau_{\vec{\eta}}$ at $r{+}1$ points. After modular $S$ transformation, one can obtain any convergent
$\omega(n_1,\ldots,n_r|{-}\tfrac{1}{\tau})$ by isolating suitable $\eta_j$-orders in
the $s_{ij} \rightarrow 0$ limit of
$\sum_{\rho \in S_r}B^\tau_{\vec{\eta}}(1,2,\ldots,r{+}1|1,\rho(2,3,\ldots,r{+}1))$,
where the permutation sum over the orderings of the integrands (\ref{looprev.5}) yields the
integrands $f^{(n_1)}_{21}f^{(n_2)}_{31} \ldots f^{(n_r)}_{r+1,1}$ in the definition
(\ref{newdef.1}) of eMZVs. Hence, our proposal (\ref{newBB.20}) implies that
${\rm SV} \, \omega(n_1,\ldots,n_r|{-}\tfrac{1}{\tau})$
occurs at the corresponding orders of $s_{ij}$ and $\eta_j$ in
the series $J^\tau_{\vec{\eta}}$, so their definition (\ref{newBB.11}) leads to ($n_1,n_r \neq 1$ and $z_1=0$)
\begin{align}
\label{nsvemzv.1}
{\rm SV} \, \omega(n_1,n_2,\ldots,n_r|{-}\tfrac{1}{\tau}) &=(\tau{-}\bar \tau)^{n_1+\ldots+n_r} \! \!  \int_{\mathfrak T^{r}} \! \! \Big( \prod_{j=2}^{r+1} \frac{ \dd^2 z_j }{\Im \tau}\Big)
 \overline{ V(1,2,\ldots,r{+}1|\tau) }
 f^{(n_1)}_{21}f^{(n_2)}_{31} \ldots f^{(n_r)}_{r+1,1}\, .
\end{align}
By the techniques in \cite{Gerken:2018jrq}, the integral on the right-hand side can be straightforwardly performed in terms of lattice sums over $p=m\tau {+}n \in \Lambda'$ in (\ref{newdef.2}): After expressing the $\overline{V_w}$-functions in terms of $\overline{ f^{(w)}_{jk}}$ via (\ref{shortv}) and inserting the double Fourier expansions
\begin{align}
\label{nsvemzv.2}
f^{(w)}_{jk}&= (-1)^{w-1} \sum_{p \in \Lambda'} \frac{ e^{2\pi i \langle p,z_{jk} \rangle }}{p^w}
\, , \ \ \ \ \ \ z_{jk}= u_{jk}\tau + v_{jk}  \, , \ \ \ \ \ \
 \langle p,z_{jk} \rangle = mv_{jk} - n u_{jk}
\end{align}
the integrals $\int_{\mathfrak T}\frac{ \dd^2 z_j }{\Im \tau} = \int^1_0 \dd u_j \int^1_0 \dd v_j$ lead to
momentum-conserving delta functions as seen in the dihedral MGFs (\ref{newdef.3}). When visualizing
each factor of $ f^{(w)}_{jk}$ and $\overline{ f^{(w)}_{jk}}$ in the integrand of (\ref{nsvemzv.1}) through
an edge between vertices $j$ and $k$, contributions from one-particle reducible graphs integrate to zero.
There are at most $r{-}1$ factors of $\overline{ f^{(w)}_{jk}}$ from the $\overline{V_{w \leq r-1} (1,2,\ldots,r{+}1)}$
in (\ref{nsvemzv.1}), and the admissible pairs $(j,k)$ are visualized via dashed lines in figure \ref{figsvemzv}.

One can anticipate from the example (reproducing (\ref{expls.6}))
\begin{align}
{\rm SV} \, \omega(0,0,2|{-}\tfrac{1}{\tau}) &=(\tau{-}\bar \tau)^{2} \! \!  \int_{\mathfrak T^{3}} \! \! \Big( \prod_{j=2}^{4} \frac{ \dd^2 z_j }{\Im \tau}\Big)  \bigg\{ \frac{1}{6} - \frac{  \overline{ V_1(1,2,3,4|\tau) } }{4\pi i}
+ \frac{ \overline{ V_2(1,2,3,4|\tau) } }{(2\pi i)^2}  \bigg\}
f^{(2)}_{41} \notag \\
&= \bigg( \frac{ \Im \tau}{\pi} \bigg)^2
\! \!  \int_{\mathfrak T^{3}} \! \! \Big( \prod_{j=2}^{4} \frac{ \dd^2 z_j }{\Im \tau}\Big) f^{(2)}_{41} \overline{ f^{(2)}_{41} }
=  \bigg( \frac{ \Im \tau}{\pi} \bigg)^2 \sum_{p \in \Lambda'} \frac{1}{|p|^4} = {\rm E}_2(\tau)
 \label{nsvemzv.3}
\end{align}
that only small subset of the terms in $ \overline{ V(1,2,\ldots,r{+}1|\tau) } $ contribute to generic
single-valued eMZVs -- the right-hand side of (\ref{nsvemzv.3}) entirely stems from
$\overline{ V_2(1,2,3,4|\tau)} \rightarrow \overline{ f^{(2)}_{41} }$. Apart from the restriction
to one-particle irreducible graphs, only those $ \overline{ V_w(1,2,\ldots,r{+}1|\tau) } $ with parity
$(-1)^w = (-1)^{n_1+n_2+\ldots +n_r}$ contribute since lattice sums with odd overall
modular weight vanish.

Note that the torus integrals in the expression (\ref{nsvemzv.1}) for single-valued eMZVs converge
whenever the eMZVs themselves do: The convergence criterion $n_1,n_r \neq 1$ rules out any double
pole $|z_{jk}|^{-2}$ in the integrand (and kinematic poles $s_{jk}^{-1}$ in the Koba--Nielsen integral)
since the only overlap between the solid and dashed lines
in figure \ref{figsvemzv} occurs via $f^{(n_1)}_{12}\overline{ f^{(w)}_{12}}$ and $f^{(n_r)}_{1,r+1}\overline{ f^{(w)}_{1,r+1}}$.

Based on the conjectural relation (\ref{newBB.15}) between the Laurent polynomials,
one can use (\ref{nsvemzv.1}) to infer the asymptotics of the MGFs on the right-hand side
by importing the Laurent polynomials of the B-cycle eMZVs from \cite{WWWbcyc} and applying
the single-valued map. Moreover, any relation among eMZVs induces a relation among the MGFs
through the lattice-sum representation of their single-valued images. Hence, the database of
MGF relations \cite{Gerken:2020aju} can be complemented by applying (\ref{nsvemzv.1})
to the eMZV relations on the website \cite{WWWe}. The lattice sums contributing to
${\rm SV} \, \omega(n_1,\ldots,n_r|{-}\tfrac{1}{\tau}) $ have (anti-)holomorphic modular weights $(w,\bar w)$ subject to $w=n_1+\ldots+n_r$ and $\bar w \leq r{-}1$. Accordingly, the Laurent polynomials of B-cycle eMZVs of ${\rm length}+{\rm weight} \leq 16$ \cite{WWWbcyc} give access to those of various combinations of MGFs with $w+\bar w\leq 15$.

In the remainder of this section, we will study the single-valued eMZVs in (\ref{nsvemzv.1})
at fixed length $r$ and comment on consistency checks with the properties of their SV preimages \cite{Broedel:2015hia}. The condition $n_1,n_r\neq 1$ for convergence is taken to hold throughout.
At length $r=1$, for instance, $\omega(2k|{-}\frac{1}{\tau})=-2 \zeta_{2k}$ and $\omega(2k{-}1|{-}\frac{1}{\tau})=0$
with $k\geq 1$ are annihilated by SV, in lines with the vanishing of the torus integral over a single $f^{(w)}_{12}$
at $w \neq 0$.

\subsubsection{Length-two examples $\omega(n_1,n_2)$}
\label{sec:5.8.1}

Single-valued eMZVs of length $r=2$ take the form (\ref{nsvemzv.1})
\begin{align}
\label{nsvemzv.4}
{\rm SV} \, \omega(n_1,n_2|{-}\tfrac{1}{\tau}) &=(\tau{-}\bar \tau)^{n_1+n_2}   \int_{\mathfrak T^{2}}  \frac{ \dd^2 z_2 }{\Im \tau} \frac{ \dd^2 z_3 }{\Im \tau}
 f^{(n_1)}_{21}f^{(n_2)}_{31} \times \left\{ \begin{array}{cl} \displaystyle
\Bigg. \frac{1}{2}  &: \ n_1{+}n_2 \ {\rm even} \\ \displaystyle
  -  \frac{ \overline{ V_1(1,2,3|\tau) } }{2\pi i}  &: \ n_1{+}n_2 \ {\rm odd}
\end{array} \right. \, ,
\end{align}
where the distinction between even and odd weight $n_1{+}n_2$ stems from the vanishing of
lattice sums with odd modular weight. At even $n_1{+}n_2 >0$, the integrand of
(\ref{nsvemzv.4}) is proportional to $f^{(n_1)}_{21}f^{(n_2)}_{31}$
which corresponds to a one-particle reducible graph with
\begin{align}
\label{nsvemzv.5}
{\rm SV} \, \omega(n_1,n_2|{-}\tfrac{1}{\tau}) \, \big|_{n_1+n_2 >0 \ \rm{even}}&=0 \, ,
\end{align}
in agreement with $\omega(2k_1,2k_2|{-}\frac{1}{\tau})=2 \zeta_{2k_1} \zeta_{2k_2}$ and $\omega(2k_1{-}1,2k_2{-}1|{-}\frac{1}{\tau})=0$ \cite{Broedel:2015hia}. For odd weight $n_1+n_2$, we keep $n_2 \neq 0$ and distinguish the two cases $n_1=0$ and $n_1 \neq 0$,
where the only contributions of $\overline{ V_1(1,2,3|\tau)}$ to the integral (\ref{nsvemzv.4}) stem from
$\overline{f^{(1)}_{31}}$ and $\overline{f^{(1)}_{23}}$, respectively:
\begin{align}
{\rm SV} \, \omega(0,n_2|{-}\tfrac{1}{\tau}) \, \big|_{n_2 \ {\rm odd}}&={-} \frac{(\tau{-}\bar \tau)^{n_2}}{2\pi i} \int_{\mathfrak T} \frac{ \dd^2 z_3}{\Im \tau}  f^{(n_2)}_{31} \overline{f^{(1)}_{31}} = \frac{i}{2\pi}(\tau{-}\bar \tau)^{n_2} \cform{ n_2 &0 \\  1 &0 }  \notag \\
{\rm SV} \, \omega(n_1,n_2|{-}\tfrac{1}{\tau})\, \big|^{n_1,n_2 \, \neq \, 0}_{n_1+n_2 \ {\rm odd}} &={-} \frac{(\tau{-}\bar \tau)^{n_1+n_2}}{2\pi i} \int_{\mathfrak T} \frac{ \dd^2 z_2}{\Im \tau}  \frac{ \dd^2 z_3}{\Im \tau} f^{(n_1)}_{21} f^{(n_2)}_{31} \overline{f^{(1)}_{23}}  \label{nsvemzv.6}\\
&=  (-1)^{n_1} \frac{ i }{2\pi}  (\tau{-}\bar \tau)^{n_1+n_2}\cform{n_1+ n_2 &0 \\  1 &0 }  \notag
\end{align}
The resulting relation ${\rm SV} \, \omega(n_1,n_2|{-}\tfrac{1}{\tau})
= (-1)^{n_1}{\rm SV} \, \omega(0,n_1{+}n_2|{-}\tfrac{1}{\tau})$ is consistent with (2.33) of \cite{Broedel:2015hia} after discarding any ${\rm SV} \, \zeta_{2k}$ with $k\geq 1$ from the equation of the reference. In combination with the vanishing of ${\rm SV} \, \omega(n_1,n_2)\, \big|_{n_1+n_2 \ {\rm even}} $, we conclude that
all single-valued eMZVs of length two do not exceed one-loop MGFs, in lines with
$\omega(0,2k{+}1)$ being iterated Eisenstein integrals of depth one.

\subsubsection{Length-three examples $\omega(n_1,n_2,n_3)$}
\label{sec:5.8.2}

Single-valued eMZVs of length three can be written as
\begin{align}
\label{nsvemzv.11}
{\rm SV} \, \omega(n_1,n_2,n_3|{-}\tfrac{1}{\tau}) &=(\tau{-}\bar \tau)^{n_1+n_2+n_3} \! \!  \int_{\mathfrak T^{3}} \! \! \Big( \prod_{j=2}^{4} \frac{ \dd^2 z_j }{\Im \tau}\Big)
 f^{(n_1)}_{21}f^{(n_2)}_{31}  f^{(n_3)}_{41}  \notag \\
 & \ \ \times \left\{ \begin{array}{cl} \displaystyle
\Bigg. \frac{1}{6} +  \frac{ \overline{ V_2(1,2,3,4|\tau) } }{(2\pi i)^2}  &: \ n_1{+}n_2{+}n_3 \ {\rm even} \\ \displaystyle
  -  \frac{ \overline{ V_1(1,2,3,4|\tau) } }{4\pi i}  &: \ n_1{+}n_2{+}n_3 \ {\rm odd}
\end{array} \right.
\end{align}
after discarding lattice sums of odd modular weight. For ${\rm SV} \, \omega(0,0,2k{+}2|{-}\tfrac{1}{\tau})$,
this results in the one-loop MGFs in the second line of (\ref{nicesum.1}). Starting from weight 8, the
bases of $\omega(n_1,n_2,n_3)$ require representatives with two non-zero entries such as $\omega(0,3,5),\omega(0,3,7)$ \cite{Broedel:2015hia}, and their single-valued versions correspond to MGFs of depth two as exemplified in section \ref{sec:5.6}. For any combination of three non-vanishing entries, the torus integral
in (\ref{nsvemzv.11}) can be expressed in terms of two-loop MGFs $\cform{a_1&a_2 &a_3 \\  b_1 &b_2 &b_3}$ in (\ref{newdef.3}) and products of one-loop MGFs,
\begin{align}
{\rm SV} \, \omega(n_1,n_2,n_3|{-}\tfrac{1}{\tau})  \, \big|^{n_1,n_2,n_3 \neq 0}_{n_1+n_2+n_3 \ {\rm odd}} &=0
 \notag \\
{\rm SV} \, \omega(n_1,n_2,n_3|{-}\tfrac{1}{\tau})  \, \big|^{n_1,n_2,n_3 \neq 0}_{n_1+n_2+n_3 \ {\rm even}} &= \frac{ (\tau{-}\bar \tau)^{n_1+n_2+n_3} }{(2\pi i)^2} \bigg\{ (-1)^{n_3} \cform{n_1&0 \\  1 &0 } \cform{n_2 +n_3 &0 \\  1 &0 }
\label{nsvemzv.12} \\
& \ \ \ \
+(-1)^{n_1} \cform{n_3&0 \\  1 &0 } \cform{n_1 +n_2 &0 \\  1 &0 } + \cform{n_1&n_2 &n_3 \\  1 &0 &1}
\bigg\} \, , \notag
\end{align}
leading to iterated Eisenstein integrals of depth two.

\subsubsection{Length four and beyond}
\label{sec:5.8.3}

Starting from single-valued eMZVs of length four
\begin{align}
\label{nsvemzv.21}
{\rm SV} \, \omega(n_1,n_2,n_3,n_4|{-}\tfrac{1}{\tau}) &=(\tau{-}\bar \tau)^{n_1+n_2+n_3+n_4} \! \!  \int_{\mathfrak T^{4}} \! \! \Big( \prod_{j=2}^{5} \frac{ \dd^2 z_j }{\Im \tau}\Big)
 f^{(n_1)}_{21}f^{(n_2)}_{31}  f^{(n_3)}_{41} f^{(n_4)}_{51}
\\
 & \ \ \times \left\{ \begin{array}{cl} \displaystyle
\Bigg. \frac{1}{24} +  \frac{ \overline{ V_2(1,2,3,4,5|\tau) } }{ 2 (2\pi i)^2}  &: \ n_1{+}n_2{+}n_3{+}n_4 \ {\rm even} \\ \displaystyle
  -  \frac{ \overline{ V_1(1,2,3,4,5|\tau) } }{12\pi i}
  -  \frac{ \overline{ V_3(1,2,3,4,5|\tau) } }{(2\pi i)^3}  &: \ n_1{+}n_2{+}n_3{+}n_4 \ {\rm odd}
\end{array} \right.  \, ,  \notag
\end{align}
MGFs of different modular weights may mix through the contributions of $\overline{V_1}$ and $\overline{V_3}$
to
\beq
{\rm SV} \,\omega(0,0,0,n_4|{-}\tfrac{1}{\tau}) \, \big|_{n_4 \ {\rm odd}} = - \frac{ (\tau{-}\bar \tau)^n }{12\pi i} \cform{n_4&0 \\  1 &0 } - \frac{ (\tau{-}\bar \tau)^n }{(2\pi i)^3} \cform{n_4&0 \\  3 &0 }
\label{nsvemzv.22}
\eeq
The first term drops out when adding $- \frac{1}{6} {\rm SV} \,\omega(0,n_4|{-}\tfrac{1}{\tau}) \, \big|_{n_4 \ {\rm odd}}$ in (\ref{nsvemzv.6}) and explains the combinations of eMZVs of different length in the third line of (\ref{nicesum.1}). Similarly, the more general combinations (\ref{alldepth}) of single-valued eMZVs of different length
that isolate one-loop MGFs can be understood from the combinations of $\overline{V_w}$ that
contribute to higher-point $\overline{V}$-functions (\ref{newBB.6}).

With multiple non-zero entries in ${\rm SV} \, \omega(n_1,n_2,n_3,n_4|{-}\tfrac{1}{\tau}) $
of odd weight, the lattice sums from integrating $\overline{V_3}$ correspond to MGFs of trihedral topology. Similarly,
single-valued eMZVs at length five introduce four-point MGFs of kite topology introduced in section 4.3 of \cite{Gerken:2020aju}. We hope that their identification with single-valued eMZVs will facilitate the study of MGFs beyond
the dihedral topology and result in efficient methods to determine their Laurent polynomials and relations at arbitrary weight.

\section{Conclusions and outlook}
\label{sec:9}

In this work, we have studied generating series of configuration-space integrals
that arise in open- and closed-string amplitudes at genus one. The differential equations
and $\tau \rightarrow i \infty$ degenerations of these generating series served as a
framework to propose the explicit form of an elliptic single-valued map. Our construction is
based on a tentative genus-one uplift of the Betti--deRham duality between integration
cycles on a disk boundary and antiholomorphic Parke--Taylor integrands which
drives the relation between closed-string and single-valued open-string tree
amplitudes \cite{Schlotterer:2012ny,Stieberger:2013wea,Stieberger:2014hba,Schlotterer:2018abc,Vanhove:2018elu,Brown:2019wna}. These considerations lead us to construct closed-string genus-one integrals
over specific antielliptic functions which are thought of as Betti--deRham dual to
open-string integration cycles in view of their singularities at $z_i {\rightarrow} z_j$
and their degeneration at $\tau {\rightarrow} i\infty$.

Most importantly, the differential equations of the open- and closed-string integrals
under investigation only differ by $\tau^j {\rm G}_k(\tau)$ vs.\ $(\tau{-}\bar \tau)^j {\rm G}_k(\tau)$
in the respective differential operators with holomorphic Eisenstein series ${\rm G}_k$.
Accordingly, we generate the elliptic multiple zeta values and modular graph forms
in their $\alpha'$-expansions via path ordered exponentials with the same polynomial
structures in kinematic invariants and formal expansion variables. The $\tau$-dependent
building blocks are iterated Eisenstein integrals in both cases -- holomorphic
ones with kernels $\tau^j {\rm G}_k(\tau)$ for the open-string integrals and their single-valued
versions involving kernels $(\tau{-}\bar \tau)^j {\rm G}_k(\tau)$ for closed strings.

Our proposal for an elliptic single-valued map is defined through the relation between
the generating series of open- and closed-string integrals.
By their respective $\alpha'$-expansion, we obtain the single-valued map for all iterated Eisenstein integrals occurring in the open-string series. This in turn determines the single-valued map of any convergent
elliptic multiple zeta value in terms of modular graph forms.

This construction hinges on the compatibility of the initial values at $\tau \rightarrow i\infty$
under the single-valued map of multiple zeta values \cite{Schnetz:2013hqa, Brown:2013gia}. We have given evidence for their compatibility
by identifying the key building blocks of genus-zero integrals at the cusp -- appropriate pairs of disk orderings and Parke--Taylor integrands. However, the detailed expressions for the asymptotic expansions
beyond two points in terms of genus-zero integrals is left for future work.
At present, the procedure also relies on the reality properties of a generating series $Y^\tau_{\vec\eta}$
of a more general class of closed-string integrals. Our method does not yet provide a direct construction of single-valued iterated Eisenstein integrals solely from open-string data.

\subsection{Genus-one integrals versus string amplitudes}
\label{sec:9.1}

The results of this work concern infinite families of configuration-space integrals
at genus one, and their application to genus-one string amplitudes requires the following
leftover steps:

For both open and closed strings, it remains to integrate over the modular
parameter $\tau$ of the respective surface. In the closed-string case, $\tau$-integrals over
modular graph forms are typically performed on the basis of their Laplace equations
\cite{DHoker:2015gmr, Basu:2016xrt, Kleinschmidt:2017ege, Basu:2017nhs, Basu:2019idd, DHoker:2019blr} and Poincar\'e-series
representations \cite{Angelantonj:2011br, Angelantonj:2012gw, Ahlen:2018wng, DHoker:2019txf, Dorigoni:2019yoq, DHoker:2019mib, Dorigoni:2020oon}.
The $\tau$-integration of open-string integrals has for instance been discussed in
\cite{Green:1981ya, Angelantonj:2002ct, Lust:2003ky, Hohenegger:2017kqy}, and a general method applicable to
arbitrary depth may be based on the representation of elliptic multiple zeta values in terms of iterated
Eisenstein integrals (including their ``over-integrated'' instances with kernels $\tau^j {\rm G}_k$
at $j >k{-}2$ \cite{Dorigoni:2020oon}) and properties of multiple modular values \cite{Brown:mmv}.
It would be particularly interesting to relate closed-string and single-valued open-string integrals
at genus one {\it after} integration over $\tau$.

For open strings, the $Z^\tau_{\vec{\eta}}$- or $B^\tau_{\vec{\eta}}$-series
are claimed to exhaust all the configuration-space integrands built from $f^{(k)}(z_i{-}z_j,\tau)$
that are inequivalent under Fay identities and integration by parts. Similarly, the $Y^\tau_{\vec{\eta}}$-series
built from double copies of the open-string integrands is expected to contain all torus integrals of this type.
Hence, by the arguments of \cite{Dolan:2007eh, Broedel:2014vla, Gerken:2018jrq},
$B^\tau_{\vec{\eta}}$ and $Y^\tau_{\vec{\eta}}$ should\footnote{It has been shown in \cite{Gerken:2018jrq}
that the integrands of massless genus-one amplitudes in bosonic, heterotic and type-II theories
are expressible in terms of products of $f^{(k)}(z_i{-}z_j,\tau)$ and their $z_i$-derivatives. The
conjectural part is that arbitrary products of $f^{(k)}(z_i{-}z_j,\tau)$ (possibly including derivatives)
are expressible in terms of the $\varphi^\tau_{\vec{\eta}}$ in (\ref{looprev.5}) with their specific chain structure
via repeated use of Fay identities and integration by parts \cite{Mafra:2019ddf, Mafra:2019xms}.}
capture the conformal-field-theory correlators
in the integrands of $n$-point genus-one amplitudes of massless states
(and possibly also of massive states) in bosonic, heterotic and type-II string theories.
In all cases, the component integrals in the $\eta_j$-expansions of the
$Z^\tau_{\vec{\eta}}$-, $B^\tau_{\vec{\eta}}$ or $Y^\tau_{\vec{\eta}}$ series need to be
dressed with kinematic factors that are determined by the correlators and carry the polarization
dependence of the respective string amplitude.

The integrands of $J^\tau_{\vec{\eta}}$ only involve antielliptic combinations of
$\overline{f^{(k)}(z_i{-}z_j,\tau)}$ and omit infinite classes of component integrals of $Y^\tau_{\vec{\eta}}$.
For a given genus-one closed-string amplitude, it is therefore not a priori clear if its correlator is generated
by the integrand of $J^\tau_{\vec{\eta}}$. Still, the correlators for the four- and five-point type-II amplitudes
can be recovered from the subsectors $J^\tau_{w,\vec{\eta}}$ of the $J^\tau_{\vec{\eta}}$-series
at fixed modular weights: The four-point correlator of \cite{Green:1982sw} resides at the $\eta^{-3}$ order
of $J^\tau_{0,\eta_2,\eta_3,\eta_4}$, and the five-point correlators of \cite{Richards:2008jg, Green:2013bza} can be assembled
from the most singular $\eta$-orders of $J^\tau_{w,\eta_2,\eta_3,\eta_4,\eta_5}$ at $w=0,1$.
Similarly, the four- and five-point amplitudes of gluons and gravitons in heterotic string theories can in principle be extracted from the same $J^\tau_{w,\vec{\eta}}$ which also appear in type II, where higher orders in $\eta_j$ are needed to capture the bosonic sectors.
It would be interesting to see if this pattern persists at higher points in supersymmetric amplitudes,
and whether the $J^\tau_{\vec{\eta}}$ are sufficient to generate bosonic-string
amplitudes at low multiplicity.

\subsection{Further directions}
\label{sec:9.2}

This work spawns a variety of further directions and open questions of relevance to both physicists and mathematicians:

The single-valued image of elliptic multiple zeta values is proposed to contain combinations of
holomorphic iterated Eisenstein integrals and their complex conjugates denoted
by $\beta^{\rm sv}$ and constructed from the $\alpha'$-expansion of closed-string
integrals in \cite{Gerken:2020yii}. It would be important to work out their detailed
relation to Brown's earlier construction of single-valued iterated Eisenstein
integrals \cite{Brown:2017qwo, Brown:2017qwo2}. In particular, it remains to relate the MZVs in the antiholomorphic
contributions to $\beta^{\rm sv}$ (fixed from reality properties of Koba--Nielsen integrals
in \cite{Gerken:2020yii}) to the combinations of multiple modular values entering
Brown's construction. This will hopefully bypass the need to use these reality properties as independent input for the construction of $\beta^{\rm sv}$ as done so far.

Several aspects of our construction are based on conjectures with
strong support from a variety of non-trivial examples. As pointed out in the relevant passages in earlier
parts of this work, it would be desirable to find mathematically rigorous proofs that
\begin{itemize}
\item any Koba--Nielsen integral at genus one involving products and derivatives of
Kronecker--Eisenstein coefficients $f^{(k)}(z,\tau)$ can be expanded in the coefficients of
the series $Z^\tau_{\vec{\eta}}$ and~$Y^\tau_{\vec{\eta}}$
\item the matrices $r_{\vec{\eta}}(\epsilon_k)$ and $R_{\vec{\eta}}(\epsilon_k)$ in open- and closed-string
differential equations (\ref{looprev.11}) preserve the commutation relations of Tsunogai's derivations $\epsilon_k$
\item the single-valued images $\beta^{\rm sv}$ of iterated
Eisenstein integrals satisfy shuffle relations, i.e.\ that the
antiholomorphic integration constants $\alpha$ do not introduce any obstructions
\item the coefficients of the $s_{ij}$- and $\eta_j$- expansion of the initial values $\Jhat^{i\infty}_{\vec{\eta}}$
and $\Yhat^{i\infty}_{\vec{\eta}}$ are single-valued multiple zeta values
\end{itemize}

The proposal of the present work concerns single-valued
integration \cite{Schnetz:2013hqa, Brown:2018omk} in the modular parameter $\tau$. An alternative approach
is to recover modular graph forms from single-valued functions of torus punctures \cite{DHoker:2015wxz, Panzertalk}.
In this context, it would be rewarding to find an explicit realization of single-valued integration
in $z$ for elliptic polylogarithms and their complex conjugates, for instance by building
upon the ideas of \cite{Panzertalk} and the depth-one results in \cite{Broedel:2019tlz}.

At genus zero, the identification of sphere integrals as single-valued disk integrals is
equivalent to the Kawai--Lewellen--Tye (KLT) relations between closed-string and
squares of open-string tree-level amplitudes \cite{Kawai:1985xq}. Accordingly, one could
wonder if the combinations of holomorphic and
antiholomorphic iterated Eisenstein integrals in the $\beta^{\rm sv}$ or modular graph forms can arise from products
of open-string type generating functions and their complex conjugates.
If such a genus-one echo of KLT relations exists, then one can expect a close connection to the monodromy
relations among open-string integrals \cite{Tourkine:2016bak, Hohenegger:2017kqy, Ochirov:2017jby, Casali:2020knc} and in particular their study in the light of twisted deRham
theory \cite{Casali:2019ihm}. And it could open up a new perspective on the quest for loop-level KLT relations
to revisit the generating functions of closed-string integrals in the framework of chiral splitting
\cite{DHoker:1988pdl, DHoker:1989cxq}, by performing the $\alpha'$-expansion at the level of the loop integrand.

A particularly burning question concerns a higher-genus realization of single-valued integration
and the associated relations between open- and closed-string amplitudes. A promising first step
could be to identify suitable holomorphic open-string analogues of the modular graph forms
\cite{DHoker:2017pvk,DHoker:2018mys} and modular graph tensors \cite{DHoker:2020uid} at higher genus. More generally, the simplified correlators of maximally supersymmetric genus-two amplitudes
at four points \cite{DHoker:2005vch, Berkovits:2005df} and five points \cite{DHoker:2020prr, DHoker:2020tcq}
provide valuable showcases of Koba--Nielsen integrals relevant to open- and closed-string scattering.
Furthermore, the construction of the generating series in this work was inspired by extended
families of genus-one Koba--Nielsen integrals that arise from heterotic or bosonic strings
\cite{Gerken:2018jrq}. Hence, the genus-two correlators of the heterotic string and the combinations
of theta functions studied in \cite{Tsuchiya:2012nf, Tsuchiya:2017joo} could give
important clues on higher-genus versions of the elliptic functions and
generating series in this work.


\medskip
\subsection*{Acknowledgments}
We would like to thank Johannes Broedel, Eric D'Hoker, Daniele Dorigoni, Cl\'ement Dupont, Andr\'e Kaderli, Erik Panzer, Oliver Schnetz,
Federico Zerbini and in particular Nils Matthes for combinations of
valuable discussions and collaboration on related topics. Moreover, we are grateful to Johannes Broedel, Eric D'Hoker, Andr\'e Kaderli and an anonymous referee for valuable
comments on a draft. OS thanks AEI Potsdam and AK \& JG thank Uppsala University for
hospitality during various stages of this work. JG, CRM and OS are grateful to the
organizers of the program ``Modular forms, periods and scattering amplitudes'' at the ETH Institute
for Theoretical Studies for providing a stimulating atmosphere and financial support during early
stages of this project. JG was supported by the International Max
Planck Research School for Mathematical and Physical Aspects of Gravitation, Cosmology
and Quantum Field Theory during most stages of this work. CRM is supported by a University Research Fellowship from the Royal Society. OS and BV are supported by the European Research Council under ERC-STG-804286 UNISCAMP.

\appendix

\section{Relations among the elliptic $V$ and $V_w$ functions}
\label{appcws}

In this appendix, we spell out a method to determine the rational coefficients $c_{w,\gamma}$ in
the expansion (\ref{mix.71}) of elliptic functions $V_w(\ldots)$ of fixed modular weights in terms of
the $V(\ldots)$ functions in (\ref{newBB.6}). This will be done by exploiting the
$\tau \rightarrow i \infty$ degeneration (\ref{newBB.14}) of the $V(\ldots)$ which
fixes the $c_{w,\gamma}$ in the ansatz (\ref{mix.71}) via
\begin{align}
\lim_{\tau \rightarrow i \infty} \frac{ V_w(1,2,\ldots,n|\tau) }{(2\pi i)^w \sigma_1 \sigma_2 \ldots \sigma_n}
&=  \sum_{\gamma \in S_{n-1}} c_{w,\gamma} \, {\rm PT}^{(1)}(1,\gamma(2,3,\ldots,n))   \label{appt.1} \\
{\rm PT}^{(1)}(1,2,\ldots,n) &= (-1)^{n-1} \lim_{\sigma_- \rightarrow \infty} |\sigma_-|^2 \Big( {\rm PT}(+,n,n{-}1,\ldots,2,1,-) + {\rm cyc}(1,2,\ldots,n) \Big)\notag
\\
&= \frac{1}{\sigma_{12} \sigma_{23} \ldots \sigma_{n-1,n} \sigma_n} + {\rm cyc}(1,2,\ldots,n)\, .
 \label{appt.2}
\end{align}
The combinations ${\rm PT}^{(1)}$ are known as {\rm one-loop Parke--Taylor factors}
from an ambitwistor-string context \cite{Geyer:2015bja},
and we have used $\sigma_+ =0$ in passing to the last line. In order to determine the degeneration of
the left-hand side of (\ref{appt.1}), we expand the elliptic functions
\begin{align}
V_w(1,2,\ldots,n|\tau) &= \sum_{a_1,a_2,\ldots,a_n \geq 0 \atop{a_1+a_2+\ldots+a_n = w}}  g^{(a_1)}_{12}
g^{(a_2)}_{23} \ldots g^{(a_{n-1})}_{n-1,n} g^{(a_n)}_{n,1}  \notag \\
\frac{ \theta_1'(0,\tau) \theta_1(z_{ij}{+}\eta,\tau) }{\theta_1(z_{ij},\tau)
\theta_1(\eta,\tau)} &= \sum_{a=0}^{\infty} \eta^{a-1} g^{(a)}_{ij}
 \label{appt.3}
\end{align}
in terms of the meromorphic Kronecker--Eisenstein coefficients $g^{(a)}_{ij}$ instead
of the $f^{(a)}_{ij}$ in (\ref{shortv}), starting with $g^{(0)}_{ij}=1$ and
$g^{(1)}_{ij}=\partial_{z_i} \log \theta_1(z_{ij},\tau)$. Their $\tau \rightarrow i \infty$ limits
\cite{Enriquez:Emzv, Broedel:2014vla}
\beq
\lim_{\tau \rightarrow i \infty}  g^{(a)}_{jk} = \left\{ \begin{array}{cl}
1&: \ a=0 \\
i\pi \frac{ \sigma_{j}{+}\sigma_k }{\sigma_j {-} \sigma_k} &: \ a=1 \\
-2\zeta_a &: \ a \in 2\mathbb N \\
0 &: \ a \in 2\mathbb N {+}1
\end{array} \right.
 \label{appt.4}
\eeq
generated by (\ref{newBB.13})
ensure that the combination $(2\pi i)^{-w}V_w(1,2,\ldots,n|\tau)$ in (\ref{appt.1})
degenerates to a rational function of the $\sigma_j$, where all factors of $i\pi$ cancel.
Hence, the only $\sigma_j$ dependence of $V_w(\ldots |\tau \rightarrow i \infty)$
occurs via $\lim_{\tau \rightarrow i \infty}  g^{(1)}_{jk}=  i\pi \frac{ \sigma_{j}{+}\sigma_k }{\sigma_j {-} \sigma_k}$.

By applying the degeneration (\ref{appt.4}) to the elliptic function $V_w$ in (\ref{appt.3}), the leftover
challenge in determining the $c_{w,\gamma}$ in (\ref{appt.1}) is to expand the
terms of the form $(\sigma_1 \sigma_2 \ldots \sigma_n)^{-1} \prod_{i=1}^r \frac{ \sigma_{j_i}{+}\sigma_{k_i} }{\sigma_{j_i} {-} \sigma_{k_i}}$ on the left-hand side in terms of Parke--Taylor factors.
For the choices of $\sigma_{j_i} ,\sigma_{k_i} $ that arise from the degeneration of $V_{w\leq n-2}$,
these Parke--Taylor decompositions can be performed by the methods of \cite{He:2017spx}: As
explained in section 3 of the reference, the net effect of the rational factor $\frac{ \sigma_{j_i}{+}\sigma_{k_i} }{\sigma_{j_i} {-} \sigma_{k_i}}$ is to modify the signs of the Parke--Taylor factors on the right-hand side of
\beq
\frac{1}{\sigma_1 \sigma_2 \ldots \sigma_n} = (-1)^{n-1} \lim_{\sigma_- \rightarrow \infty}
|\sigma_-|^2 \sum_{\rho \in S_{n}} {\rm PT}(+,\rho(1,2,\ldots,n),-)
 \label{appt.5}
\eeq
More specifically, with the notation
\beq
{\rm sgn}_{jk}^\rho = \left\{ \begin{array}{rl}
+1 &: \ j \ \text{is on the right of $k$ in} \ \rho(1,2,\ldots,n) \\
-1 &: \ j \ \text{is on the left of $k$ in} \ \rho(1,2,\ldots,n)
\end{array} \right.
 \label{appt.6}
\eeq
the modification of (\ref{appt.5}) by degenerations of $g^{(1)}_{j_i k_i}$ can be written as \cite{He:2017spx}
\begin{align}
&\lim_{\tau \rightarrow i\infty} \frac{g^{(1)}_{j_1 k_1} g^{(1)}_{j_2 k_2}\ldots g^{(1)}_{j_r k_r}  }{(2\pi i)^r \sigma_1 \sigma_2 \ldots \sigma_n} = \frac{(-1)^{n-1}}{2^r} \lim_{\sigma_- \rightarrow \infty}
|\sigma_-|^2  \label{appt.7} \\
&\ \ \ \times \sum_{\rho \in S_{n}} {\rm sgn}_{j_1k_1}^\rho {\rm sgn}_{j_2k_2}^\rho \ldots {\rm sgn}_{j_rk_r}^\rho {\rm PT}(+,\rho(1,2,\ldots,n),-)  \, .
\notag
\end{align}
The contributions of $(2\pi i )^{-2k} g^{(2k)}_{jk}$ in turn degenerate to rational constants by (\ref{appt.4}) which multiply the overall
sum over permutations $\rho$. Hence, (\ref{appt.7}) allows to straightforwardly expand the left-hand side
of (\ref{appt.1}) in terms of Parke--Taylor factors in an $n!$-element basis of ${\rm PT}(+,\ldots,-)$. Matching
the Parke--Taylor coefficients with those on the right-hand side determines the $c_{w,\gamma}$ in (\ref{mix.71}).
It is a special property of the elliptic functions $V_w$ that their degeneration conspires to the cyclic
combinations ${\rm PT}^{(1)}$ in (\ref{appt.2}), i.e.\ that the $(n{-}1)!$ independent $c_{w,\gamma}$
are sufficient to accommodate the $n!$ permutations of ${\rm PT}(+,1,2,\ldots,n,-)$ in $1,2,\ldots,n$.

For instance, the decompositions in (\ref{mix.3}) to (\ref{mix.4}) follow from the special cases of (\ref{appt.7})
\begin{align}
\lim_{\tau \rightarrow i\infty} \frac{V_1(1,2,3|\tau) }{(2\pi i) \sigma_1 \sigma_2  \sigma_3} &= \frac{1}{2} \lim_{\sigma_- \rightarrow \infty} |\sigma_-|^2 \sum_{\rho \in S_3} {\rm PT}(+,\rho(1,2,3),-) ( {\rm sgn}^\rho_{12}
+ {\rm sgn}^\rho_{23} + {\rm sgn}^\rho_{31}) \notag \\
\lim_{\tau \rightarrow i\infty} \frac{V_1(1,2,3,4|\tau) }{(2\pi i) \sigma_1 \sigma_2  \sigma_3 \sigma_4} &= - \frac{1}{2} \lim_{\sigma_- \rightarrow \infty} |\sigma_-|^2 \sum_{\rho \in S_4} {\rm PT}(+,\rho(1,2,3,4),-) ( {\rm sgn}^\rho_{12}
{+} {\rm sgn}^\rho_{23} {+} {\rm sgn}^\rho_{34}{+} {\rm sgn}^\rho_{41}) \notag \\
\lim_{\tau \rightarrow i\infty} \frac{V_2(1,2,3,4|\tau) }{(2\pi i)^2 \sigma_1 \sigma_2  \sigma_3 \sigma_4} &=  {-} \! \! \lim_{\sigma_- \rightarrow \infty} |\sigma_-|^2 \sum_{\rho \in S_4} {\rm PT}(+,\rho(1,2,3,4),-) \Big\{ \frac{1}{3} + \frac{1}{4}  ( {\rm sgn}^\rho_{12}{\rm sgn}^\rho_{34}{+}{\rm sgn}^\rho_{23}{\rm sgn}^\rho_{41}) \notag \\
& \ \ \ \ \ \ \ +\frac{1}{4} (
 {\rm sgn}^\rho_{12}{\rm sgn}^\rho_{23}+
  {\rm sgn}^\rho_{23}{\rm sgn}^\rho_{34}+
   {\rm sgn}^\rho_{34}{\rm sgn}^\rho_{41}+
    {\rm sgn}^\rho_{41}{\rm sgn}^\rho_{12}) \Big\}
    \end{align}
once the right-hand sides are matched with the combinations of one-loop Parke--Taylor
factors ${\rm PT}^{(1)}$ in (\ref{appt.1}) and (\ref{appt.2}).

\section{The initial value $\widehat{B}^{i\infty}_{\eta_2,\eta_3}$ at three points}
\label{app:bhat}

This appendix gathers the three-point initial values
$\widehat{B}^{i\infty}_{\eta_2,\eta_3}(2,3|\rho(2,3))$ for the
$\alpha'$-expansion (\ref{svEMZV.8}) of B-cycle integrals up to and including
weight four. The corresponding
orders of $\widehat{J}^{i\infty}_{\eta_2,\eta_3}(2,3|\rho(2,3))$
relevant to the $\alpha'$-expansion (\ref{svEMZV.80}) of $J$-integrals
are obtained from the single-valued map $(\zeta_2,\zeta_3,\zeta_4)\rightarrow (0,2\zeta_3,0)$.
Since even (odd) orders in the $\eta_j$-expansion integrate to zero on the odd (even) integration
cycles $\mathfrak B(2,3) \pm \mathfrak B(3,2)$, we will separate the two types of contributions in order to
infer $\widehat{B}^{i\infty}_{\eta_2,\eta_3}(2,3|3,2)$ from a relabeling
of $\widehat{B}^{i\infty}_{\eta_2,\eta_3}(2,3|2,3)$.

The expressions in this appendix along with various higher-order terms in the
$s_{ij}$- and $\eta_j$-expansions can be found in an ancillary file within
the arXiv submission of this article, also see (\ref{appbb.4}) for the appearance of $\zeta_{3,5}$.

\subsection{Even orders in $\eta_j$}
\label{app:bhat.1}

The terms of even orders in $\eta_j$ in the three-point initial values are given by
\begin{align}
&\widehat{B}^{i\infty}_{\eta_2,\eta_3}(2,3|2,3) \, \big|_{\rm even} =
\frac{1}{ \eta_{23} \eta_{3} } \Big( \frac{1}{2} +  \frac{\zeta_{2}}{12} (s_{12}^2 + s_{13}^2 + s_{23}^2) +
   \frac{ \zeta_{3}}{24} (s_{12}^3 + s_{13}^3 + s_{23}^3)  \notag \\
   &\ \ \ \ \ \ \ \ +
    \zeta_{4} \Big[ \frac{ 131}{1440}  (s_{12}^4 + s_{13}^4 + s_{23}^4) +
       \frac{5}{144} (s_{12}^2 s_{13}^2 + s_{12}^2 s_{23}^2 + s_{23}^2 s_{13}^2) +
        \frac{1}{18} s_{12} s_{13} s_{23} s_{123} \Big] +\ldots \Big)  \notag \\
& \ \  + \frac{ \eta_{23}}{\eta_{3}}  \Big({-}\zeta_{2} - \frac{\zeta_{3}}{2} s_{12}
 -    \zeta_{4} \Big[ \frac{29  s_{12}^2}{24} + \frac{5  s_{13}^2}{12} +
 \frac{ s_{13} s_{23}}{3} + \frac{5 s_{23}^2}{ 12} \Big] + \ldots \Big)  \notag \\
&\ \ + \frac{ \eta_{3}}{  \eta_{23}} \Big({-}\zeta_{2} - \frac{\zeta_{3}}{2}  s_{23} -
    \zeta_{4} \Big[ \frac{29 s_{23}^2}{24} + \frac{5   s_{12}^2}{12}
    + \frac{s_{12} s_{13}}{3} + \frac{5  s_{13}^2}{  12} \Big] +\ldots \Big)  \notag \\
&\ \ +\Big( \frac{ 3 \zeta_{2}  s_{13}}{s_{123}} + \frac{ \zeta_{3} }{2} s_{13}+
 \frac{ 15 \zeta_{4} s_{13} }{ 4 s_{123} } (s_{12}^2 + s_{12} s_{23} + s_{23}^2)
  - \frac{ 41}{24} \zeta_{4} (s_{12} +s_{23})s_{13} + \frac{ 49 \zeta_{4} s_{13}^2}{24 }
+\ldots \Big) \notag  \\
 &\ \   + \eta_{23} \eta_{3} (5 \zeta_{4})
    - \frac{ \eta_{23}^3 }{\eta_{3} } \zeta_{4}
    - \frac{ \eta_{3}^3 }{\eta_{23} } \zeta_{4} + \ldots
    \label{appbb.2}
\end{align}
and
\begin{align}
&\widehat{B}^{i\infty}_{\eta_2,\eta_3}(2,3|3,2) \, \big|_{\rm even} = \widehat{B}^{i\infty}_{\eta_2,\eta_3}(2,3|2,3) \, \big|_{\rm even}  \, \big|^{\eta_2 \leftrightarrow \eta_3}_{s_{12}\leftrightarrow s_{13}}
\label{appbb.2a}
\end{align}
with MZVs of weight $\geq 5$ in the ellipsis.

\subsection{Odd orders in $\eta_j$}
\label{app:bhat.2}

The terms of odd orders in $\eta_j$ in the three-point initial values are given by
\begin{align}
\widehat{B}^{i\infty}_{\eta_2,\eta_3}(2,3|2,3) \, \big|_{\rm odd} &=
\frac{1}{\eta_{3}} \Big(\frac{1}{s_{12} }+ \frac{ \zeta_{2} s_{123}^2}{6 s_{12}}
+   \zeta_{3} \Big[ \frac{ s_{13} s_{23}}{4} + \frac{ s_{123}^3}{12 s_{12}} \Big]
+\zeta_4 \Big[ \frac{131 s_{123}^4}{720 s_{12}} - \frac{s_{23} s_{123} s_{13}}{20} \Big]+\ldots \Big)  \notag \\
& \ \ + \frac{ 1}{\eta_{23}} \Big( \frac{1}{s_{23}} + \frac{ \zeta_{2} s_{123}^2}{6 s_{23}}
+     \zeta_{3} \Big[  \frac{ s_{12} s_{13}}{4} + \frac{ s_{123}^3}{12 s_{23}} \Big]
+\zeta_4 \Big[  \frac{131 s_{123}^4}{720 s_{23}} - \frac{s_{12} s_{123} s_{13}}{20} \Big] +\ldots \Big)  \notag \\
& \ \
+  \eta_{3} \Big({-}\frac{2 \zeta_{2}}{s_{12}} - \frac{ \zeta_{3} s_{123} }{s_{12}} +\zeta_4 \Big[
{-} \frac{29 s_{123}^2}{12 s_{12}} + \frac{s_{13}}{4} \Big]+ \ldots \Big)  \notag \\
&\ \ +
 \eta_{23} \Big({-} \frac{2 \zeta_{2}}{s_{23}} - \frac{ \zeta_{3} s_{123}}{s_{23}}+\zeta_4 \Big[
 {-} \frac{29 s_{123}^2}{12 s_{23}} +  \frac{s_{13}}{4} \Big] +\ldots \Big)
 \notag \\
 &\ \  + \frac{ \eta_{23}^2 }{ \eta_{3}} \Big({-} \zeta_{3} + \zeta_4 \Big[  \frac{2}{3}  s_{123} +  \frac{ s_{12}}{4} \Big] + \ldots \Big) + \frac{ \eta_{3}^2}{\eta_{23}} \Big( {-} \zeta_{3}+
 \zeta_4 \Big[ \frac{2}{3}  s_{123} + \frac{  s_{23}}{4} \Big]+\ldots \Big) \notag \\
 &\ \ + \eta_{23}^3
 \Big(  {-} \frac{ 2 \zeta_4 }{s_{23}} + \ldots \Big) + \eta_3^3
 \Big( {-} \frac{ 2 \zeta_4}{s_{12}} + \ldots \Big)  + \ldots
 \label{appbb.3}
\end{align}
and
\begin{align}
\widehat{B}^{i\infty}_{\eta_2,\eta_3}(2,3|3,2) \, \big|_{\rm odd} &= - \widehat{B}^{i\infty}_{\eta_2,\eta_3}(2,3|2,3) \, \big|_{\rm odd}  \, \big|^{\eta_2 \leftrightarrow \eta_3}_{s_{12}\leftrightarrow s_{13}}\, ,
 \label{appbb.3b}
\end{align}
again with MZVs of weight $\geq 5$ in the ellipsis.

\section{Examples of single-valued eMZVs}
\label{app:2pt}

\subsection{Systematics at depth one}
\label{app:2pt.1}

The simplest examples (\ref{expls.6}) and (\ref{SVex04}) of single-valued eMZVs
extracted from the two-point integrals (\ref{comp2p}) are special cases of the
SV map (\ref{eesv.1}) on holomorphic iterated Eisenstein integrals. For their depth-one
combinations $\betaBRno{j }{k }$ in (\ref{svEMZV.9}), the SV image $\bsvBRno{j }{k }$ yields the
following Cauchy--Riemann derivatives $\nabla = 2i(\Im \tau)^2 \partial_\tau$ of non-holomorphic Eisenstein series \cite{Gerken:2020yii}
\begin{align}
\betasv{ k-1 \\2k} &= -\frac{[(k{-}1)!]^2}{(2k{-}1)!} {\rm E}_k + \frac{2\zeta_{2k-1}}{(2k{-}1) (4y)^{k-1}}
\notag \\
\betasv{ k-1+m \\ 2k} &=  - \frac{(-4)^m (k{-}1)!\,(k{-}1{-}m)! \, (\pi \nabla)^m {\rm E}_k}{(2k{-}1)!} + \frac{2\zeta_{2k-1}}{(2k{-}1) (4y)^{k-1-m}}
\label{eq:eksol.1} \\
\betasv{ k-1-m \\ 2k} &=  - \frac{ (k{-}1)!\,(k{-}1{-}m)! \, (\pi \overline\nabla)^m {\rm E}_k}{(-4)^m(2k{-}1)! y^{2m}} + \frac{2\zeta_{2k-1}}{(2k{-}1) (4y)^{k-1+m}}\, ,
\notag
\end{align}
also see (\ref{eq:EE}).
While the objects on the right-hand side are expressible in terms of the lattice sums $\cform{ a &0 \\ b &0 } $
in (\ref{newdef.4}) via (\ref{newdef.5}), the $\betaBRno{j }{k }$ are simple combinations of B-cycle eMZVs
$\omega(0^p,k|{-}\frac{1}{\tau})$, where $0^p$ stands for a sequence $0,0,\ldots,0$ of $p$ successive zeros.
On these grounds, $\bsvBRno{j }{k } = {\rm SV} \, \betaBRno{j }{k }$ translates into simple relations such as
\begin{align}
{\rm SV}\,  \omega(0,2k{+}1 | {-}\tfrac{1}{\tau}) &= -\frac{ (\tau{-}\bar \tau)^{2k+1}}{2\pi i}
\cform{ 2k+1 &0 \\ 1 &0 }  \, , &&k\geq 1
 \notag\\
{\rm SV}\,  \omega(0,0,2k{+}2 | {-}\tfrac{1}{\tau}) &= \frac{ (\tau{-}\bar \tau)^{2k+2} }{(2\pi i)^2}
\cform{ 2k+2 &0 \\ 2 &0 }\, , &&k\geq 0
 \label{nicesum.1}\\
{\rm SV}\,\Big( \omega(0,0,0,2k{+}3 | {-}\tfrac{1}{\tau})  - \frac{1}{6} \omega(0,2k{+}3 | {-}\tfrac{1}{\tau}) \Big) &=
 - \frac{ (\tau{-}\bar \tau)^{2k+3} }{(2\pi i)^3}
\cform{ 2k+3 &0 \\ 3 &0 } \, , &&k\geq {-}1
\notag\\
{\rm SV}\,\Big( \omega(0,0,0,0,2k{+}4 | {-}\tfrac{1}{\tau})  - \frac{1}{6} \omega(0,0,2k{+}4 | {-}\tfrac{1}{\tau}) \Big) &=   \frac{ (\tau{-}\bar \tau)^{2k+4} }{(2\pi i)^4}
\cform{ 2k+4 &0 \\ 4 &0 }\, , &&k\geq {-}1
\notag
\end{align}
as well as
\begin{align}
&{\rm SV}\,\Big( \omega(0,0,0,0,0,2k{+}5 | {-}\tfrac{1}{\tau})  - \frac{1}{6} \omega(0,0,0,2k{+}5 | {-}\tfrac{1}{\tau})
+ \frac{7}{360} \omega(0,2k{+}5 | {-}\tfrac{1}{\tau}) \Big)  \notag \\
&\ \ \ \ \ \ \ \
=  - \frac{ (\tau{-}\bar \tau)^{2k+5} }{(2\pi i)^5} \cform{ 2k+5 &0 \\ 5 &0 } \, , \ \ \  \ \ \ k\geq {-}2 \, .
 \label{nicesum.2}
\end{align}
The relative factors of $-\frac{1}{6}$ and $\frac{ 7}{360}$ among the eMZVs of different lengths are
engineered to streamline the iterated-Eisenstein-integral representation \cite{Broedel:2015hia} and generalize as follows
\cite{privNils}
\begin{align}
{\rm SV}\, \sum_{j=0}^{\ell-1} \frac{ B_j }{j!} \omega(0^{\ell-j},2k{+}\ell | {-}\tfrac{1}{\tau})
= (-1)^\ell
\frac{ (\tau{-}\bar \tau)^{2k+\ell} }{(2\pi i)^\ell} \cform{ 2k+\ell &0 \\ \ell &0 } \, , \ \ \  \ \ \ k\geq 1{-}\left \lceil \frac{ \ell}{2} \right \rceil \, , \ \ \ell \geq 1 \, .
 \label{nicesum.3}
\end{align}
In obtaining (\ref{nicesum.1}) and (\ref{nicesum.2}) from (\ref{nicesum.3}), we have
used ${\rm SV} \, \omega(m)  = 0 \ \forall \ m \geq 1$
and the following simplifications of the only eMZV $\omega(0^{\ell-1},2k{+}\ell )$ whose length and weight adds up to an even number~\cite{Broedel:2015hia},
\begin{align}
\omega(0, 0,  2 k{+}1) &= \frac{1}{2} \omega(0, 2 k{+}1)
\notag \\
  \omega(0, 0, 0,   2 k) &=
   \frac{1}{2} \omega(0, 0,   2 k) -  \frac{1}{24} \omega( 2 k)
    \label{nicesum.4}
\\
  \omega(0, 0, 0, 0, 2 k{+}1) &=
   \frac{1}{2} \omega(0, 0, 0, 2 k{+}1) - \frac{1}{24} \omega(0,  2 k{+}1) \, .
\notag
   \end{align}
Based on the dictionary (\ref{newdef.5}) between
lattice sums $\cform{ a &0 \\ b &0 } $ and non-holomorphic Eisenstein series,
one can reformulate (\ref{nicesum.3}) as
\begin{align}
{\rm SV}\, \sum_{j=0}^{\ell-1} \frac{ B_j }{j!} \omega(0^{\ell-j},2k{+}\ell | {-}\tfrac{1}{\tau})
= (-1)^\ell
\frac{ (k{+}\ell{-}1) ! }{ (2k{+}\ell{-}1) ! }  (-4 \pi \nabla)^k {\rm E}_{k+\ell}
 \, , \ \ \  \ \ \ k\geq 0 \, , \ \ \ell \geq 1 \, ,
 \label{nicesum.5}
\end{align}
where $k=0$ needs to be excluded if $\ell=1$, for instance
\begin{align}
{\rm SV}\,  \omega(0,2k{+}1 | {-}\tfrac{1}{\tau}) &= - \frac{k!}{(2k)!} (-4\pi \nabla)^k {\rm E}_{k+1}
 \, , &&\hspace{-1.2cm}k\geq 1
 \label{nicesum.6}\\
{\rm SV}\,  \omega(0,0,2k{+}2 | {-}\tfrac{1}{\tau}) &= \frac{(k{+}1)!}{(2k{+}1)!} (-4\pi \nabla)^k {\rm E}_{k+2}
\, ,&&\hspace{-1.2cm}k\geq 0\, .
\notag
 \end{align}
Moreover, by extending (\ref{nicesum.3}) to $k \rightarrow -k$ and applying the
complex conjugate of (\ref{newdef.5}), we also obtain antiholomorphic Cauchy--Riemann
derivatives as single-valued eMZVs (with $\ell - 2k >0$),
 \begin{align}
{\rm SV}\, \sum_{j=0}^{\ell-1} \frac{ B_j }{j!} \omega(0^{\ell-j},\ell{-}2k | {-}\tfrac{1}{\tau})
&= (-1)^\ell \frac{ (\tau{-}\bar \tau)^{\ell-2k} }{(2\pi i)^\ell} \cform{ \ell-2k &0 \\ \ell &0 }
\label{nicesum.9} \\
&= (-1)^{\ell+k} \frac{ (\ell{-}k{-}1) ! }{(\ell{-}1)!} \frac{ (\pi \overline \nabla)^k {\rm E}_{\ell-k} }{(2y)^{2k}} \, .
 \notag
\end{align}
The simplest examples include
\begin{align}
{\rm SV}\,\Big( \omega(0,0,0,1 | {-}\tfrac{1}{\tau})  - \frac{1}{6} \omega(0,1 | {-}\tfrac{1}{\tau})
\Big) &=  \frac{ \pi \overline \nabla  {\rm E}_2 }{8y^2}
\notag \\
{\rm SV}\,\Big( \omega(0,0,0,0,2 | {-}\tfrac{1}{\tau})  - \frac{1}{6} \omega(0,0,2 | {-}\tfrac{1}{\tau})
\Big) &= - \frac{ \pi \overline \nabla  {\rm E}_3 }{12y^2}
\label{nicesum.10}
\\
{\rm SV}\,\Big( \omega(0,0,0,0,0,1 | {-}\tfrac{1}{\tau})  - \frac{1}{6} \omega(0,0,0,1 | {-}\tfrac{1}{\tau})
+ \frac{7}{360} \omega(0,1 | {-}\tfrac{1}{\tau}) \Big) &= - \frac{ (\pi \overline \nabla)^2 {\rm E}_3 }{192y^4} \notag \\
{\rm SV}\,\Big( \omega(0,0,0,0,0,3 | {-}\tfrac{1}{\tau})  - \frac{1}{6} \omega(0,0,0,3 | {-}\tfrac{1}{\tau})
+ \frac{7}{360} \omega(0,3 | {-}\tfrac{1}{\tau}) \Big) &= \frac{ \pi \overline \nabla {\rm E}_4 }{16y^2} \, , \notag
\end{align}
and the first two lines are equivalent to those in (\ref{expls.46}).

\subsection{Examples with real MGFs at depth two}
\label{app:2pt.2}

By inspecting the $s_{ij}^4$ order of the two-point integrals $B^\tau_{(0)},J^\tau_{(0)}$
and the $s_{ij}^3$ order of $B^\tau_{(2)},J^\tau_{(2)}$, we have obtained the
representations (\ref{expls.7}) of ${\rm E}_{2,2}$ and $\pi \nabla {\rm E}_{2,2}$ as single-valued eMZVs.
One can extract similar representations for ${\rm E}_{2,3},\pi \nabla {\rm E}_{2,3}$ and $(\pi \nabla)^2 {\rm E}_{2,3}$
from the $s_{ij}^5$ order of $B^\tau_{(0)},J^\tau_{(0)}$,
the $s_{ij}^4$ order of $B^\tau_{(2)},J^\tau_{(2)}$ and the
$s_{ij}^3$ order of $B^\tau_{(4)},J^\tau_{(4)}$, respectively:
\begin{align}
{\rm E}_{2,3} &=
  {\rm SV}\,  \Big(
{-}\frac{167}{35}  \omega(0,0,0,0,0,5 | {-}\tfrac{1}{\tau}) + 2 \omega(0,0,0,0,1,4 | {-}\tfrac{1}{\tau})
+ \frac{97}{210} \omega(0,0,0,5 | {-}\tfrac{1}{\tau}) \notag \\
&\ \ \ \ \ \ \ \ \ \ - \frac{1}{3} \omega(0,0,2,3 | {-}\tfrac{1}{\tau}) +
 2 \omega(0,0,0,0,2 | {-}\tfrac{1}{\tau}) \omega(0,3 | {-}\tfrac{1}{\tau}) + \frac{7}{200} \omega(0,5 | {-}\tfrac{1}{\tau})
\Big)
\notag \\
 \pi \nabla {\rm E}_{2,3} &=
{\rm SV} \, \Big(
{-} \frac{1}{12} \omega(0, 3|{-}\tfrac{1}{\tau})^2 + \frac{13}{168} \omega(0, 0, 6|{-}\tfrac{1}{\tau}) +  \omega(0, 3|{-}\tfrac{1}{\tau}) \omega(0, 0, 0, 3|{-}\tfrac{1}{\tau}) \label{apc2.1} \\
&\ \ \ \ \ \ \ \ \ \  - \frac{41}{28} \omega(0, 0, 0, 0, 6|{-}\tfrac{1}{\tau}) + \frac{1}{2} \omega(0, 0, 0, 2, 4|{-}\tfrac{1}{\tau}) \Big) \notag \\
  ( \pi \nabla )^2 {\rm E}_{2,3} &=
 {\rm SV} \, \Big(
\frac{25}{336} \omega(0, 7|{-}\tfrac{1}{\tau}) + \frac{5}{8} \omega(0, 3|{-}\tfrac{1}{\tau}) \omega(0, 0, 4|{-}\tfrac{1}{\tau}) -
 \frac{23}{28} \omega(0, 0, 0, 7|{-}\tfrac{1}{\tau})\notag \\
&\ \ \ \ \ \ \ \ \ \  + \frac{1}{4} \omega(0, 0, 2, 5|{-}\tfrac{1}{\tau}) +
 \frac{1}{8} \omega(0, 0, 4, 3|{-}\tfrac{1}{\tau})  \Big)
 \notag
\end{align}
The corresponding lattice-sum representations \cite{Broedel:2018izr, Gerken:2020aju} and
$\beta^{\rm sv}$ representations \cite{Gerken:2020yii} are given by
\begin{align}
{\rm E}_{2,3} &=
   \bigg( \frac{ \Im \tau}{\pi} \bigg)^5  \Big( \cform{ 3 &1 &1 \\  3 &1 &1 }  - \frac{43}{35} \cform{ 5 &0 \\ 5 &0 }  \Big) \notag \\
&= {-}120 \betasv{2&1\\4&6} - 120 \betasv{3&0\\6&4} + \frac{ 12 \zeta_5}{y}  \betasv{0\\4} +
 80  \zeta_3 \betasv{1\\6} - \frac{ \zeta_5}{36}
+ \frac{ 7 \zeta_7}{16 y^2} - \frac{\zeta_3 \zeta_5}{2 y^3}
\notag \\
 \pi \nabla {\rm E}_{2,3} &=
 \frac{ (\Im \tau)^6 }{\pi^4}  \Big(3\cform{1&1&4\\1&1&2}+2\cform{1&2&3\\1&0&3}-\frac{43}{7}\cform{6&0\\4&0}\Big) \notag \\
 &= 90 \betasv{2&2\\4&6} + 60 \betasv{3&1\\6&4} +
 30 \betasv{4&0\\6&4} \label{apc2.2}\\
 &\ \ \ \ - 60  \zeta_3 \betasv{2\\6} -
 12  \zeta_5 \betasv{0\\4} - \frac{6 \zeta_5}{y}   \betasv{1\\4}
 - \frac{7 \zeta_7}{8 y}+ \frac{ 3 \zeta_3 \zeta_5}{2 y^2}
 \notag \\
  ( \pi \nabla )^2 {\rm E}_{2,3} &=
 \frac{ (\Im \tau)^7 }{\pi^3}   \Big(4\cform{0&2&5\\1&0&2}-4\cform{3&0\\1&0}\cform{4&0\\2&0}-\frac{62}{7}\cform{7&0\\3&0}\Big)
 \notag \\
&= -45 \betasv{2&3\\4&6} - 15 \betasv{3&2\\6&4} -
 30 \betasv{4&1\\6&4}  \notag \\
 &\ \ \ \ + 30  \zeta_3 \betasv{3\\6} +
 12 \zeta_5  \betasv{1\\4} + \frac{3  \zeta_5}{ 2 y}  \betasv{2\\4}
  + \frac{7 \zeta_7}{8} - \frac{3 \zeta_3 \zeta_5}{y} \, .
 \notag
\end{align}

\providecommand{\href}[2]{#2}\begingroup\raggedright\endgroup

\end{document}